\documentclass[twocolumn]{aastex631}
\pdfoutput=1 
\usepackage{amsmath,amssymb,amstext}
\usepackage{gensymb}
\usepackage{appendix}
\usepackage{tabularx}
\usepackage{xspace}
\usepackage{url}

\newcommand{\cl}{\mathrm{cl}}
\newcommand{\tcc}{\,\ifmmode{t_{\mathrm{cc}}}\else $t_{\mathrm{cc}}$\xspace\fi}
\newcommand{\tdrag}{\,\ifmmode{t_{\mathrm{drag}}}\else $t_{\mathrm{drag}}$\xspace\fi}

\submitjournal{ApJ}

\shorttitle{Survival of Magellanic Debris}
\shortauthors{Bustard \& Gronke}

\begin{document}

\title{Radiative Turbulent Mixing Layers and the Survival of Magellanic Debris}

\correspondingauthor{Chad Bustard}
\email{bustard@ucsb.edu}
\author{Chad Bustard}
\affiliation{Kavli Institute for Theoretical Physics, University of California - Santa Barbara, Kohn Hall, Santa Barbara, CA 93107, USA}

\author{Max Gronke}
\altaffiliation{Hubble Fellow}
\affiliation{Department of Physics $\&$ Astronomy, Johns Hopkins University, Bloomberg Center, 3400 N. Charles St., Baltimore, MD 21218, USA}

\begin{abstract}
The Magellanic Stream is sculpted by its infall through the Milky Way’s circumgalactic medium, but the rates and directions of mass, momentum, and energy exchange through the Stream-halo interface are relative unknowns critical for determining the origin and fate of the Stream. Complementary to large-scale simulations of LMC-SMC interactions, we apply new insights derived from idealized, high-resolution ``cloud-crushing'' and radiative turbulent mixing layer simulations to the Leading Arm and Trailing Stream. Contrary to classical expectations of fast cloud breakup, we predict that the Leading Arm and much of the Trailing Stream should be surviving infall and even gaining mass due to strong radiative cooling. Provided a sufficiently supersonic tidal swing-out from the Clouds, the present-day Leading Arm could be a series of high-density clumps in the cooling tail behind the progenitor cloud. We back up our analytic framework with a suite of converged wind-tunnel simulations, finding that previous results on cloud survival and mass growth can be extended to high Mach number ($\mathcal{M}$) flows with a modified drag time $t_{drag} \propto 1 + \mathcal{M}$ and longer growth time. We also simulate the Trailing Stream; we find that the growth time is long ($\sim$ Gyrs) compared to the infall time, and approximate H$\alpha$ emission is low on average ($\sim$few mR) but can be up to tens of mR in bright spots. Our findings also have broader extragalactic implications for e.g. galactic winds, which we discuss.

\end{abstract}

\keywords{hydrodynamics, Galaxy: halo, Magellanic Clouds, turbulence}

\section{Introduction}

The Magellanic Stream is the dominant multiphase gas structure in the Milky Way halo. As the Large and Small Magellanic Clouds (LMC and SMC, respectively) fall through our Milky Way's circumgalactic medium (CGM), their tidal forces strip loosely bound gas from their disks, forming part of the massive Magellanic Stream \citep{ElenaReview2016}. The Stream comprises the Trailing Stream, which contains at least $2 \times 10^{9} M_{\odot}$ of gas (both neutral and ionized, \citealt{Fox2014, Barger2017}) as well as the Leading Arm extending tens of kpc out in front of the Clouds and with HI mass $\sim 3 \times 10^{7} M_{\odot}$ \citep{Bruns2005}. The Magellanic System, if it collides with the Milky Way disk, promises to drastically transform the structure of our galaxy and replenish our Galaxy's fuel for star formation \citep{Cautun2019}.  

The Magellanic System, though, is not only a future life-source for our quickly gas-consuming galaxy but also a useful case-study on galaxy dynamics and evolution more broadly. The LMC and SMC, at only $\sim 50$ and $\sim 60$ kpc away, respectively, provide a birds-eye view of stellar feedback and gravitational and hydrodynamic interactions that is unmatched by observations of other, more distant galaxies \citep{Mathewson1974, MAthewson1977,Bruns2005, Stanimirovic2008, 2009ApJS..181..398M, Nidever2010, PutmanReview2012, Fox2014, Barger2017}. Consequently, matching the Magellanic System's present-day morphology, including the Leading Arm and Trailing Stream, has been a long-standing goal for simulations of galaxy evolution \citep{Gardiner1994, Connors2006, Besla2007, Besla2012, Hammer2015THECLOUDS, Pardy2018, TepperGarcia2019, Wang2019, Lucchini2020}. 

While tidal interactions between the LMC and SMC can account for much of the Stream's shape and size, enhanced further by an ionized Magellanic corona \citep{Lucchini2020}, it's also clear from the fragmented structure of the Leading Arm and Trailing Stream \citep{Bruns2005}, as well as the compressed leading edge of the LMC \citep{Salem2015RAMMEDIUM}, that the Clouds and their debris are sculpted by the surrounding Milky Way corona. Some observations have attempted to probe this in focused detail. \cite{Nigra2012} use pointed observations of an isolated cloud in the northern Magellanic Stream to probe the structure and kinematics of the Stream-halo interface. They find that the cloud properties are best explained by a turbulent mixing layer rather than a conduction dominated layer, and they argue that cooling in the mixing layer should prolong the cloud lifetime by at least a factor of 2. Similarly, \citet{Putman2011} and \citet{For2013, For2014} analyze the discrete HI clouds and filaments of the Leading Arm and Trailing Stream. In both environments, there are a large number of head-tail clouds with generally random pointing directions, suggesting a strong role played by turbulence. The number of head-tail clouds decreases along the length of the Stream, consistent with a decrease in ram pressure and hence a distance gradient along the Stream. There is similarly strong evidence for a distance variation in the Leading Arm (see e.g. Figure 5 of \citealt{AntwiDanso2020} for a compilation of data points), with some clumps believed to be within $\sim 20$ kpc of the Milky Way disk \citep{McClureGriffiths2008, venzmer2012, For2016, Richter2018} in spite of ram pressure from the hot Milky Way halo.

While these observational studies clearly demonstrate the presence of turbulent mixing layers between the Magellanic Stream and the surrounding hot halo, the rate and direction of the mass and momentum flux through these mixing layers (e.g. whether the Stream is evaporating or growing in mass) is a major uncertainty and a key determinant of the origin and fate of the Magellanic Stream. To that end, we aim to contribute an initial exploration here, motivated by significant theoretical insights gained from high-resolution simulations of cold gas structures embedded in hot, supersonic flows (e.g. \citealt{Begelman1990, Klein1994, Scannapieco2015, Armillotta2017, Gronke2018, Mandelker2020a}). A key finding is that, in the presence of efficient radiative cooling, cold gas structures can survive for much longer than the classical gas destruction time and even \emph{gain} mass by mixing and cooling the surrounding hot medium \citep{Gronke2018, Ji2019Mixing, Gronke2020a, Mandelker2020a, Fielding2020, Li2020, Sparre2020, Banda2020, kanjilal2021, Tan2021, Abruzzo2021,Farber2021}. While still a developing field, these insights have wide-ranging implications for the survival of cold clouds embedded in hot galactic winds, as well as the filaments and high-velocity cloud (HVC) complexes that comprise the Magellanic Stream.

This paper is outlined as follows. In \S \ref{sec:background}, we review some previous simulations of the interaction between the Milky Way halo and the Magellanic System, and we outline the theoretical framework for turbulent mixing. In \S \ref{sec:LATheory}, we apply this framework to the Leading Arm. As part of this effort, we present new simulations of a toy Leading Arm cylinder in a high Mach number flow (\S \ref{sec:LASimulations}); this technical piece to extend the cloud crushing/survival framework beyond mildly transonic flows is necessary given the possibly high velocity achieved by the Leading Arm upon swing-out from the Clouds. In \S \ref{sec:trailingstream}, we consider the Trailing Stream and estimate its prospects for survival and mass growth during infall. We discuss our results in \S \ref{sec:discussion}, including a discussion of our work's broader implications (\S \ref{sec:implications}) and the resolution criterion for mass flux convergence (\S \ref{sec:resolutionGuidelines}). We conclude in \S \ref{sec:conclusions}.

\section{Background}
\label{sec:background}
\subsection{Previous Studies of the Magellanic - Milky Way Corona Interaction}

To-date, only a few computational studies have probed the influence of the Milky Way halo on the Magellanic System, focusing on the dynamical role of ram pressure $P_{ram} = \rho v^{2}$ due to the Clouds' infall through the background Milky Way halo. While strong ram pressure stripping of outer, loosely bound LMC and SMC gas has been theorized by many authors as the formation mechanism for the Trailing Stream (see \citealt{ElenaReview2016} and references therein), it alone doesn't explain the existence of the Magellanic Bridge connecting the two Clouds (a sign of past tidal stripping), and the mass stripped from simulated Clouds falls far short of the observed Stream mass \citep{Salem2015RAMMEDIUM} unless the Milky Way halo is significantly denser than expected out to the virial radius \citep{Wang2019}.

Ram pressure also affects the formation and survival of the Leading Arm. Without including a Milky Way corona, simulations can more-or-less reproduce the shapes and positions of the entire Magellanic System \citep{Pardy2018}; however, when otherwise identical simulations include a gaseous Milky Way halo, the Leading Arm disappears (\citealt{TepperGarcia2019}; see also \citealt{Wang2019, Lucchini2020, Williamson2021}). A plausible alternative is that the Leading Arm is the remnant of one or more fore-runner galaxies torn apart by ram pressure \citep{Yang2014, Hammer2015THECLOUDS, TepperGarcia2019}. The recent discovery of dwarf galaxies belonging to a Magellanic Group \citep{2008ApJ...686L..61D, Koposov2015, 2016ApJ...833L...5D, Pardy2020, Patel2020}, of which the LMC and SMC are simply the largest, lend credence to this idea, but as pointed out in \cite{TepperGarcia2019}, there isn't yet a clear candidate for the Leading Arm progenitor galaxy.

There are two possible, physical reasons why the present-day Leading Arm was not seeded by tidal stripping between the Magellanic Clouds: 1) The drag force on the Leading Arm, during it's swing out from the Clouds, was too strong, thereby blowing the gas back into the Trailing Stream. Indeed, \cite{TepperGarcia2019} note that, in their runs without Leading Arm formation, the gas mass in the Trailing Stream increases relative to simulations with a clearly formed Leading Arm. 2) The Leading Arm swung out in front of the clouds but was disintegrated during infall. This scenario would seem plausible given the classical expectation: a cloud subjected to strong ram pressure would break apart within a few ``cloud-crushing" times, $\tcc \sim \chi^{1/2} r_{\cl}/v_{\rm wind}$ \citep{Klein1994} where $v_{\rm wind}$ is the velocity of the cloud relative to the background (in this case, the infall velocity), r$_{\cl}$ is the cloud radius, and $\chi$ is the overdensity of the cloud relative to the hot background. For a typical size of 1 kpc and using 250 km/s, roughly the orbital velocity of the LMC, as the cloud infall velocity, this timescale is $< 50$ Myrs. Unless the Leading Arm formed very recently, as suggested by e.g. metallicity estimates that match that of the SMC within the last 200 Myrs \citep{Richter2018}, this destruction timescale is too short to allow for Leading Arm survival.  

In this paper, we will address both the drag time and cloud destruction concerns; however, we do so using analytic arguments that scan a large parameter space and are motivated by sufficiently resolved simulations of cloud survival. In short, we find that \emph{reasonably massive Leading Arm complexes should quite easily survive infall (and possibly grow in mass)} due to radiative cooling at the turbulent mixing layer between the cold cloud and hot CGM. The only hurdle to the Leading Arm's Magellanic origin, then, is the drag force, which can be overcome provided the Leading Arm's swing-out speed due to tidal interactions was fast enough. 
Thus far, studies calibrating the swing-out systematically have not included a hot halo. 
Future, more comprehensive simulations are needed to explore whether such a large swing-out is possible while simultaneously re-creating the morphology of the Trailing Stream, but until then, we can’t definitively rule out a Magellanic origin for the Leading Arm in favor of plausible but alternative fore-runner theories. Now, we outline the framework that led us to these conclusions.

\subsection{Radiative Turbulent Mixing}
The evolution of multiphase gas in galaxy halos is governed by mass, momentum, and energy transfer at the interfaces between different phases. Interface properties are set by a combination of thermal conduction, magnetic fields, cosmic ray pressurization and heating, and shear flows that drive the Kelvin-Helmholtz instability and subsequent fluid mixing. While mixing layers are ubiquitous in astrophysics and are known to play important roles in the ISM, CGM, and IGM, most studies until recently have focused on the adiabatic Kelvin-Helmholtz instability, largely neglecting the important role played by radiative cooling, which can be very efficient in intermediate temperature gas, $T \sim 10^{5}$ K, populating interface regions. Driven largely by attempts to explain the observed abundance of cold, $T \sim 10^{4}$ K, gas embedded in hot galactic outflows, an explosion of papers have been published in recent years revisiting the interplay between turbulence and radiative cooling. Here, we will briefly review the most robust results from these studies before applying them to the Magellanic System.

Let's envision a cold $\sim 10^{4}$ K cloud with radius $r_{cl}$ moving through a hot $\sim 10^{6}$ K background at velocity $v_{wind}$. Let the overdensity of this cloud relative to the background be $\chi$, where $\chi \sim 100$ corresponds to pressure equilibrium between the cloud and background and will be the fiducial value throughout this paper\footnote{This is not guaranteed to be true for multiphase structures with long sound crossing times relative to other timescales, but it appears to be true for at least HVCs in the direction of the Trailing Stream \citep{Fox2005}}. A main quantity of interest is the ratio of shearing time (or cloud crushing time) to cooling time of the intermediate temperature \emph{mixed} gas at the interface between hot and cold phases. This cooling time depends on $T_{mix} \approx \sqrt{T_{cl}T_{h}}$ and $n_{mix} \approx \sqrt{n_{cl}n_{h}}$ \citep{Begelman1990} where $T_{cl}$ and $T_{h}$ are the cloud and hot halo temperatures, and $n_{cl}$ and $n_{h}$ are the cloud and hot halo number densities. If this cooling time, which is approximately an order of magnitude longer than the cooling time of the cold gas, is shorter than the cloud crushing time ($t_{cc} = \chi^{1/2} r_{cl}/v_{wind}$), then the mixed gas can cool as quickly as it is produced by mixing, hence increasing the cold gas mass of the cloud. The ratio of cooling time to destruction time can be re-arranged to give a critical cloud radius above which the destructive effects of shear instabilities are offset by cooling and subsequent mass flux onto the cold cloud \citep{Gronke2018}:

\begin{equation}
\label{rclEqn}
    r_{\rm cl,crit} \approx 20 \rm pc \frac{T_{cl,4}^{5/2} \mathcal{M}} {P_{h,2} \Lambda_{\rm mix, -21.4}} \frac{\chi}{100}
\end{equation}
The Mach number, $\mathcal{M}$, is defined as the ratio of velocity of the cold cloud (relative to the hot background) to the \emph{hot} gas sound speed, and other quantities are normalized by their fiducial values: the cloud temperature $T_{cl,4} = T_{cl}/(10^{4}$ K), the pressure $P_{h,2} = n_h T_h / (10^{2} \rm cm^{-3} K)$, and the cooling function $\Lambda_{\rm mix, -21.4} = \Lambda_{\rm mix}/(10^{-21.4}$ erg cm$^{3}$ s$^{-1}$) where $\Lambda \sim 10^{-21.4}$ is appropriate for solar metallicity gas at $T_{mix} \sim 10^{5}$ K. 

Clouds with $r_{cl} > r_{cl,crit}$ may undergo significant morphological changes, but the amount of cold gas should persist and grow. This process is, however, difficult to capture in the low-resolution CGM of galaxy evolution simulations. High-resolution cloud-crushing simulations demonstrate that mass growth, even for clouds with radii $r_{\cl} \gg r_{\rm cl,crit}$, is only converged when the resolution is $\gtrapprox 8$ cells per cloud radius. Unless refinement criteria specifically target the CGM \citep{Hummels2019, Peeples2019, vandeVoort2019, Nelson2020}, numerical diffusion dominates and multiphase gas is over-mixed. \emph{To our knowledge, all currently published simulations of LMC-SMC interactions lack sufficient resolution to capture cloud survival and converged mass growth.} This is also true of isolated LMC ram pressure stripping simulations \citep{Salem2015RAMMEDIUM, Bustard2018, Bustard2020}. We will discuss the implications of this further in \S \ref{sec:resolutionGuidelines}. 

The instantaneous rate at which the cold gas mass increases is given by the growth time (or entrainment time)
\begin{equation}
\label{tgrowEqn}
    t_{grow} = \frac{m}{\dot{m}} \approx \frac{\rho_{cl}r_{cl}A}{\rho_{h}v_{mix}A_{eff}}
\end{equation}
where the numerator is the cloud mass (or for streams, the mass per unit length) and the denominator is the change in mass per unit time, which depends on the mixing velocity, $v_{\rm mix}$, and $A_{eff}$, the total effective interface area where mixing occurs. From here on, we define a fudge-factor $f_{sa}$ such that $A_{eff} = f_{sa} A$ to account for any changes in surface area. For a disrupting cloud in the pre-entrainment phase, the surface area steadily increases at a rate $\propto 1/t_{cc}$ as the initial cloud is redistributed into a tail of debris, increasing the overall mixing. Therefore, the increase in cold gas mass is tied to the increase in surface area resulting from cloud stretching and disruption. 

For cold streams, the mechanism for mass growth is slightly different. For an infinitely long cylindrical stream (as modeled in e.g. \citealt{Mandelker2020a, Mandelker2020b}), there is no elongation along the stream that can increase the surface area ($A = A_{eff}$, $f_{sa} = 1$). Instead, the growth in mass per unit length is tied to stream pulsation in the radial direction. Analogous to the survival criterion for clouds (Equation \ref{rclEqn}) defined by $t_{cool,mix} = t_{cc}$, survival of streams is determined by the ratio of $t_{cool,mix}$ and an appropriate disruption time. Detailed studies define the stream disruption time as $t_{\rm shear} = r_{\rm cyl}/\alpha_{\rm cyl}v_{\rm cyl}$ where $r_{\rm cyl}$ and $v_{\rm cyl}$ are the radius and velocity of the infalling stream, and $\alpha_{\rm cyl}$ is a dimensionless quantity parameterizing the growth rate of the mixing layer due to eddy mergers (see \citealt{Padnos2018, Aung2019, Mandelker2019} for more background). From \cite{Mandelker2020a}, the stream survival criterion is 

\begin{equation}
    r_{\rm cyl,crit} \approx 20 \rm pc \left(\frac{\chi}{100}\right)^{3/2} \frac{\alpha_{\rm cyl, 0.1} T_{cyl,4}^{7/2} \mathcal{M}}{P_{h,2} \Lambda_{mix,-21.4}}
    \label{rcyleqn}
\end{equation}
For a $T \sim 10^{4}$ K, $\chi \sim 100$ stream in pressure equilibrium with the hot halo, Equation \ref{rcyleqn} gives an identical result to Equation \ref{rclEqn} when $\alpha_{\rm cyl} = 0.1$, which is approximately satisfied for stream overdensities of $\chi \approx 100$ and transonic or supersonic infall speeds \citep{Mandelker2019, Mandelker2020a}.

The growth time as defined in Equation \ref{tgrowEqn} is applicable to both clouds and streams and can be combined with an equation for $v_{\rm mix}$ (Equation 8 from \citealt{Gronke2020a}) to give 
\begin{equation}
\label{tgrowEqn2}
    t_{grow} \approx \frac{\chi t_{sc}}{f_{sa}} \left(\frac{t_{cool,cold}}{t_{sc,cold}}\right)^{1/4}
\end{equation}
Note that the dependence on $t_{cool,cold}/t_{sc,cold}$ is weak, meaning the growth time is largely insensitive to the value of the cooling curve at $10^{4}$ K. Therefore, we expect (and simulations show) that these equations hold even when the metallicity is very low or when photoionization from the UV background is included (Section \ref{sec:Stream_sims} and \citealp{Ji2019Mixing, Mandelker2020a}).  In our mildly transonic simulations (\S \ref{sec:LASimulations}) and consistent with previous literature, we find growth times as low as $\sim 20 t_{sc}$ in the early phases of mass growth, while they asymptote to values of order $\chi t_{sc} = 100 t_{sc}$ at late times when the cold gas is co-moving with the hot background.

Equations \ref{rclEqn} and \ref{tgrowEqn2} describe the survival parameter space and rate of growth, respectively. Of course, ram pressure and interface mixing both impart a drag force, as well. The drag time for a cloud of radius $r_{cl}$ is $t_{drag} \sim \chi r_{cl}/v_{wind}$. Under the assumption that the cloud maintains its radius, $t_{drag} \sim \chi^{1/2} t_{cc}$, which is approximately true for a transonic headwind; however, supersonic flows change the cloud geometry, and in the strong cooling regime, momentum transfer also occurs due to mixing. What we find in \S \ref{sec:LASimulations} is an increased drag time for higher Mach number flows $\propto 1+\mathcal{M}$ (see also \citealt{Scannapieco2015, Gronke2020a}) and an additional dependence on cooling efficiency $t_{cool,mix}/t_{cc}$ (see Figure \ref{fig:M45_varytcc}) that we subsume in a parameter $\alpha$: 
\begin{equation}
    t_{\rm drag} \sim \alpha (1 + \mathcal{M}) \chi^{1/2} t_{cc}
\end{equation}

For our fiducial simulations, we find $\alpha \sim 0.2$, which appears to be the strong cooling limit; $\alpha$ is larger when $t_{\rm cool,mix}/t_{\rm cc}$ is larger. This is consistent with \cite{kanjilal2021}, which found that the simulated drag time was $\approx 0.4 \chi^{1/2} t_{cc}$ for their $\mathcal{M} \sim 1$ simulations. This may also depend on the form of the cooling curve near $T \sim T_{cl}$ \citep{Abruzzo2021}, though we don't explicitly test this.

\subsection{Equations for the Survival and Growth of Cold Gas}

For applications to the Milky Way CGM, maybe the most powerful piece of this formulism is that, if we reasonably assume the fiducial parameters of Equation \ref{rclEqn} ($\chi = 100$, $T_{cl} = 10^{4}$ K, $T_{h} = 10^{6}$ K), we can reduce the equations for critical radius, growth time, and drag time to just three free parameters: the density of the hot Milky Way halo, $n_{h}$, the cold gas column density, $N_{cl} \sim 2 r_{cl} n_{cl} = 200 r_{cl} n_{h}$ and the Mach number, $\mathcal{M}$. With these fiducial parameters, the critical radius criteria from Equations \ref{rclEqn} and \ref{rcyleqn} give identical results for streams and clouds and can be transformed into a critical column density criterion:

\begin{equation}
\begin{split}
    N_{crit} & \sim 1.2 \times 10^{18} \mathcal{M} \quad \rm cm^{-2} \qquad (Z = Z_{\odot}) \\
    N_{crit} & \sim 4.8 \times 10^{18} \mathcal{M} \quad \rm cm^{-2} \qquad (Z = 0.1 Z_{\odot})
\end{split}
\label{eqn:Ncrit}
\end{equation}
and it can be shown that $N_{cl}/N_{crit} = t_{cc}/t_{\rm cool,mix}$. The difference between the two critical column densities above is our estimate for $\Lambda (T_{mix})$, which we take to be $10^{-21.4}$ erg cm$^{3}$ s$^{-1}$ for $Z = Z_{\odot}$ and $10^{-22}$ erg cm$^{3}$ s$^{-1}$ for $Z = 0.1 Z_{\odot}$. The latter value of 0.1 $Z_{\odot}$, which we take as our fiducial value for all analytic estimates in this paper, is more appropriate for both the Leading Arm, with inferred metallicities generally in the 0.05-0.3 $Z_{\odot}$ range \citep{Fox2018, Richter2018}, and the main SMC filament of the Trailing Stream with $Z \sim 0.1 Z_{\odot}$ \citep{Fox2013_COS1}. The LMC filament is more metal-enriched with $Z \sim 0.5 Z_{\odot}$ \citep{Richter2013_COS2}, which would decrease $N_{\rm crit}$ by a factor of $\sim 2$ compared to our estimate for $Z = 0.1 Z_{\odot}$.

We can subsequently write the cloud crushing time, drag time, and instantaneous growth time\footnote{As written, there is no Mach number dependence here, but our simulations \emph{do} show evidence for a $\sim \mathcal{M}^{3}$ dependence of $t_{grow}$ that warrants more detailed study (cf. Fig.~\ref{fig:tgrow_area}). We neglect it here because, as we'll see, $t_{grow}$ is already long compared to the lookback time for plausible Leading Arm formation scenarios.} as 

\begin{equation}
    t_{cc} \sim \frac{10 r_{cl}}{v_{cl}} \sim 14.4 \rm Myrs \left(\frac{N_{19}}{\mathcal{M} n_{-4}}\right)
\end{equation}

\begin{equation}
\begin{split}
    t_{drag} & \sim \alpha (1+\mathcal{M})\chi^{1/2} t_{cc} \\
    & \sim 28.8 \rm Myrs (\alpha_{0.2}) \left(\frac{N_{19}}{n_{-4}}\right) \left(1+\frac{1}{\mathcal{M}}\right) 
    \label{eqn:tdrag}
\end{split}
\end{equation}

\begin{equation}
\begin{split}
    t_{grow} &= \frac{m}{\dot{m}} \sim \frac{\chi t_{sc}}{f_{sa}} \left(\frac{t_{cool,s}}{t_{sc}}\right)^{1/4} \\
    & \sim 739 \rm Myrs \left(\frac{N_{19}^{3/4}}{n_{-4} \Lambda_{-23}^{1/4} f_{sa}}\right)
\end{split}
\label{eqn:tgrow}
\end{equation}

where we use $t_{sc} \sim \mathcal{M}t_{cc} \sim r_{cl}/c_{s,cl}$ and we define the normalized cloud column density $N_{19} = N_{cl}/(10^{19}$ cm$^{-2}$), background halo density $n_{-4} =  n_{h}/(10^{-4}$ cm$^{-3}$), $\alpha_{0.2} = \alpha / 0.2$ and for an estimate of the stream cooling time, the cooling function $\Lambda_{-23} = \Lambda (1.5 T_{s})/10^{-23}$ erg cm$^{3}$ s$^{-1}$, where $T_{s}$ is the cold stream temperature. We fiducially evaluate the stream cooling time at $T \sim 1.5 T_{s}$, approximately where the cooling time is minimized. Note that the growth time has only a very weak dependence $\propto \Lambda^{-1/4}$ on this choice, and therefore, Equations \ref{eqn:tdrag} and \ref{eqn:tgrow} are largely independent of metallicity, though the onset of growth may be affected by the low-temperature form of the cooling curve \citep{Abruzzo2021}.

\begin{figure*}
\centering
\includegraphics[width=0.85\textwidth]{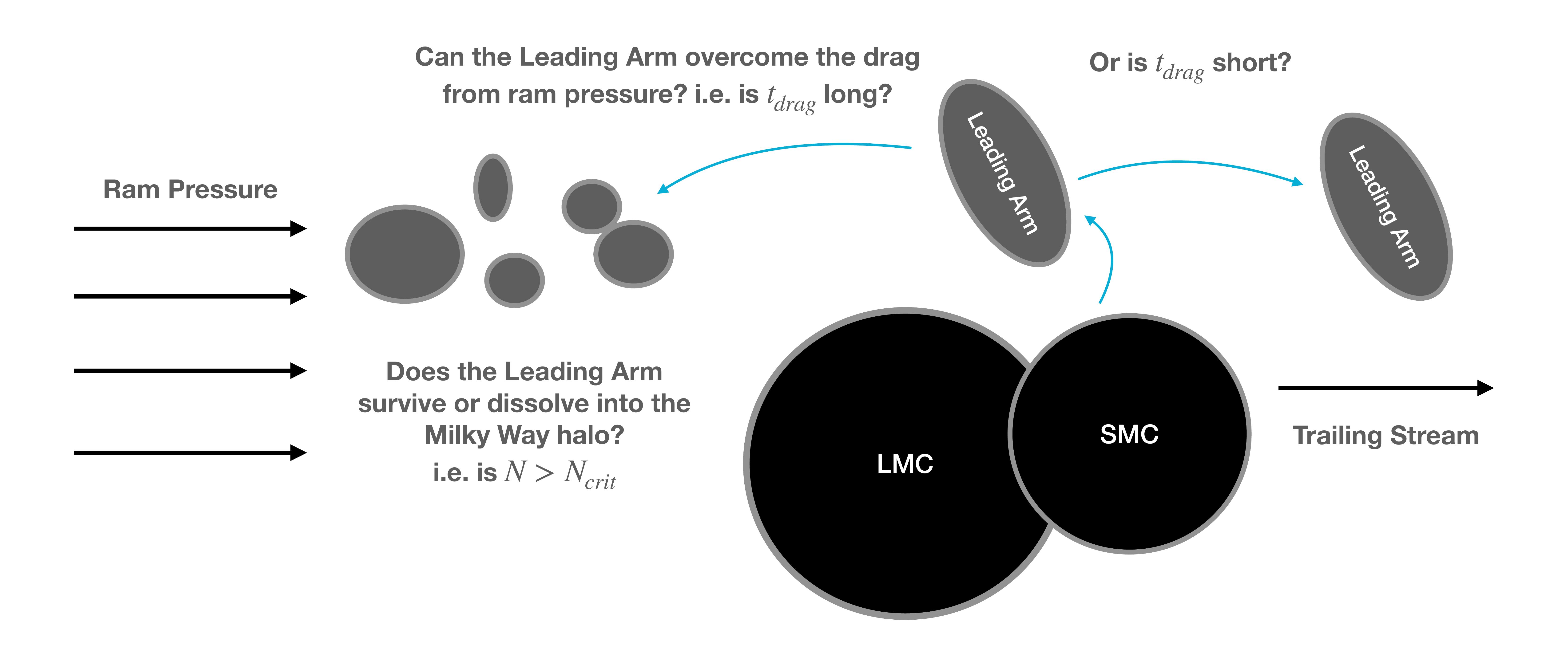}
\caption{Sketch of our simplified theoretical experiment, in which we envision the Leading Arm swinging out in front of the LMC and SMC at a velocity exceeding the infall velocity of the Clouds. To recreate the present-day position of the Leading Arm, the Leading Arm must survive infall through the Milky Way halo (with survival criterion $N > N_{crit}$) and maintain its position ahead of the Clouds while being pushed back by ram pressure (i.e. the drag time $t_{drag}$ must be sufficiently long).}
\label{fig:LA_Sketch}
\end{figure*}

\section{The Leading Arm: Analytic Estimates}
\label{sec:LATheory}

Figure \ref{fig:LA_Sketch} sketches the two main questions we have concerning the origin and evolution of the Leading Arm:

\begin{enumerate}
    \item Can the Leading Arm survive the (possibly quite strong) ram pressure upon swing-out from the Magellanic Clouds in the outer Milky Way halo? Or will the Leading Arm break apart and irreversibly mix into the hot halo medium? 
    
    \item Can the Leading Arm overcome drag from the background Milky Way halo and separate itself from the LMC/SMC system to a reasonably large spatial extent to put it near it's present-day position in front of the Clouds?
\end{enumerate}

\begin{figure}
\centering
\includegraphics[width=0.48\textwidth]{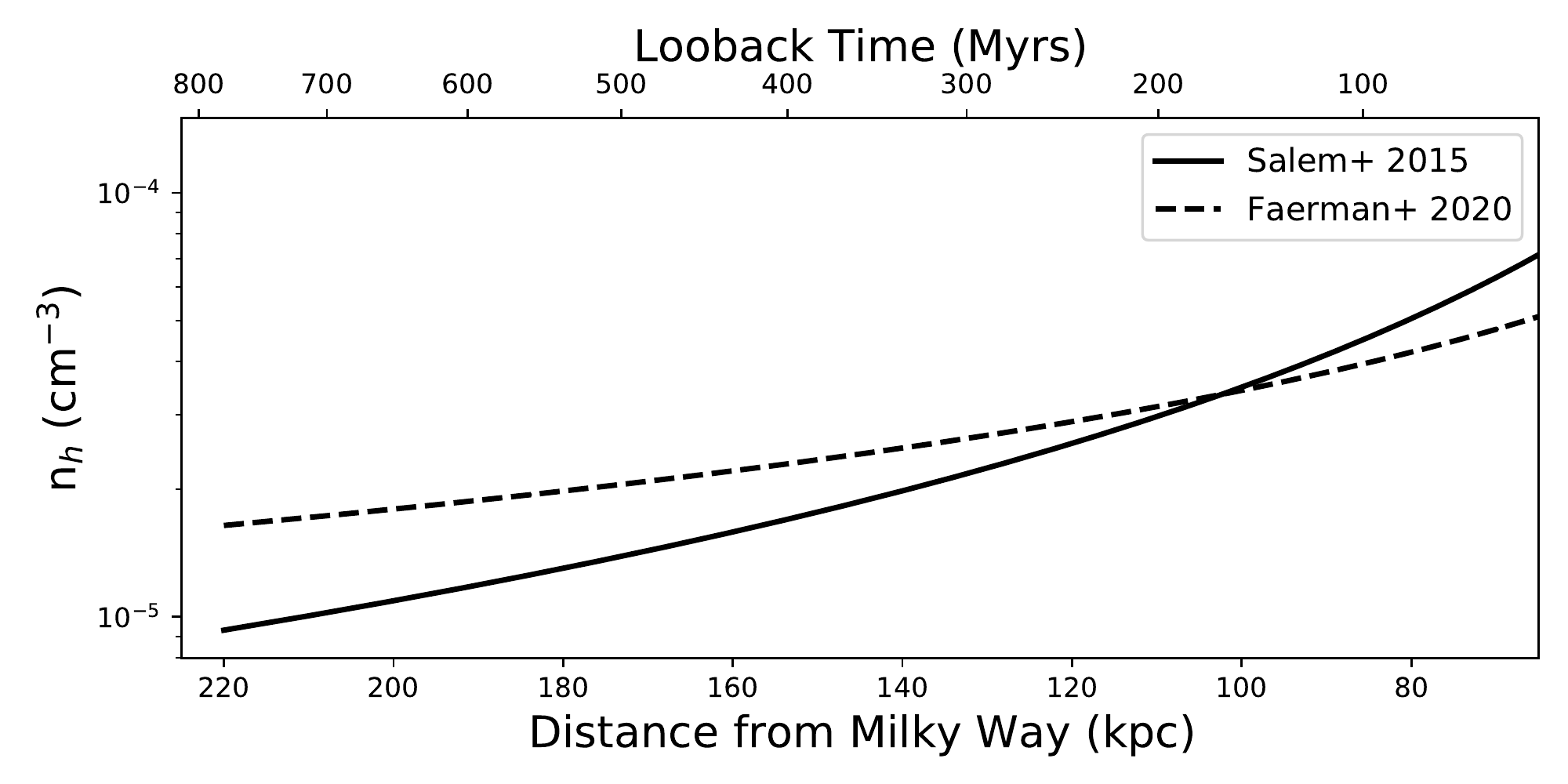}
\includegraphics[width=0.48\textwidth]{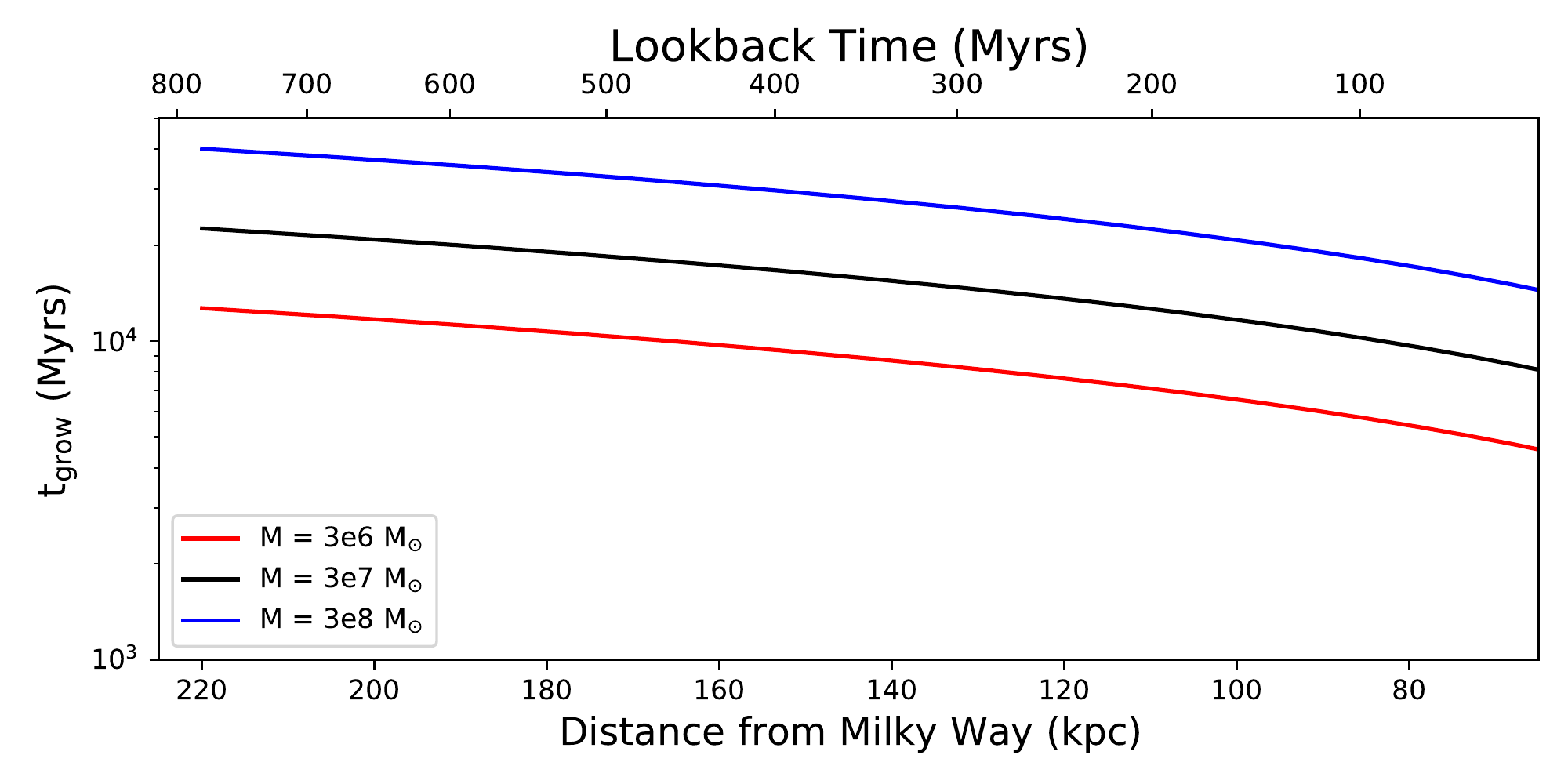}
\includegraphics[width=0.48\textwidth]{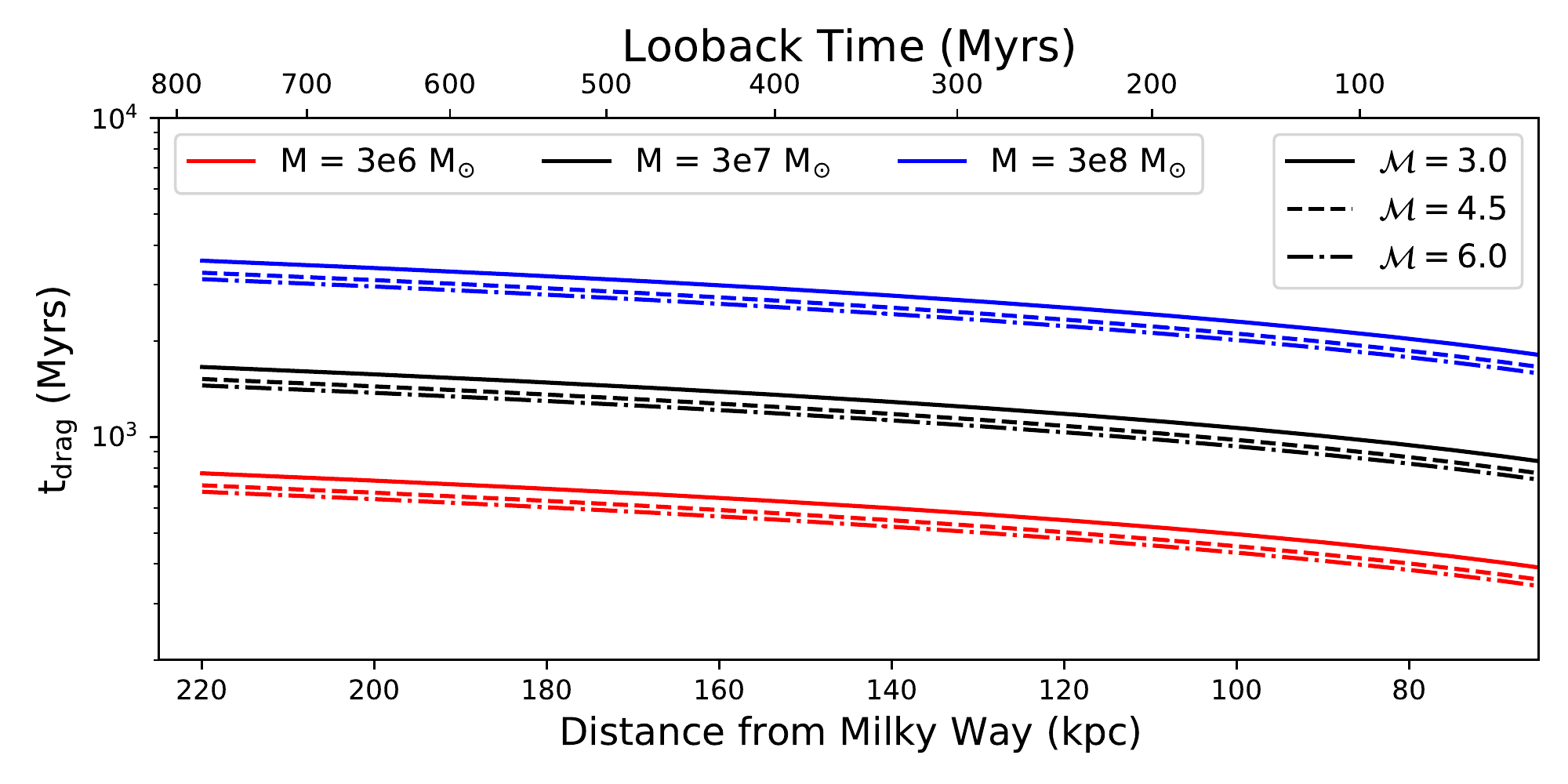}
\includegraphics[width=0.48\textwidth]{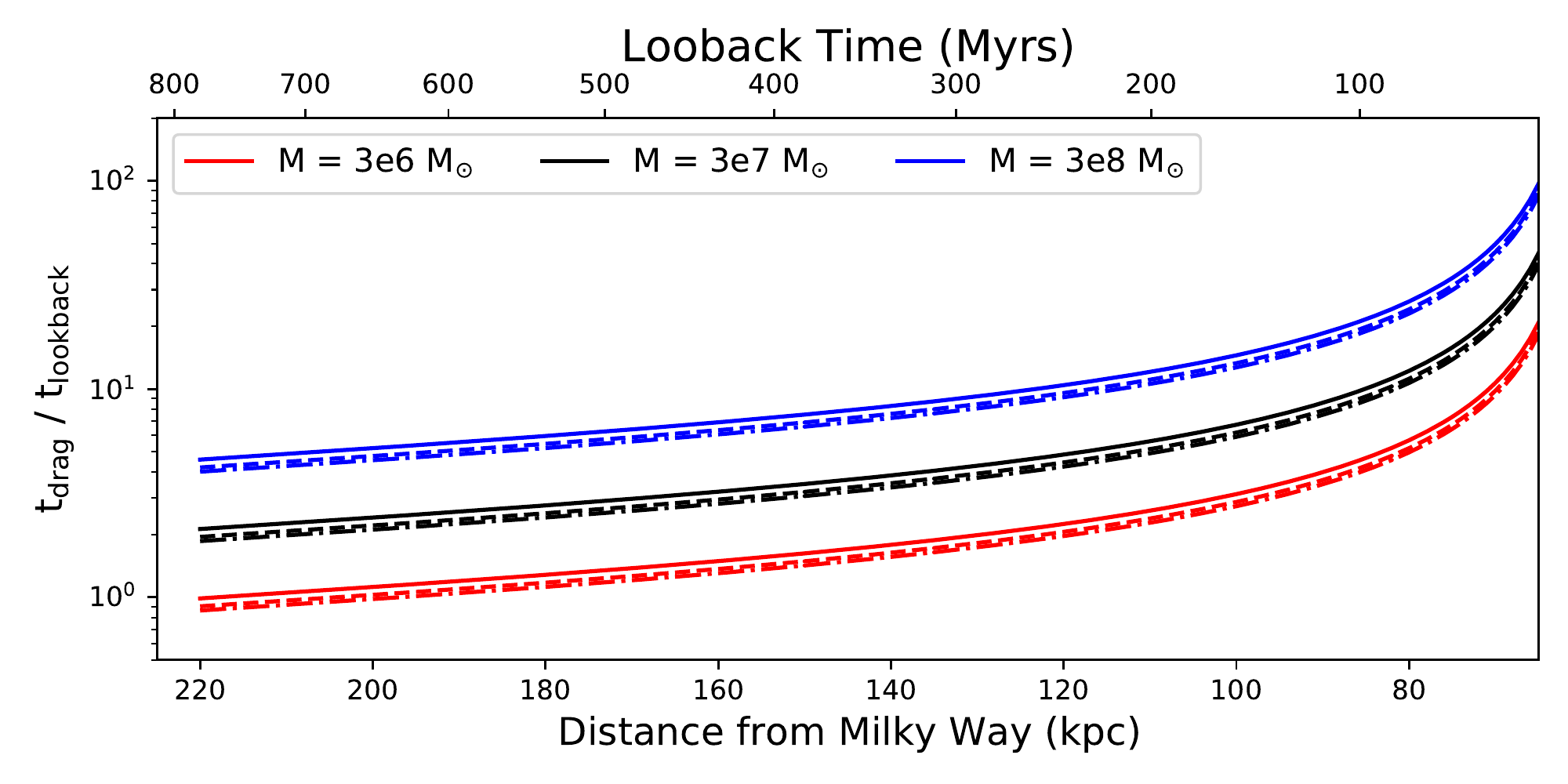}
\caption{From top to bottom row, \emph{First:} Milky Way halo number density vs distance from \cite{Salem2015RAMMEDIUM} and \cite{Faerman2020}. To relate the lookback time to approximate distance from the Milky Way, we use the LMC orbit provided by \cite{Kallivayalil2013}, more specifically a polynomial fit as a function of time as was used in \cite{Bustard2020}. \emph{Second:} Growth time (Equation \ref{eqn:tgrow}) for Leading Arms of constant masses $3 \times 10^{6} M_{\odot}$, $3 \times 10^{7} M_{\odot}$ (present-day estimate of HI mass from \citealp{Bruns2005}), and $3 \times 10^{8} M_{\odot}$. We've here used the \cite{Salem2015RAMMEDIUM} halo profile, and we've assumed spherical symmetry to relate the Leading Arm parameters to a total mass. Different linestyles denote varying $\mathcal{M}$. \emph{Third:} Drag time (Equation \ref{eqn:tdrag} assuming $\alpha = 0.2$)  \emph{Fourth:} Ratio of drag time to lookback time, which for all reasonable Leading Arm masses, is $> 1$, meaning the present-day Leading Arm is in the non-entrained phase. }
\label{fig:LA_profile}
\end{figure}

\begin{figure*}
\centering
\includegraphics[width=0.45\textwidth]{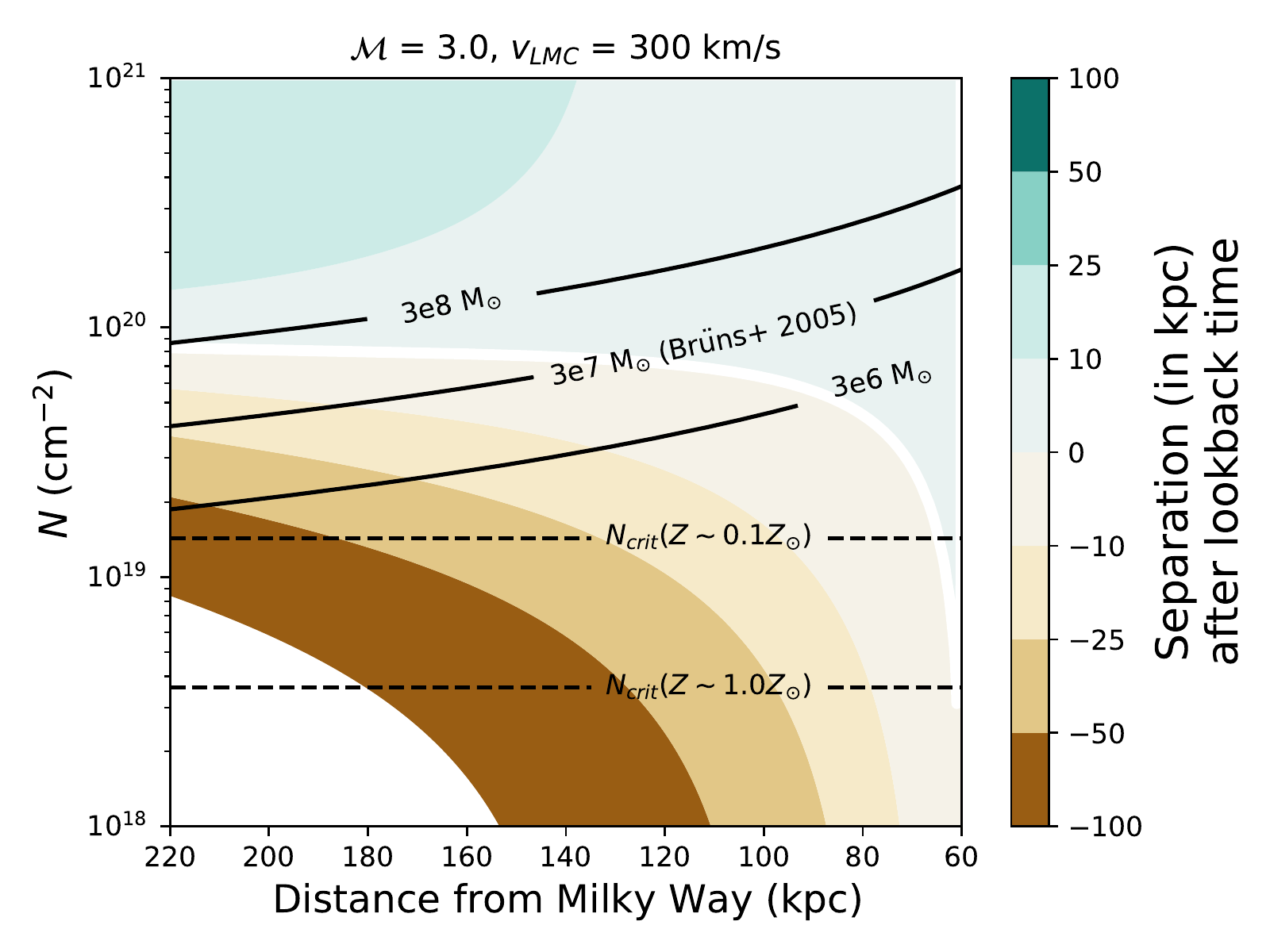}
\includegraphics[width=0.45\textwidth]{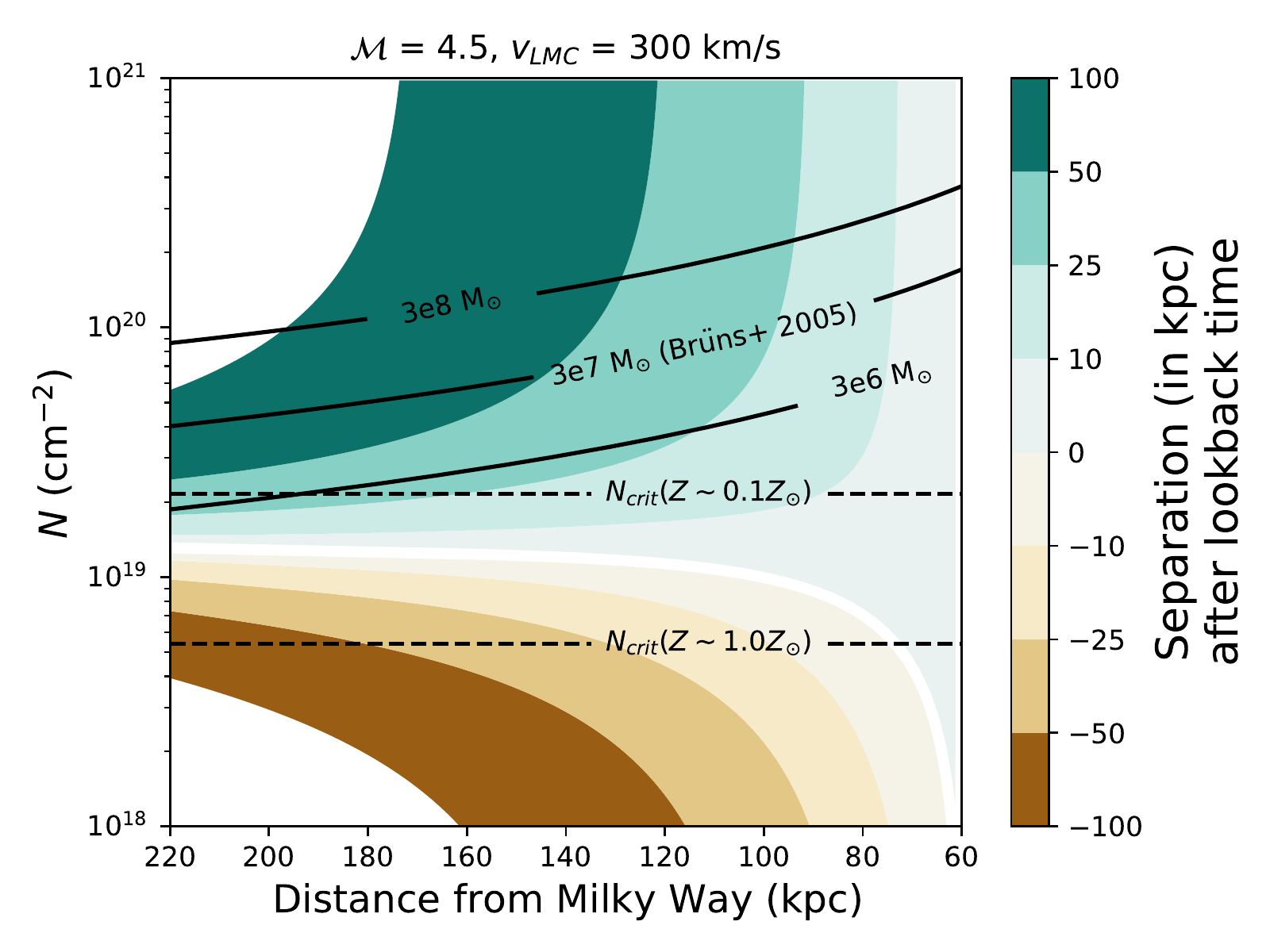}
\caption{Final separation between the Leading Arm and LMC/SMC (Equation \ref{eqn:drag}) if the Leading Arm with column density N (y-axis) is swung-out from the Clouds at a certain distance from the Milky Way (x-axis) for two different swing-out velocities $v = \mathcal{M} c_{s,hot}$ (with $\mathcal{M}$ denoted in the title of the left and right panel). Mappings between distance, lookback time, and Milky Way halo density are shown in Figure \ref{fig:LA_profile}. Here, we assume the \cite{Salem2015RAMMEDIUM} halo density profile, and in the Appendix, we show nearly identical plots using the profile from \cite{Faerman2020}. We also overplot solid black lines of constant Leading Arm mass, including the present-day HI estimate \citep{Bruns2005}, assuming spherical symmetry to relate the mass to column density. Dashed lines denote the critical column densities for the Leading Arm to survive swing-out (Equation \ref{eqn:Ncrit}). Note that for all reasonable Leading Arm masses $\sim 3 \times 10^{7} M_{\odot}$, the column density is above the survival criterion. For a $\mathcal{M} = 4.5$ swing-out, corresponding to an initial Leading Arm velocity of $\sim 500$ km/s relative to a static Milky Way halo, a large parameter space exists where the Leading Arm could out-run the Clouds instead of being blown back into the Trailing Stream. For a gentler swing-out (e.g. $\mathcal{M} = 3$), the separation is everywhere reduced. Leading Arm formation in the inner Milky Way halo especially requires larger Mach number swing-outs.}
\label{fig:LA_analytic}
\end{figure*}

The scenario we envision is a race between the LMC/SMC complex moving at velocity $v_{\rm LMC} \sim 300$ km/s and a Leading Arm cloud moving at velocity $v_{\rm LA}$ relative to the Milky Way halo\footnote{We assume for simplicity that the Milky Way halo is non-rotating}. In this section, we will assume that the Leading Arm is initially infalling faster than the Clouds at lookback time $t_{\rm lookback}$, but it's velocity can decrease dramatically due to drag, thereby allowing the Clouds to pass the Leading Arm. Realistic scenarios, then, are those in which the Leading Arm is out in front of the Clouds at present-day, i.e. after time $t_{\rm lookback}$. Eventually, full hydrodynamic simulations of the Magellanic System at sufficient resolution (\S \ref{sec:discussion}) will be needed, but for now, this exercise serves to illuminate the parameter space of allowable scenarios to zeroth order, neglecting other dynamical effects such as gravity from the Milky Way. Most notably, a few key points/assumptions of this simplified exercise are in order: 
\begin{enumerate}
    \item No specific orbits for the LMC/SMC or Leading Arm are assumed here. Instead, the medium in which this ``race" occurs is fixed at $t_{\rm lookback}$: $n_{h}(t)$ and $t_{drag}(t)$ are both assumed to be constant, fixed to their values when the Leading Arm is ejected from the Clouds at $t = t_{\rm lookback}$. In reality, both will change throughout the Leading Arm's infall; however, we don't know the path of the Leading Arm through the Milky Way halo. Should the Leading Arm follow the same path as the LMC/SMC, which is itself uncertain? As we will show in Figure \ref{fig:LA_profile}, for plausible Leading Arm masses, $t_{\rm drag}$ is longer than the lookback time, implying that the Leading Arm is not entrained during its infall. While not modeled here, two competing effects will happen in a realistic scenario: gravity will pull the Leading Arm towards the Milky Way, and a gradual increase in density will decrease the drag time. Since $t_{\rm drag} \propto n_{h}^{-1/3}$ if we consider a constant mass cloud maintaining pressure equilibrium with the halo, the decrease in drag time is expected to be small, and changes to our conclusions are minor. 
    \item Similarly, the Leading Arm mass is assumed to be constant throughout its evolution. Given long growth times compared to infall times for the Leading Arm (see Figure \ref{fig:LA_profile}), this assumption seems quite reasonable.
    \item Finally, to calculate the Leading Arm mass from the column density and background halo density, we assume that $n_{cl} = \chi n_{h}$ with $\chi = 100$, and we assume the Leading Arm has a spherical mass distribution for simplicity. It may, instead, be that the Leading Arm is more of an elongated spheroid or cylinder (as we simulate in Section \ref{sec:LASimulations}); in either case, what matters for Leading Arm survival and dynamical evolution is the radius $r_{min}$ of the object along the minor axis. If the mass of the true Leading Arm is spread out over a severely elongated object, $t_{cc} \propto r_{min}$ and $t_{drag} \propto r_{min}$ will both decrease, and Leading Arm destruction is increasingly likely compared to the optimistic estimates here assuming a spherical cloud.
\end{enumerate}

Aware of the assumptions and caveats above, we proceed to solve for the distance between the Leading Arm and LMC.

\begin{equation}
    d = \int_{0}^{t_{\rm lookback}} (v_{\rm LA}(t) - v_{\rm LMC})dt
    \label{eqn:drag}
\end{equation}
where we assume $v_{\rm LMC} = 300$ km/s is a constant, and $v_{LA}(t)$ follows from the drag equation

\begin{equation}
    \frac{dv_{\rm LA}}{dt} = - \frac{v_{\rm LA}^{2}}{t_{\rm drag}v_{\rm LA,0}}
\end{equation}
which is derived by assuming that the ram pressure force $\rho_{h} v_{\rm LA}^{2}$ decelerates the cloud. The solution to Equation \ref{eqn:drag} is 
\begin{equation}
    d = v_{\rm LA,0} t_{\rm drag} ln\left(1+\frac{t_{\rm lookback}}{t_{\rm drag}}\right) - v_{\rm LMC}t_{\rm lookback}
\end{equation}
where $v_{\rm LA,0} = \mathcal{M}c_{s,h}$ is the initial swing-out velocity of the Leading Arm, $\mathcal{M}$ is the Mach number of that swing-out, and $c_{s,h} \sim 111$ km/s is the sound speed in the $\sim 10^{6}$ K Milky Way halo. 

To obtain the lookback time, we created a polynomial fit to the LMC distance as a function of time (as was done in \citealt{Bustard2020}) taking the LMC orbit from \cite{Kallivayalil2013, Salem2015RAMMEDIUM}. While other orbits are possible, we choose this one because it is motivated by recent proper motion estimates and prevalently used in first-passage simulations where many of the Magellanic System features are well-recreated (e.g. \citealt{Besla2012, Pardy2018, Lucchini2020}). One can easily go through this same exercise for other orbits. 

The Leading Arm distance is then mapped to a given halo density, assuming either the best-fit $\beta$-profile from \cite{Salem2015RAMMEDIUM}

\begin{equation}
    n_{h}(r) = n_{0} \left(1 + \left(\frac{r}{r_{c}}\right)^{2}\right)^{-3\beta/2}
    \label{eqn:Salem}
\end{equation}
with $n_{0} = 0.46$ cm$^{-3}$, $r_{c}$ = 0.35 kpc, and $\beta = 0.559$, or the profile from \cite{Faerman2020}

\begin{equation}
    n_{h}(r) = n_{0} \left(\frac{r}{r_{\rm CGM}}\right)^{-a}
    \label{eqn:Faerman}
\end{equation}
with $n_{0} = 1.3 \times 10^{-5}$ cm$^{-3}$, $r_{\rm CGM}$ = 283 kpc, and a = 0.93.

These density profiles are plotted together in Figure \ref{fig:LA_profile}, where we've reversed the x-axes to show progression over time going from left to right. Both profiles are similar to other recent models \citep{Bregman2018} but differ most notably at large radii where observational constraints are somewhat less certain.\footnote{The \cite{Faerman2020} profile is isentropic and includes non-thermal pressure contributions. How cosmic rays and magnetic fields affect the mass entrainment and deceleration of the Leading Arm are fascinating topics beyond the scope of this paper.} For the remainder of this section, we will assume the profile from \cite{Salem2015RAMMEDIUM}, and in the Appendix, we show our results are not sensitive to this choice.

We consider Leading Arms with a range of masses, related to the column density and cloud radius (assuming a spherical mass distribution and $n_{cl} = 100 n_{h}$) by 
\begin{equation}
    M_{cl} \sim 4/3 \pi \bar{m} n_{cl} r_{cl}^{3} \sim 4 \times 10^{3} \frac{N_{19}^{3}}{n_{-4}^{2}} M_{\odot}
\end{equation}
For reference, the present-day HI gas mass of the Leading Arm is estimated to be $3 \times 10^{7} M_{\odot}$ \citep{Bruns2005}. While there must be some additional mass in ionized hydrogen, the Leading Arm has been notoriously difficult to observe in H$\alpha$ \citep{AntwiDanso2020}, and there is no clear estimate of the ionized hydrogen mass. We consider plausible Leading Arm masses at swing-out to be those within a factor of 10 of the present-day HI mass from \cite{Bruns2005}. This allows for the possibility that the Leading Arm grew by a factor up to 10 compared to its swing-out mass, or that the observed Leading Arm is only one piece of the otherwise hidden, full structure, e.g. a set of clumps in the Leading Arm's stretched, cooling tail.

The second panel of Figure \ref{fig:LA_profile} shows the growth time for Leading Arms with masses between $3 \times 10^{6}$ and $3 \times 10^{8} M_{\odot}$. The growth times (Gyrs or longer) are significantly greater than the considered lookback times, so it's unlikely that the Leading Arm mass has grown since its swing-out from the Clouds. The third and fourth panels of Figure \ref{fig:LA_profile} show the drag time and ratio of drag to lookback times as a function of distance in the halo. Note the very weak dependence on Mach number, following from Equation \ref{eqn:tdrag}. Like the growth times, drag times are consistently longer than lookback times, meaning the present-day Leading Arm is in a pre-entrainment, stretching phase during which it either gains mass very slowly or is in a transient period of cold mass reduction. \emph{This leads us to conclude that the Leading Arm must have been at least as massive at swing-out as it is today.} 

Nevertheless, while significant mass growth during infall appears unlikely, plausible Leading Arms fit comfortably within the strong cooling (survival) regime. Figure \ref{fig:LA_analytic} shows the separation $d$ (Equation \ref{eqn:drag}) as a function of initial Leading Arm column density $N$ and distance from the Milky Way, using Equation \ref{eqn:tdrag} to calculate $t_{drag}$. Overplotted are lines of constant Leading Arm cloud mass for a given column density and halo density. We also overplot lines corresponding to the critical column density $N_{crit}$ (Equation \ref{eqn:Ncrit}) for both solar metallicity ($Z = Z_{\odot}$) and $Z = 0.1 Z_{\odot}$. These lines fall well below the regime of plausible Leading Arm masses, meaning that $t_{cool,mix}/t_{cc} = N_{crit}/N_{cl} < 1$. For instance, taking $Z = 0.1 Z_{\odot}$ and a Leading Arm of mass $3 \times 10^{7} M_{\odot}$ ejected at a halo distance of 150 kpc, $t_{cc}/t_{\rm cool,mix} \approx 3-4$ for $\mathcal{M} = 3$. \emph{Instead of dissolving into the Milky Way halo, the Leading Arm should survive infall, with its mass redistributed into a clumpy, cooling tail (as in Figure \ref{fig:ProjectionPlots})}.

Consistent with \cite{TepperGarcia2019}, however, the Leading Arm may be instead blown into the Trailing Stream. We consider the Leading Arm formation scenario to roughly ``work" when the swing-out mass is close to the \cite{Bruns2005} estimate and when the final separation between Leading Arm and LMC/SMC is at least tens of kpc. More separation is better, since Equation \ref{eqn:drag} actually gives us the distance traveled by the head of the Leading Arm, while the long cooling tail could trail behind for tens or hundreds of kpc. In our strong cooling simulations, we especially find this to be true, as the distance traveled (Equation \ref{eqn:drag}) is actually quite comparable to the total tail length since the tail grows as fast as the cloud is destroyed: $l_{tail}(t)/r_{cl,0} \sim \alpha (1+\mathcal{M}) \chi ln(1+t/t_{drag})$ (cf. \S \ref{sec:tail}).

Some regions of the parameter space in Figure \ref{fig:LA_analytic} clearly forbid a sufficient separation, resulting in just enough Leading Arm slow-down to allow the Clouds to ``catch up''. This is especially true for a $\mathcal{M} = 3$ swing-out, corresponding to $v_{0} \approx 333$ km/s, where earlier formation scenarios further out in the Milky Way halo are preferred, as the drag force is lower in the outer halo. On the other hand, even drag in the inner 75-100 kpc halo can be overcome if the swing-out is fast enough. A $\mathcal{M} = 4.5$ swing-out, corresponding to $v_{0} \approx 500$ km/s, is sufficient even if the Leading Arm was formed only $\approx 200$ Myrs ago.

So can the Leading Arm originate in the Clouds? Yes, as long as the swing-out is fast enough, and the initial Leading Arm mass is at least comparable to the present-day mass \citep{Bruns2005}. If in the strong cooling regime, which seems applicable, the present-day Leading Arm is likely a series of clumps in the long cooling tail behind the progenitor cloud. This is consistent with observations, which find a clear distance variation along the Leading Arm \citep{venzmer2012, For2016, AntwiDanso2020}. The clumpy, fragmented nature of the Leading Arm, rather than a coherent tail, could be due to cut-offs in observational sensitivity, i.e. that we observe only the highest column density clumps formed through thermal instability (see the distinctly clumpy morphology in Figure \ref{fig:ProjectionPlots}) or disruptive turbulence in the Milky Way halo. The latter may affect future mass growth \citep{Gronke2021}, especially for high-$\mathcal{M}$ flows which redistribute cloud mass into longer, more diffuse tails that could be more easily disrupted by external turbulence. 

Why does the Leading Arm disappear in LMC-SMC simulations? It's likely that the swing-out speed is simply too low. This is to be expected since the same simulations run without a Milky Way corona, hence no drag force, roughly reproduce the present-day Leading Arm configuration \citep{Pardy2018, TepperGarcia2019}. Another possibility is that refinement schemes in LMC-SMC simulations lead to over-mixing and over-destruction of the Leading Arm. Cooling, cloud coagulation, and subsequent mass growth are most important in the wake behind the Leading Arm, which may be under-resolved if traditional density-based refinement criteria are used (see \S \ref{sec:resolutionGuidelines}).  

A task for future simulations, we believe, is to test whether tidal interactions can simultaneously produce a large enough swing-out speed while still reproducing the morphology of the Trailing Stream. The simulations of \cite{Williamson2021}, for instance, suggest that Leading Arm features can generally be formed by tidal interactions even when including a gaseous Milky Way halo. While those simulations are not direct LMC-SMC analogs, we echo their sentiment that more simulations are needed before invoking alternate Leading Arm formation scenarios.

\section{The Leading Arm: Simulations}
\label{sec:LASimulations}

\begin{table*}
  \centering
  \caption{Simulation parameters. For all simulations, we choose $T_{cl} = 4 \times 10^{4}$ K, $T_{hot} = 4 \times 10^{6}$ K, and $\chi = 100$ (following \cite{Gronke2020a}). All conversions to physical units assume $\Lambda (T_{mix}) \sim 10^{-21.4}$ erg cm$^{3}$ s$^{-1}$ appropriate for gas that is not too far from solar metallicity.}
  \begin{tabularx}{0.73\textwidth}{cccccccc}
  \toprule
  \textbf{$\frac{t_{\rm cool,mix}}{t_{\rm cc}}$} & t$_{cc}$ (Myrs) & \textbf{$\mathcal{M}$} & $\frac{r_{\rm cl}}{\Delta x}$ & $\frac{r_{\rm cl}  (\rm pc)}{n_{-4}}$ & $\frac{m_{\rm res} (M_{\odot})}{10^{3} n_{-4}^{2}}$ & $\frac{N_{\rm cl} (\rm cm^{-2})}{10^{19}}$ & $\frac{N_{\rm crit} (\rm cm^{-2})}{10^{19}}$ \\
  \hline
  0.1 & 15 & 1.5 & 8, 16, 32 & 291 & 0.73, 0.36, 0.18 & 1.8 & 0.18 \\
  0.1 & 15 & 3 & 8, 16, 32 & 582 & 0.73, 0.36, 0.18 & 3.6 & 0.36  \\
  0.1 & 15 & 4.5 & 8, 16, 32 & 873 & 0.73, 0.36, 0.18 & 5.4 & 0.54  \\
  0.1 & 15 & 6.0 & 8, 16 & 1164 & 0.73, 0.36 & 7.2 & 0.72  \\
  1.0 & 1.5 & 1.5 & 16 & 29 & 0.36 & 0.18 & 0.18  \\
  2.0 & 0.75 & 1.5 & 16 & 15 & 0.36 & 0.09 & 0.18  \\
  0.2 & 7.5 & 4.5 & 16 & 437 & 0.36 & 2.7 & 0.54 \\
  0.4 & 3.75 & 4.5 & 16 & 218 & 0.36 & 1.4 & 0.54  \\
  1.0 & 1.5 & 4.5 & 16 & 87 & 0.36 & 0.54 & 0.54 \\
  2.0 & 0.75 & 4.5 & 16 & 44 & 0.36 & 0.27 & 0.54 \\
  10.0 & 0.15 & 4.5 & 16 & 8.7 & 0.36 & 0.054 & 0.54 \\
  \hline
  
  \end{tabularx}
\label{table1}
\end{table*}

In \S \ref{sec:LATheory}, we estimated the survival criterion, growth time, and drag time for the Leading Arm assuming that the theoretical framework of cloud crushing and radiative turbulent mixing is applicable for higher Mach number flows, which has not been rigorously tested. Here, we present a suite of idealized, hydrodynamic simulations of a 2.5-dimensional cylinder akin to an ``arm" of gas in a hot, laminar flow. We demonstrate that the equations of \S \ref{sec:LATheory} do apply for high Mach number flows (we test up to $\mathcal{M} = 6$) and demonstrate reasonable convergence, which is not \emph{a priori} clear. We do, however, find evidence for a Mach number dependence on $t_{grow}$, which is omitted from Equation \ref{tgrowEqn2}. While not of significant consequence for the Leading Arm scenario we study ($t_{grow}$ is already large compared to $t_{lookback}$), this additional dependence may have broader implications, especially for galactic winds (\S \ref{sec:implications}). Those interested primarily in implications for the Magellanic system can skip directly to \S \ref{sec:trailingstream}. 

\subsection{Numerical Methods}

Our simulations largely follow the setup from \cite{Gronke2018, Gronke2020a} albeit with critical differences. We used Athena 4.0 \citep{Athena2008} to solve the coupled equations of hydrodynamics on a 3D cartesian grid using the HLLC Riemann solver, second-order reconstruction, and the van Leer unsplit integrator \citep{Stone2009}. For radiative cooling, we employ the \cite{Townsend2009} ``exact" cooling routine, which is efficient and accurate. While the cooling curve we use assumes solar abundances, the survival and mass growth picture we probe is robust to different cooling curves (see \citealt{Gronke2018, Ji2019Mixing}), and our results can be scaled to a given $t_{cool,mix}/t_{cc}$ -- we fiducially assume $\Lambda (T_{mix}) \sim 10^{-21.4}$ erg cm$^{3}$ s$^{-1}$ for $Z = Z_{\odot}$ and $\Lambda (T_{mix}) \sim 10^{-22}$ erg cm$^{3}$ s$^{-1}$ for $Z = 0.1 Z_{\odot}$. Following the setup in \cite{Gronke2018, Gronke2020a}, we assume the background medium and cloud are at T = $4 \times 10^{6}$ K and T = $4 \times 10^{4}$ K, respectively. To ensure that the background medium doesn't cool over the long timescales of our simulations and to emulate the effect of heating, we turn off cooling for temperatures above $2.4 \times 10^{6}$ K; previous studies have verified that this has no appreciable effect on the cloud evolution \citep{Gronke2020a, kanjilal2021, Abruzzo2021}. We also enforce a temperature floor at the cloud temperature of $4 \times 10^{4}$ K to mimic the effects of photoionization from the extragalactic UV background; again, the results are not overly sensitive to this choice. 

To mimic infall through the Milky Way halo, we place ourselves in the frame of the Leading Arm and initialize a hot background with velocity ${\bf v} = v_{x}\hat{x}$. To reduce computational cost, we employ a cloud-tracking routine that shifts the reference frame to track the cold gas center of mass. The left boundary ($-\hat{x}$) is an inflow or ``wind-tunnel" boundary condition that enforces a prescribed density and pressure but varies the velocity as the Leading Arm is accelerated by the headwind, i.e. as the reference frame shifts. Along the cylinder axis ($\hat{y}$), we use periodic boundary conditions, while at all other boundaries (the $\pm \hat{z}$ and $+\hat{x}$ directions) we use outflow conditions where interior quantities are copied over to the ghost cells and allows for inflow if the last active zones have inflowing velocities.

We choose a 2.5D setup for the Leading Arm for two reasons. First and foremost, we do this out of computational necessity; shrinking the box along the axis of the cylinder allows us to extend the box along the wind direction. As we'll see, for high-$\mathcal{M}$ flows, the debris tail is significantly longer than in previously simulated regimes that focused on $\mathcal{M} \sim 1$ flows. Second, in LMC-SMC simulations, the Leading Arm is tidally stretched perpendicular to the infall direction. However, this arm shouldn't be much longer than the Clouds themselves, which makes it reasonable to consider either a cylinder or a spherical cloud. For now, we focus on a cylinder, but future work will simulate spherical setups, as well.

Two very technical points are in order here. First, simulations with a smooth object such as this (or a more typically simulated sphere) aligned with the grid show clear signs of the carbuncle instability, whereby strong planar shocks propagating along a grid become unstable and exhibit numerical artefacts on the grid scale. To counter this, we tried the ``H-correction" in Athena, which adds numerical dissipation in the direction perpendicular to the shock; however, we found that this correction changed the mass growth in our test cases. Instead, we found that a more robust fix was to break the cylinder's symmetry by applying lognormally-distributed perturbations to the cylinder's shape. These small perturbations, in some sense, make the simulations less idealized and more realistic, while they also appear to mitigate the carbuncle issue. 

\begin{figure*}
\centering
\includegraphics[width=0.8\textwidth]{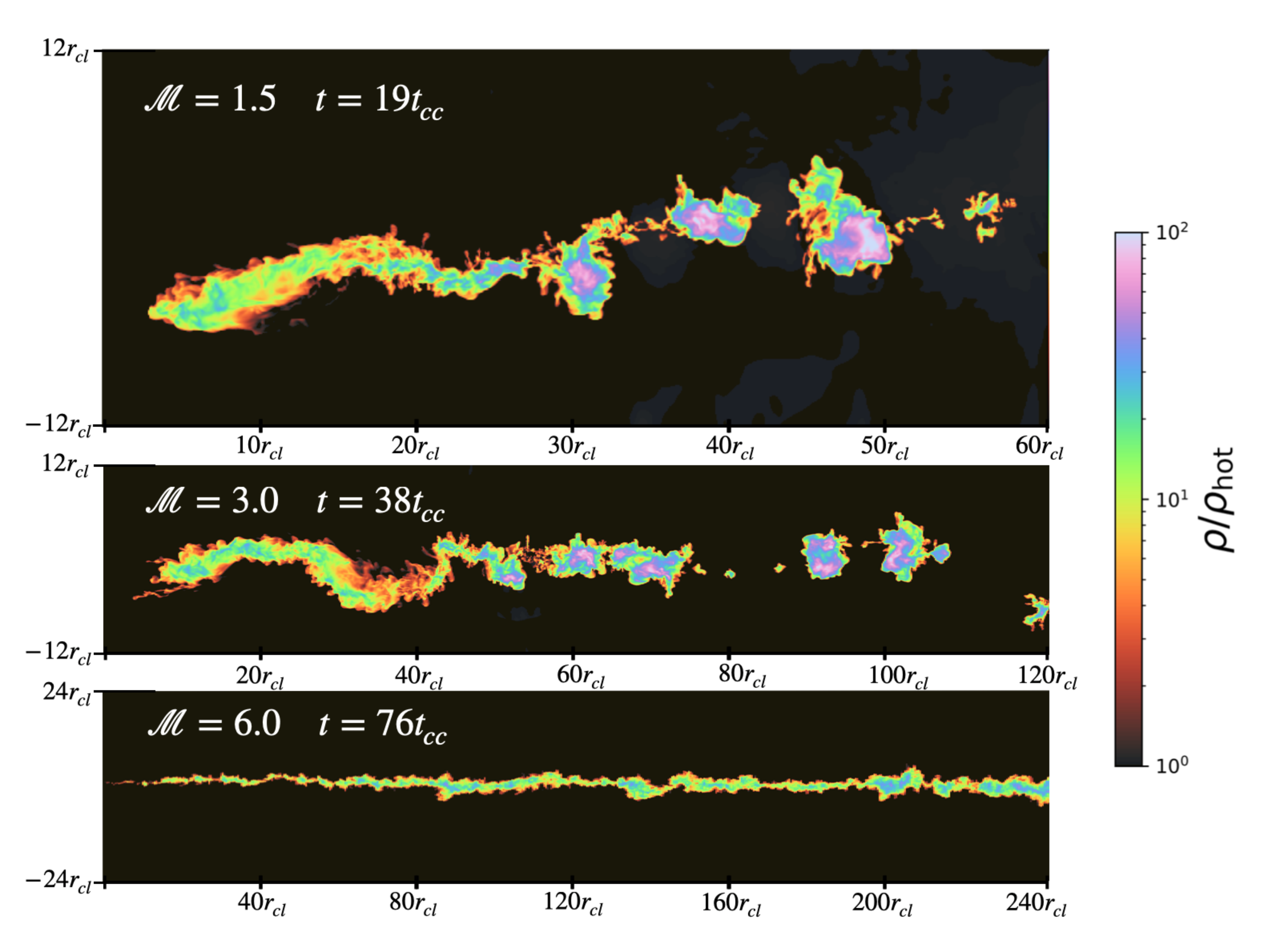}
\caption{Snapshots of average overdensity projected along the axis of the cylinder (y-axis) at the end of each simulation when the cold gas is effectively entrained in the hot flow. Snapshots are shown for $\mathcal{M} = 1.5,3,6$ simulations with $t_{cool,mix}/t_{cc} = 0.1$. Length scales are shown to highlight that higher Mach number flows lead to longer cool tails, hence the need for longer simulation boxes. Note that our simulation box for the $\mathcal{M} = 6$ run is actually longer than $240 r_{cl}$; we cut it here for presentation. }
\label{fig:ProjectionPlots}
\end{figure*}

\begin{figure}
\centering
\includegraphics[width=0.44\textwidth]{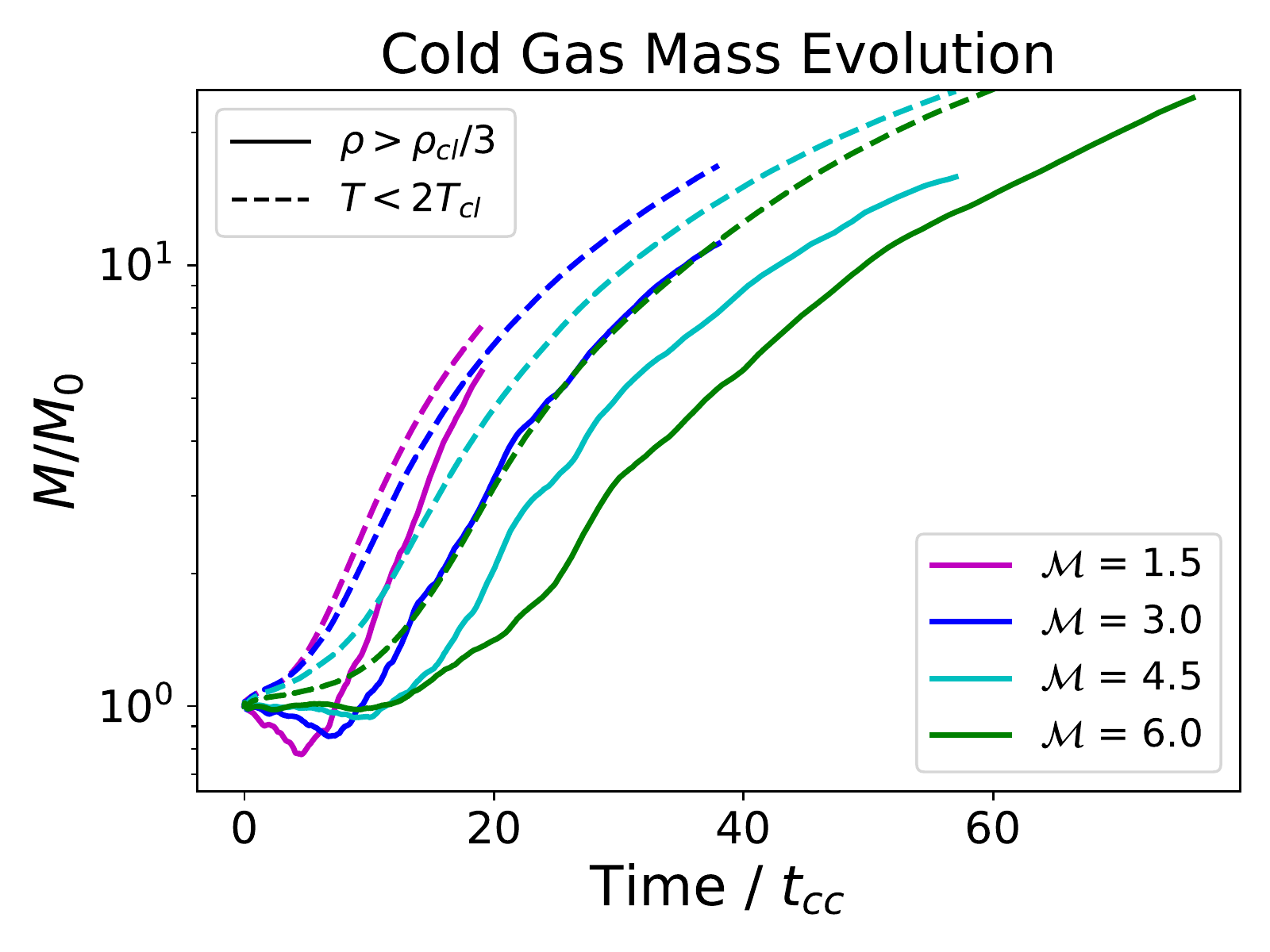}
\includegraphics[width=0.44\textwidth]{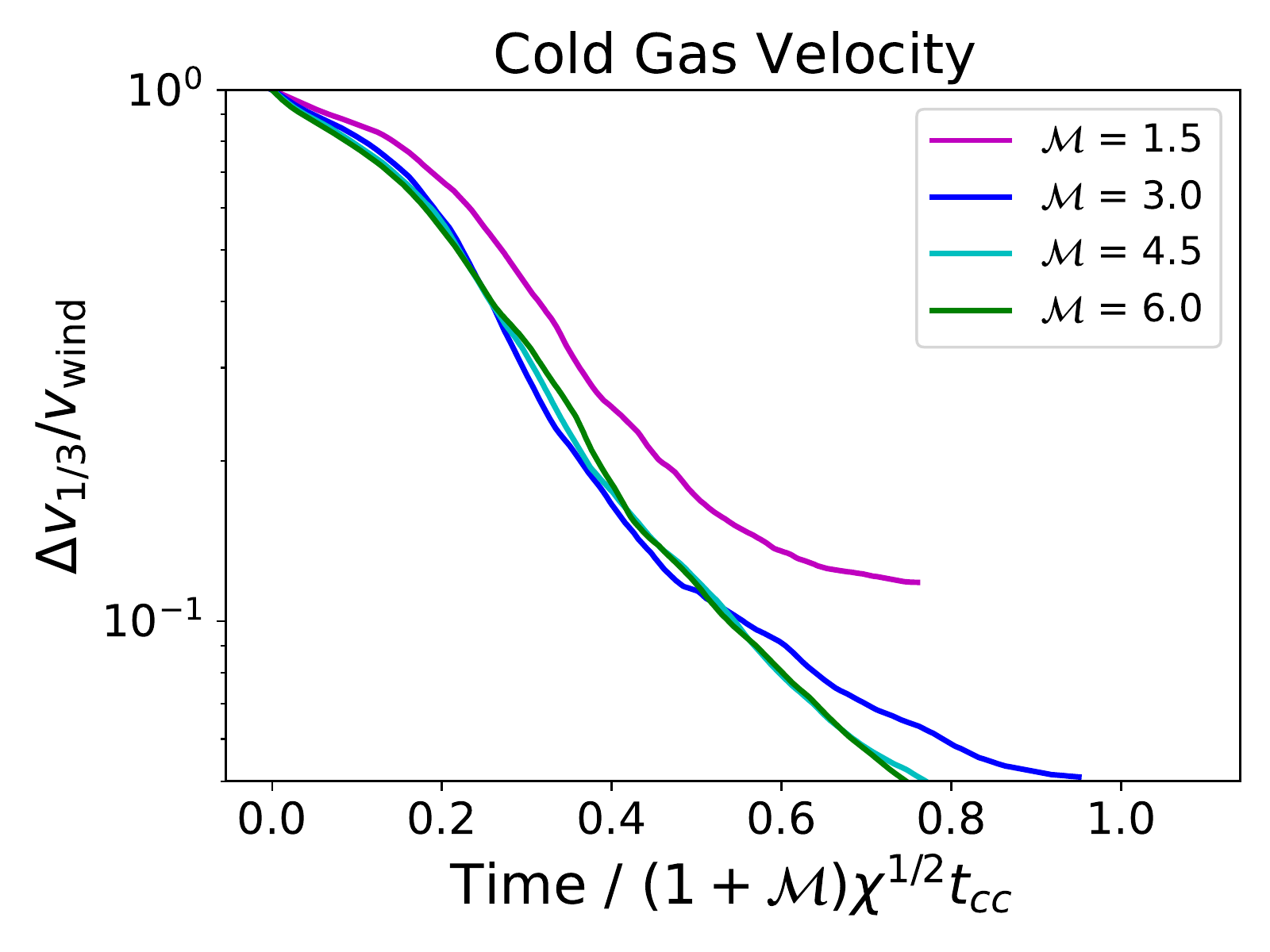}
\includegraphics[width=0.42\textwidth]{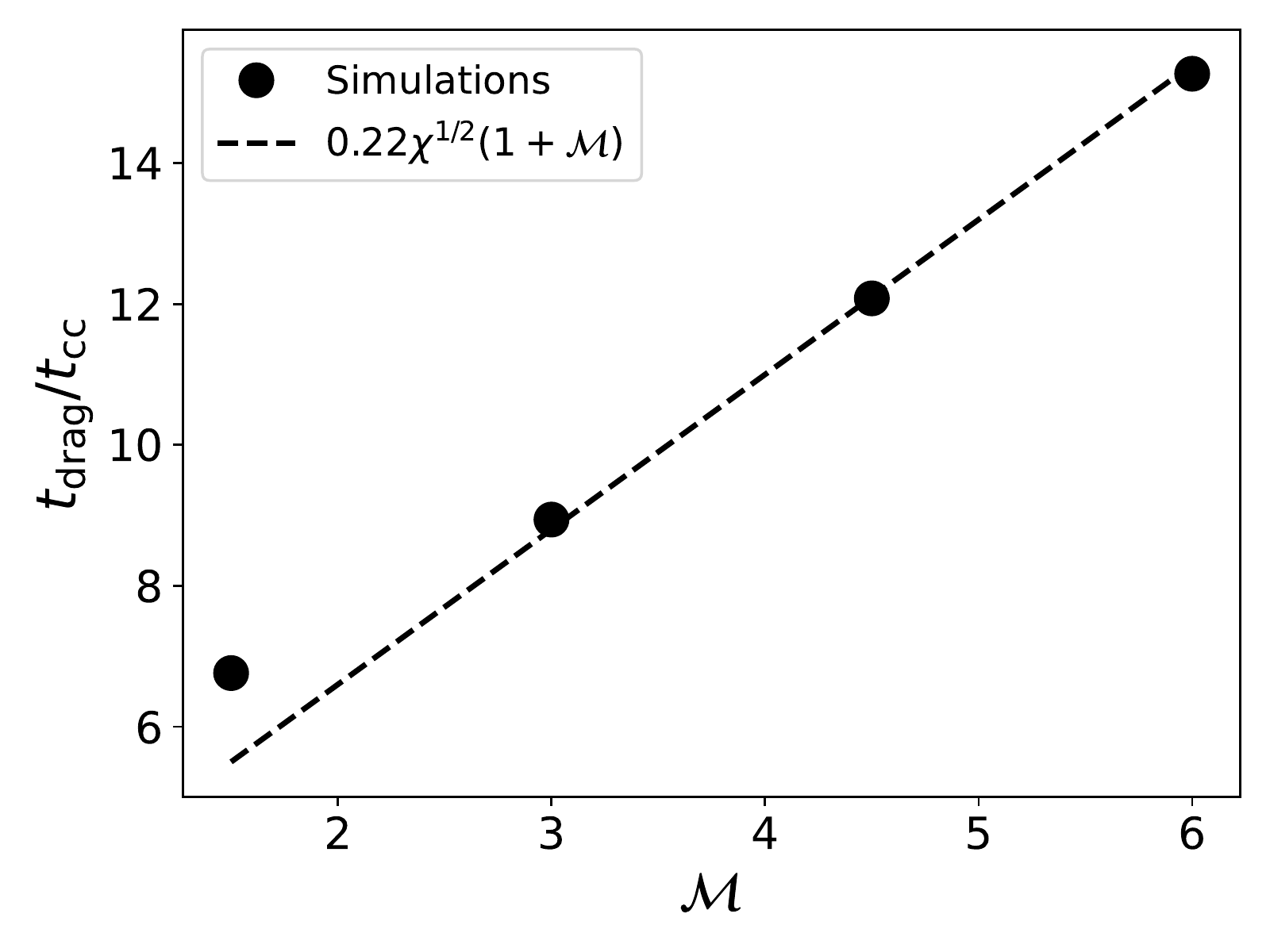}
\caption{Varying $\mathcal{M}$ and keeping $t_{cool,mix}/t_{cc} = 0.1$. \emph{Top:} Cold gas mass as a function of time normalized to the initial cold mass. Both $M(\rho > \rho_{cl}/3)$ (sold lines) and $M (T< 2T_{cl})$ (dashed lines) are shown. \emph{Middle:} Average velocity of cold gas (with $\rho > \rho_{cl}/3$) normalized to the wind velocity. The black dashed line shows when the velocity has dropped to half it's initial velocity. \emph{Bottom:} The drag time, defined as the time when the velocity has dropped by half, compared to the line $0.22 (1+\mathcal{M}) \chi^{1/2} t_{cc}$. For these strong cooling simulations with $t_{cool,mix}/t_{cc} \sim 0.1$, it appears $\alpha = 0.2$ is an appropriate choice for the drag time (Equation \ref{eqn:tdrag}).}
\label{fig:varyM}
\end{figure}

Second, we tested a number of grid dimensions and found that the grid length along the cylinder axis generally didn't matter; the simulations were converged in mass growth, drag time, etc. all the way down to a grid length of just two cells. Even a purely 2D simulation showed similar trends, albeit with slower mass growth than in 3D, but was morphologically very different. We settled on a domain width of $3 \times r_{cl}$ in the $\hat{y}$ direction. In the $\hat{z}$ direction, we found good convergence using a width of $24 \times r_{cl}$ (twice as wide as in \citealt{Gronke2020a}), which ensured that all cloud fragments stayed within the domain for the $\mathcal{M} =$ 1.5 and 3 simulations. For the $\mathcal{M} =$ 4.5 and 6 simulations, we noticed a wider dispersal of cloud fragments in the $\hat{z}$ direction and therefore extended the box width to $36 r_{cl}$ and $48 r_{cl}$, respectively, in those runs. It was also important to use a sufficiently long box in the stream-wise $\hat{x}$ direction.  Since $\tdrag/t_{\rm cc} \propto \mathcal{M}$, the box length also needs to scale $\propto \mathcal{M}$ in order to enclose the full gaseous tail that forms behind the Leading Arm. The box length for the $\mathcal{M}$ = 1.5, 3, 4.5, 6 runs are 60, 120, 180, and 360 cloud radii, where we found through low-resolution runs that the $\mathcal{M} = 6$ run produced a tail longer than $240 r_{cl}$, necessitating a somewhat longer box.

The suite of simulations we analyzed is shown in Table \ref{table1}, along with conversions to physical units for the cloud radius, cloud crushing time, mass resolution, cloud column density, and critical column density for cloud survival. Unless otherwise noted, all results shown are for simulations with resolution of 16 cells per cloud radius.

\subsection{Results}
\label{LAResults}

\begin{figure*}
\centering
\includegraphics[width=0.96\textwidth]{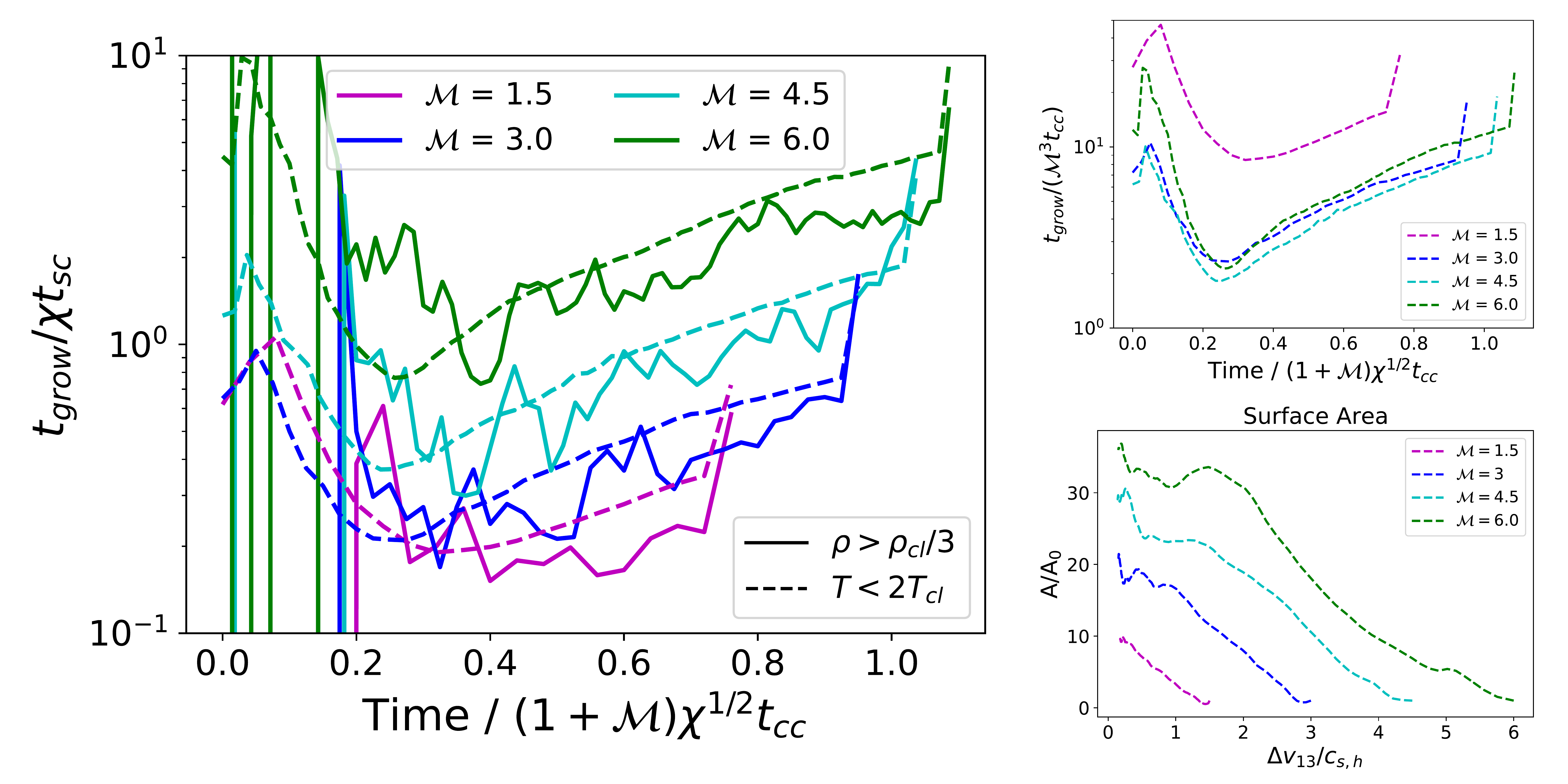}
\caption{Varying $\mathcal{M}$ and keeping $t_{cool,mix}/t_{cc} = 0.1$. \emph{Left:} Cold gas growth time normalized to the sound crossing time, $t_{sc}$, defining cold gas as $\rho > \rho_{cl}/3$ (solid lines) or T $< 2T_{cl}$ (dashed lines). The expectation from Equation \ref{tgrowEqn2} is that $t_{grow} \approx 100 t_{sc}$, which holds approximately true in the late-time entrained phase, but the growth time is much shorter in the initial expansion phase when the cloud surface area grows rapidly. \emph{Lower right:} The change in cold ($T < 2 T_{cl}$) gas surface area as a function of cold gas velocity normalized by the initial wind speed. This shows that higher Mach number flows lead to proportionally greater increases in surface area since the cloud is stretched length-wise for a longer period of time (t$_{drag} \propto \mathcal{M}$). By entrainment (when $\Delta v_{13} \sim 0$), $A/A_{0} \propto \mathcal{M}$, as the $\mathcal{M} = 3$ and 4.5 runs with 16 cells per cloud radius have surface areas $\approx$ 2 and 3 times greater than the $\mathcal{M} = 1.5$ run, although we note the absolute values of the surface area are not converged even at 32 cells per cloud radius. \emph{Upper right:} $t_{\rm grow}$ normalized by $\mathcal{M}^{3} t_{cc}$. Only the dashed curves, using $T < 2 T_{cl}$, are plotted for clarity. Aside from the $\mathcal{M} = 1.5$ simulation, $t_{\rm grow} \propto \mathcal{M}^{3}$.}
\label{fig:tgrow_area}
\end{figure*}

Figure \ref{fig:ProjectionPlots} shows the average projected overdensity $\rho/\rho_{hot}$ for the $\mathcal{M} = 1.5, 3, 6$ simulations, each with $t_{cool,mix}/t_{cc} = 0.1$. Snapshots are viewed down the axis of the cylinder (y-axis) and taken at the end of each simulation, when the cold gas is approximately entrained in the hot flow. Each simulation results in clouds that lose their morphology to strong Kelvin-Helmholtz instabilities after a few cloud crushing times. The overall cold gas mass grows significantly, though, as the cloud debris increases the surface area at which the hot gas shearing past the cold gas produces a radiative turbulent mixing layer, facilitating mass transfer from the hot to cold phases. After entrainment, as shown in each panel of Figure \ref{fig:ProjectionPlots}, there is a clear tail of cold gas in the leading cloud's wake.

In Figure \ref{fig:varyM}, we quantify the mass growth (top panel) and relative velocity $\Delta v_{13}$ (middle panel) of the cold gas (where the subscript denotes $\rho > \rho_{cl}/3$) for each of the $t_{cool,mix}/t_{cc} = 0.1$ simulations with varying Mach number. In the bottom panel, we define the drag time as the time where the velocity has dropped to half it's initial value and plot this as a function of Mach number. The drag time, with the slight exception of the $\mathcal{M} = 1.5$ simulation, is well-fit by $0.22 (1+\mathcal{M}) \chi^{1/2} t_{cc}$, which motivates our choice of $\alpha = 0.2$ in Equation \ref{eqn:tdrag}. We will explore the reason for this $t_{drag}$ scaling in more detail in \S \ref{sec:dragExplanation}, but we note that $\alpha$ is dependent on cooling strength (see Figure \ref{fig:M45_varytcc}) with $\alpha = 0.2$ being the saturated value in the strong cooling limit. To quantify the total cold gas mass, we use two definitions: $\rho > \rho_{cl}/3$ (solid lines in Figure \ref{fig:varyM}) $T < 2T_{cl}$ (dashed lines). Both show clear increases in mass over time, with $M/M_{0} \approx$ 10-20 by the time the cloud is entrained in the hot flow. The cold mass measured from the temperature criterion ($T < 2T_{cl}$) is consistently higher than that using the density criterion ($\rho > \rho_{cl}/3$), which reflects that the cooling tail lacks thermal pressure support. This appears to be more prominent at higher Mach number, likely due to increased support from turbulent pressure \citep{Ji2019Mixing}.

To unravel these behaviors further, we quantify the growth time $t_{grow} = m(t)/\dot{m}(t)$ and the surface area increase of the cold gas, $A(t)/A_{0}$ for each of these four runs. We calculate the surface area using a marching cubes algorithm and a threshold for `cold gas' defined as $T < 2T_{cl}$. The results are shown in Figure \ref{fig:tgrow_area}. From the left panel, we see that $t_{grow}$ is greatest at late times when the cloud is more-or-less entrained, but during the initial cloud stretching phase, the surface area increases rapidly, leading to shorter growth times. For $\mathcal{M} = 1.5$, $t_{grow}$ is as short as 20$t_{sc}$, but there is clearly a Mach number dependence that we find to approximately scale $\propto \mathcal{M}^{3}$ (upper right panel). We'll revisit the origins of this scaling in more detail in \S \ref{sec:tgrowScaling}. 

The lower right panel of Figure \ref{fig:tgrow_area} shows the change in surface area vs $\Delta v_{13}/c_{s,h}$, i.e. the average cold gas velocity divided by the hot gas sound speed. At early times, d(A/A$_{0}$)/dt is approximately the same for each Mach number, but the curves begin to diverge around 10-20t$_{cc}$ as the simulations reach entrainment at different rates. The higher Mach number simulations reach entrainment the slowest, hence the surface area continues to increase for a longer duration. Qualitatively, the late-time cold gas morphologies are very similar, but stretched stream-wise to lengths $\propto \mathcal{M}$ when measured in units of the cloud radius. By entrainment ($\Delta v_{13}/c_{s,h} \approx 0$), $A/A_{0} \propto \mathcal{M}$. The picture, then, is that mixing and mass growth occur regardless of Mach number, but it takes longer for the destructive effects of high-$\mathcal{M}$ flows to be offset by the increased surface area (due to increased $t_{drag} \propto 1+ \mathcal{M}$).

\subsubsection{Tail Length}
\label{sec:tail}

\begin{figure}
\centering
\includegraphics[width=0.48\textwidth]{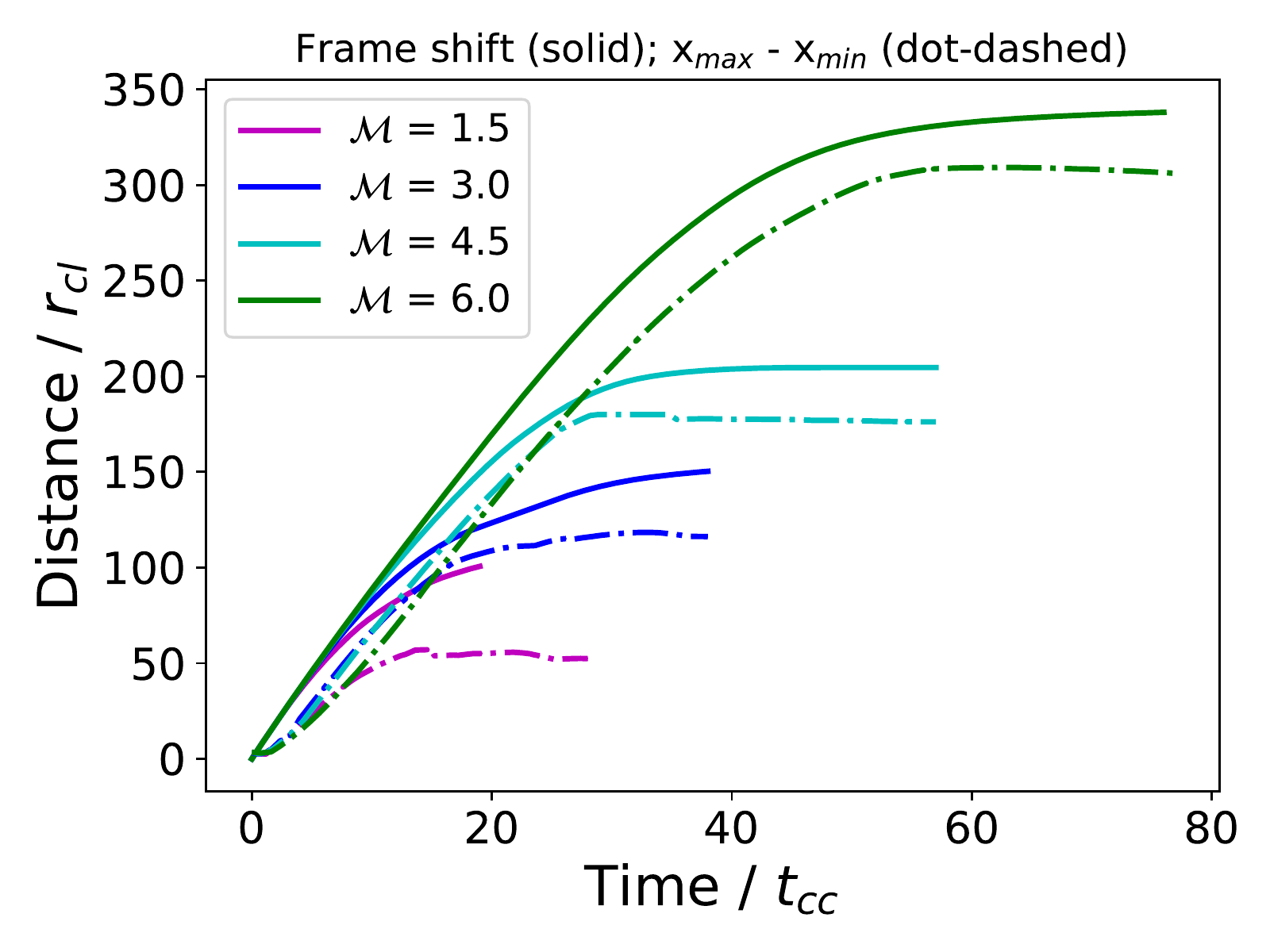}
\caption{Solid lines: The simulation frame shift, representing the distance traveled by the front of the cold cloud. Dot-dashed lines: The distance between the front and back of the cold tail, which is only marginally smaller than the full distance traveled by the cloud head in these strongly cooling simulations ($t_{cool,mix}/t_{cc} = 0.1$). }
\label{fig:frameshift}
\end{figure}

A consequence of the increased drag time is that the duration of `stretching' is longer. This implies
an increased tail length, and thus, increased surface area for higher Mach number runs. One can immediately see this in Figure \ref{fig:ProjectionPlots}, where we mark the scale on each panel.

For these simulations in the strong cooling regime ($t_{cool,mix}/t_{cc} = 0.1$), cooling happens so rapidly that the tail forms immediately upon cloud breakup. We show a rough quantification of this in Figure \ref{fig:frameshift}, which plots the amount the simulation frame has shifted over time (solid lines), as well as the largest distance between cold ($\rho > \rho/3$) gas cells (dot-dashed lines). The former tells us how far the front of the cloud has traveled, which we find to fit the drag law distance very well. The latter gives a rough estimate of the tail length, which is only marginally less than the distance traveled by the cloud head (Equation \ref{eqn:drag}). 
Since from the initial shock onwards, cloudlets peel off the cold gas cloud and are quickly entrained in the hot halo, the 
tail length is approximately the full distance traveled by the cloud head, which could be hundreds of kpc for the Leading Arm infall scenario. 

One relevant consequence of this is that, in the pre-entrainment phase ($t < t_{drag}$), when the total cold gas mass has not yet significantly increased, the mass per unit length $\sim M_{cl,0}/(v_{wind} t)$ is decreasing. Since this transient phase is prolonged for higher Mach number flows, the low-density tail may be disrupted by external turbulence before mass growth can begin. Simulations with a turbulent rather than a laminar flow are needed to assess the implications of this.

\begin{figure*}
\centering
\includegraphics[width=0.9\textwidth]{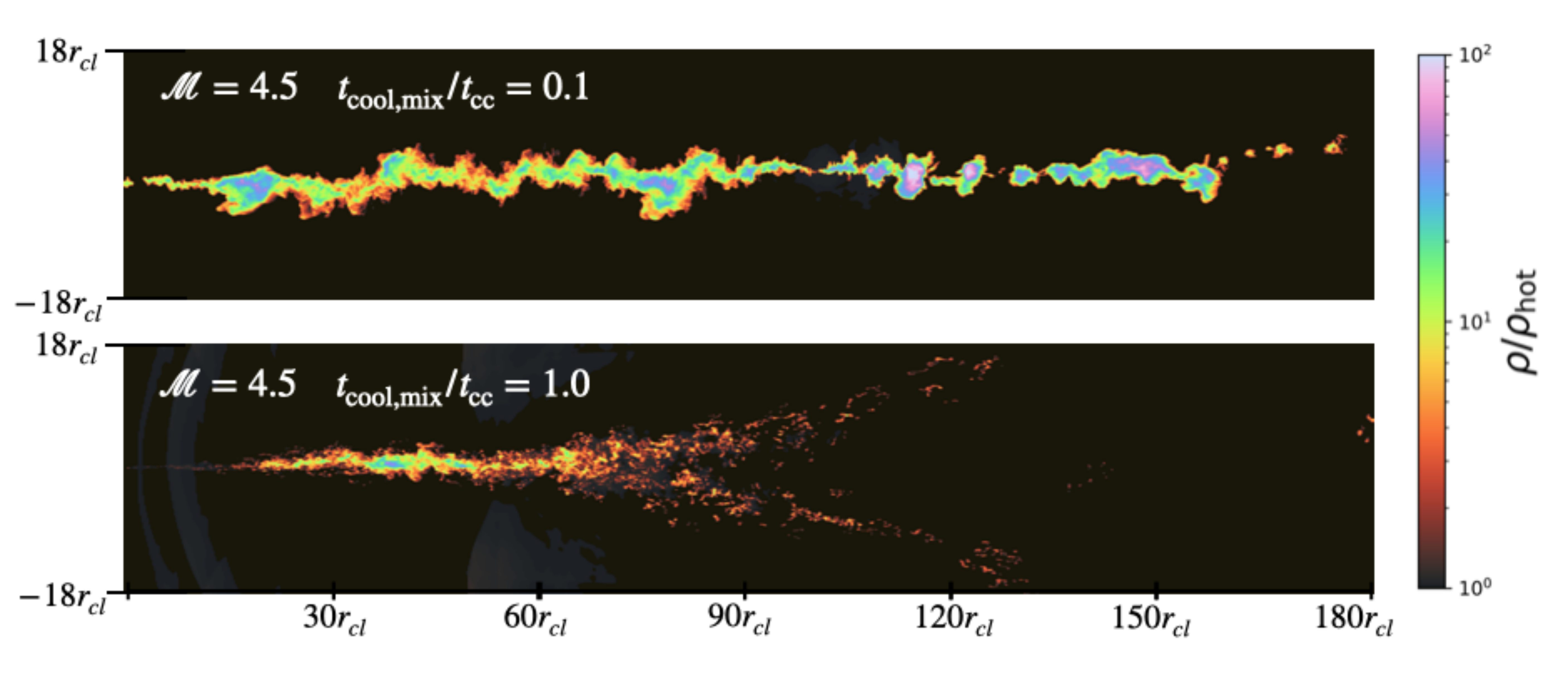}
\caption{Snapshots of average overdensity projected along the axis of the cylinder (y-axis) at the end of each simulation when the cold gas is effectively entrained in the hot flow ($\Delta v_{13}/v_{wind} \approx 0$). Snapshots are shown for $\mathcal{M} = 4.5$ simulations with $t_{cool,mix}/t_{cc} = 0.1$ and $t_{cool,mix}/t_{cc} = 1$. Simulations in the weak or marginal growth regime ($t_{cool,mix}/t_{cc} \sim 1$) show more cloud dispersal, a delay in cloud coagulation and mass growth, and therefore a less coherent tail by the time the cold gas is entrained.}
\label{fig:Mach45_Projections}
\end{figure*}

\begin{figure*}
\centering
\includegraphics[width=0.32\textwidth]{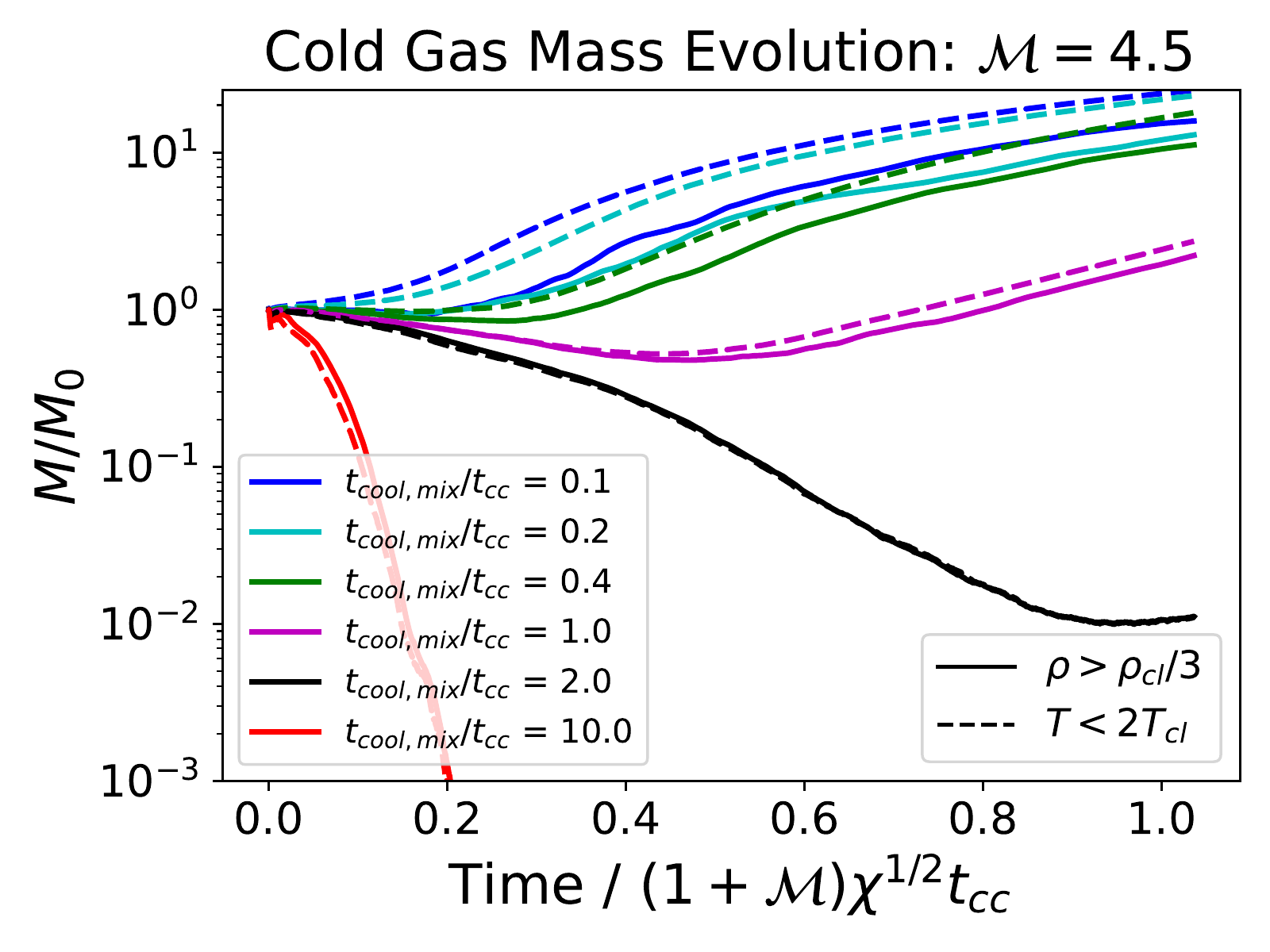}
\includegraphics[width=0.32\textwidth]{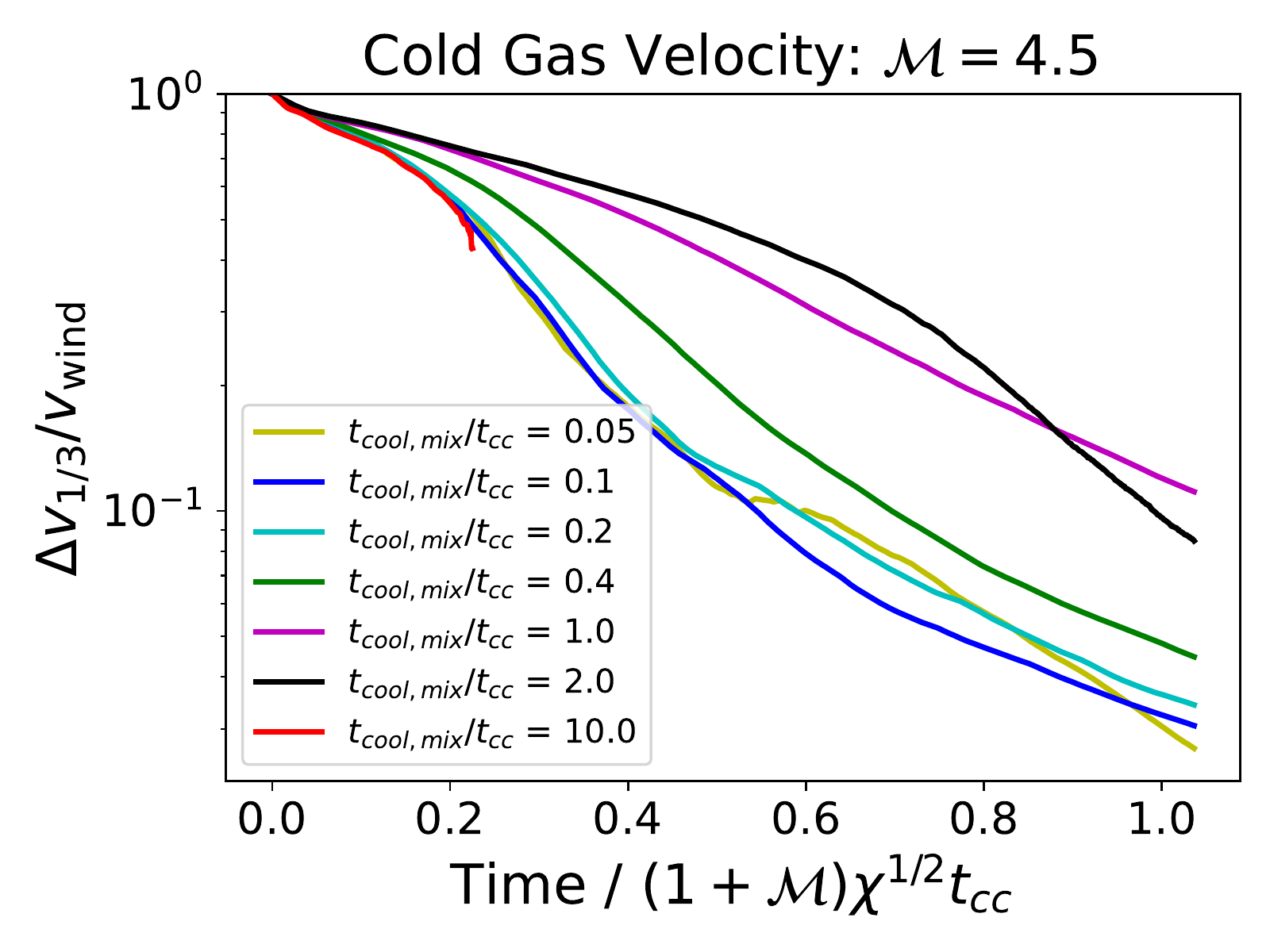}
\includegraphics[width=0.32\textwidth]{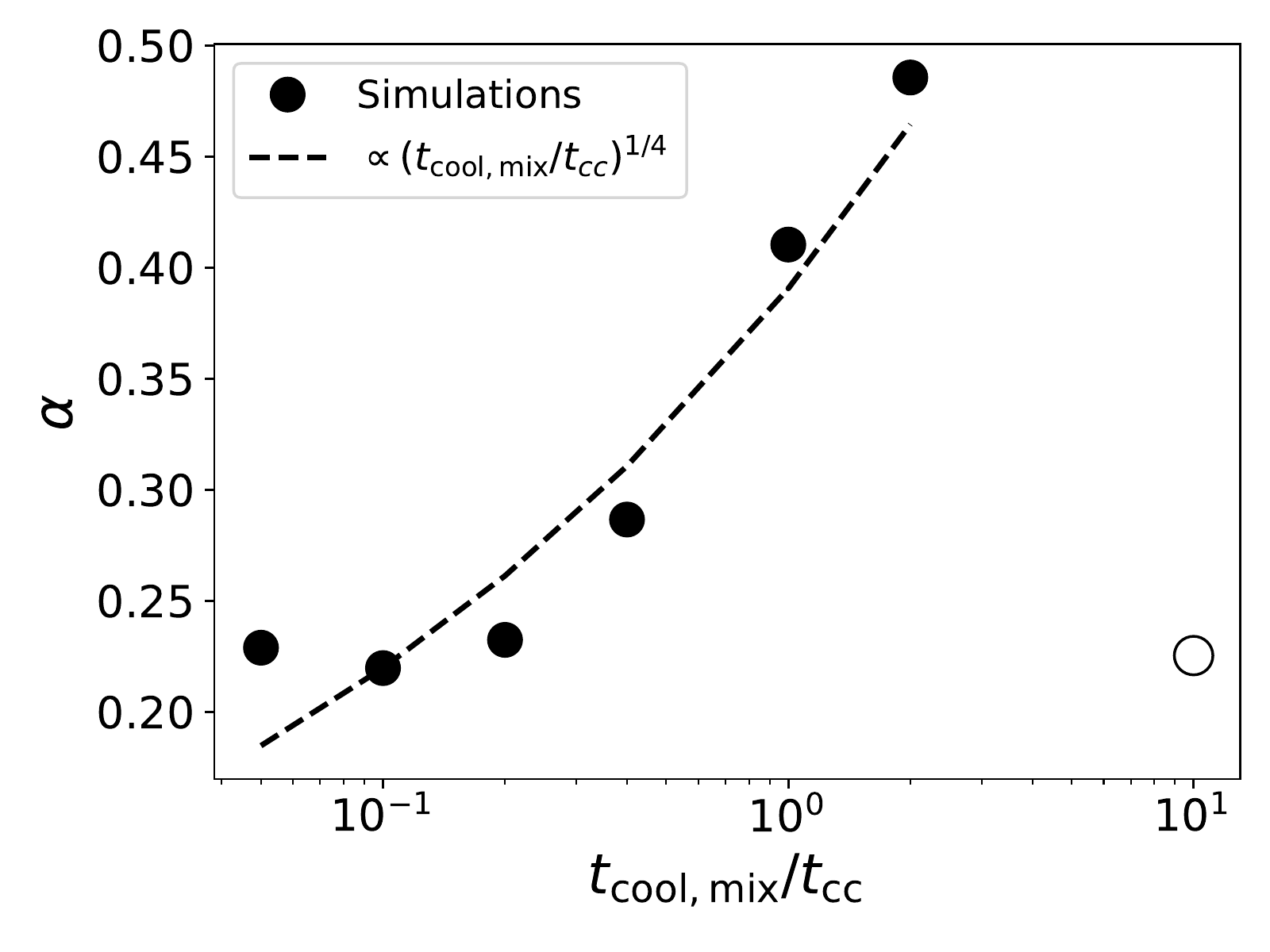}
 \caption{\emph{Left:} Cold gas mass as a function of time normalized to the initial cold mass. \emph{Middle:} Average velocity of cold gas (with $\rho > \rho_{cl}/3$) normalized to the wind velocity. \emph{Right:} $\alpha = t_{\rm drag}/(1+\mathcal{M})\chi^{1/2} t_{cc}$, where $t_{\rm drag}$ is defined as the time where the cold gas velocity has dropped by half. In the left panel, both M($\rho > \rho_{cl}/3$) (sold lines) and M(T $< 2T_{cl}$) (dashed lines) are shown. In all cases, $\mathcal{M} = 4.5$, but the cloud size, hence $t_{cc}$, is varied. $t_{cool,mix}/t_{cc} \lessapprox 1.0$ appears to be a suitable cut-off for mass growth, as the $t_{cool,mix}/t_{cc} = 10$ simulation shows a precipitous drop in cold gas mass, while the $t_{cool,mix}/t_{cc} = 1$ simulation is marginal. The middle and right panels show that the velocity drops to $\sim 1/2$ its initial value after a time $\sim 0.2 (1+\mathcal{M}) \chi^{1/2} t_{cc}$ but monotonically trends towards higher drag times as $t_{cool,mix}/t_{cc}$ increases from 0.1 to 2.0. Small $t_{\rm drag}$ for the $t_{cool,mix}/t_{cc} \sim 10$ simulation reflects the fast loss of cold gas mass and, therefore, fast entrainment in the hot flow; we mark this point with an open circle to separate it from the others. Clearly the drag time is correlated with cooling strength, as cooling leads to momentum transfer beyond that due to ram pressure. As expected in the mass growth regime, $t_{\rm drag} \propto t_{\rm grow} \propto (t_{\rm cool,mix}/t_{cc})^{1/4}$ (shown by the dashed line), with $\alpha = 0.2$ appearing to be the minimum value in the strong cooling limit.}
\label{fig:M45_varytcc}
\end{figure*}

\subsubsection{Varying Cooling Efficiency}
Finally, having analyzed a set of simulations where the parameters are comfortably in the survival / mass growth regime of $t_{cool,mix}/t_{cc} < 1$, we explore a set of $\mathcal{M} = 4.5$ simulations with varying $t_{cool,mix}/t_{cc}$ to test whether the cloud survival criterion changes at high Mach number. To run these simulations, we uniformly scale the cloud size and box dimensions to change $t_{cc}$, while $t_{cool,mix}$ is the same for each simulation. Figure \ref{fig:Mach45_Projections} compares snapshots at the end of the simulation for the $t_{cool,mix}/t_{cc} = 0.1, 1.0$ simulations. For the smaller cloud ($t_{cool,mix}/t_{cc} \sim 1.0$) not in the strong cooling (strong survival) regime, the downstream tail is less coherent, as the mass fluxes into and out of the cold phase are roughly balanced. The result is that the cold gas mass stays roughly constant, with only a small uptick at late times (see Figure \ref{fig:M45_varytcc}), and the cold gas is more dispersed rather than concentrated in a monolithic stream.  

In Figure \ref{fig:M45_varytcc}, we show the mass (left) and velocity (middle) evolution for $t_{cool,mix}/t_{cc}$ varying from 0.1 to 10.0. The behavior follows the survival criterion well, with the $t_{cool,mix}/t_{cc} = 10.0$ simulation showing a clear decrease in cold gas mass, while the $t_{cool,mix}/t_{cc} < 1.0$ simulations all grow at a rate that monotonically decreases with increasing $t_{cool,mix}/t_{cc}$. Interestingly, even the $t_{cool,mix}/t_{cc} = 1.0$ simulation shows a slight up-tick in cold gas mass at late times when the shear velocity has fallen below $\sim 1/2$ its initial value. In this case, the dispersed chunks of cold gas coagulate back together and, by the time they are entrained, have even increased in total mass compared to the initial cloud mass. This, as well as the finite growth times seen even at late times in Figure \ref{fig:tgrow_area}, would appear to be consistent with mass entrainment by mixing and cooling due to continuous `pulsations' \citep{Gronke2020a}, which occurs at a mixing rate independent of Mach number.

\subsubsection{Drag Time Dependence on Cooling and Mach Number}
\label{sec:dragExplanation}

\begin{figure}
\centering
\includegraphics[width=0.48\textwidth]{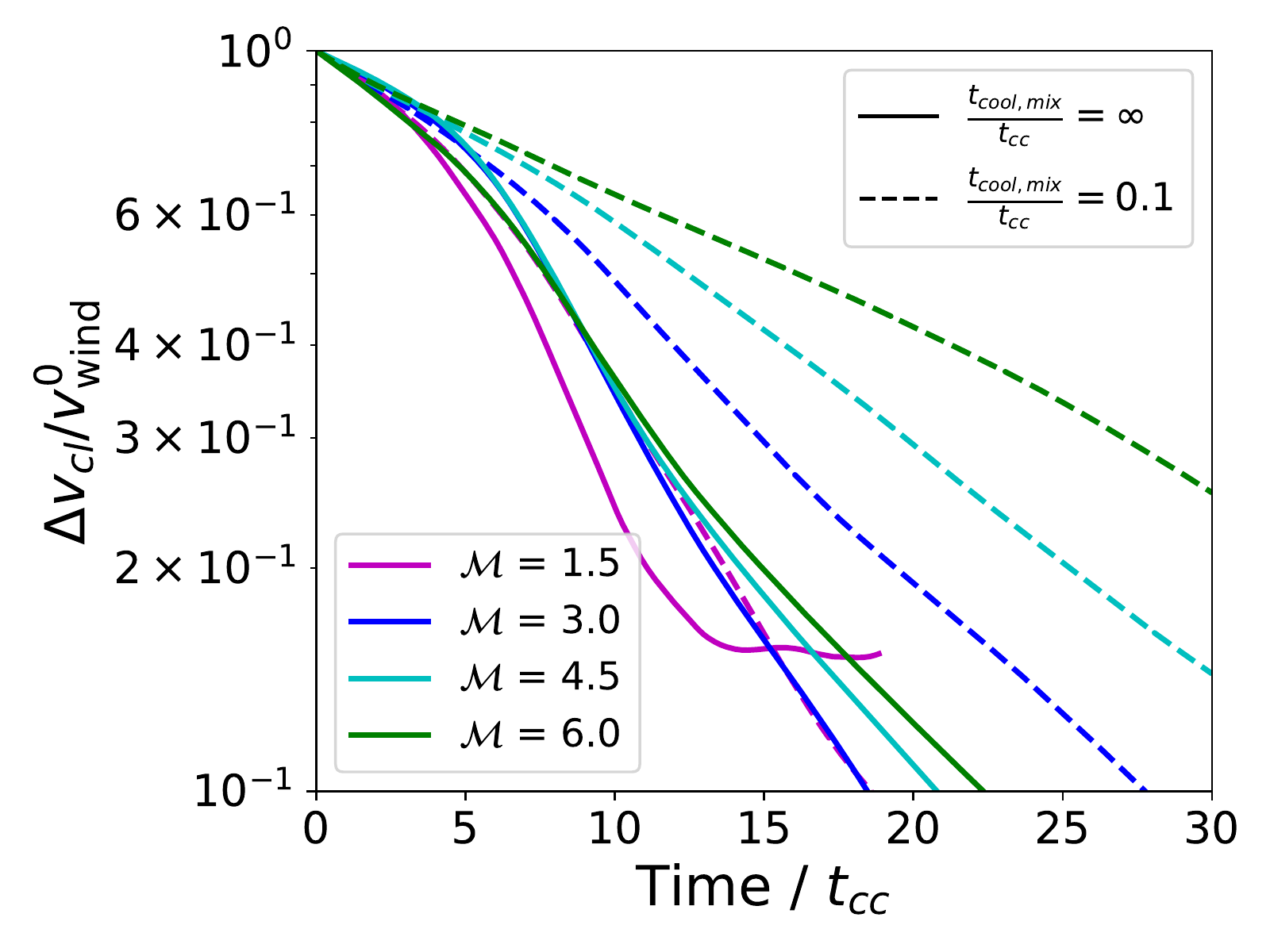}
\includegraphics[width=0.48\textwidth]{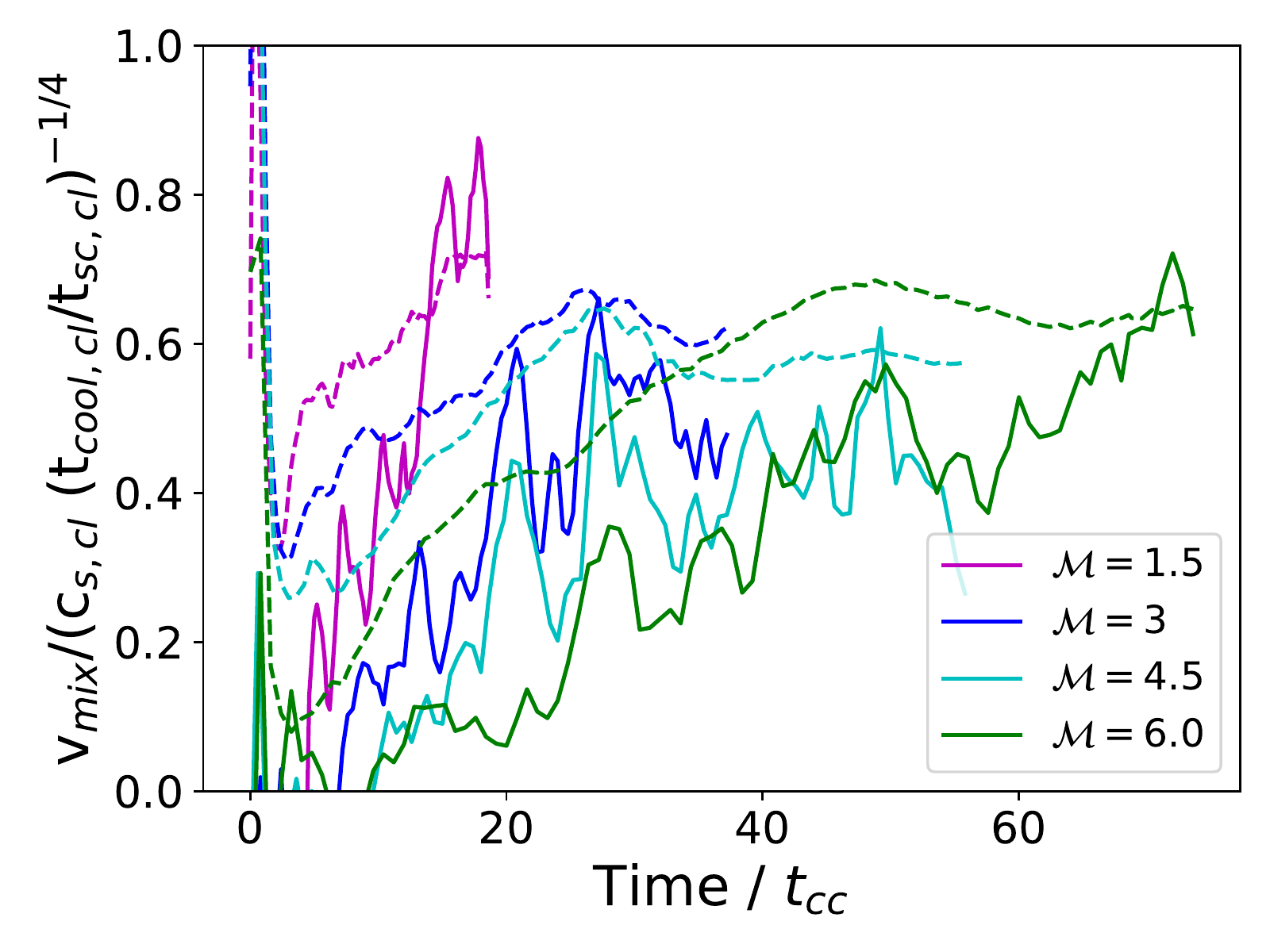}
\caption{\emph{Top:} Average gas velocity, weighted by a cloud tracer, for our strongly cooling simulations $t_{cool,mix}/t_{cc} = 0.1$ and identical simulations run without cooling $t_{cool,mix}/t_{cc} = \infty$. The adiabatic simulations show similar drag times independent of Mach number. \emph{Bottom:} Mixing velocity $v_{mix} = \dot{m}/A\rho_{h}$ normalized to its expected scaling (Equation 8 from \citealt{Gronke2020a}). $v_{mix}$ shows a clear inverse Mach number dependence, only asymptoting to similar values after entrainment, which takes longer for higher Mach number flows. }
\label{fig:velTracer_vary}
\end{figure}

The middle and right panels of Figure \ref{fig:M45_varytcc} show that the drag time depends on the cooling efficiency $t_{cool,mix}/t_{cc}$. To understand this better, we also ran a set of simulations with the same parameters as our $t_{cool,mix}/t_{cc} = 0.1$ simulations but without cooling. The velocity profiles are shown in Figure \ref{fig:velTracer_vary}. Since the adiabatic ($t_{cool,mix}/t_{cc} = \infty$) runs destroy the cloud and mix all cold gas into the hot background within a few $t_{cc}$, we no longer quantify the \emph{cold} gas velocity but instead use a passively advected tracer fluid that is initially set to 1 inside the cloud and 0 elsewhere. The average cloud velocity is then $v_{cl} = \langle c v_{x}\rangle/\langle c\rangle$, where $c$ is the tracer concentration and $c v_{x}$ is the concentration-weighted velocity along the headwind direction. With the slight exception of the $\mathcal{M} = 1.5$ run, the drag times for adiabatic simulations are independent of $\mathcal{M}$, consistent with other adiabatic cloud-crushing simulations (e.g. \citealt{GoldsmithPittard2017}). 

The cooling simulations, however, have drag times $\propto 1+\mathcal{M}$, and interestingly, the drag times are longer than those for the adiabatic runs. This initially seems to be in conflict with Figure \ref{fig:M45_varytcc}, which shows stronger cooling generally leads to faster entrainment. However, cooling plays a dual role: compared to an adiabatic simulation, it increases the drag time by inhibiting cloud dispersal \citep{Scannapieco2015}, but it also decreases the drag time by facilitating momentum transfer. Which effect dominates depends on the strength of cooling. Figure \ref{fig:M45_varytcc} suggests an inflection point between $t_{cool,mix}/t_{cc} =$ 2 and 10 where the increased drag time due to compression becomes offset by the decreased drag time due to momentum transfer in the growing cloud tails. Naively, we expect $t_{drag} \propto t_{grow} \propto (t_{cool,cl}/t_{sc})^{1/4}$ (see Equation \ref{eqn:tgrow}). In Figure \ref{fig:M45_varytcc}, the only difference for these simulations is an increased cloud radius (and sound crossing time), hence $t_{drag}$ should vary by a factor of $(t_{cool,mix}/t_{cc})^{1/4} \propto (t_{cool,cl}/t_{sc})^{1/4}$. The dashed line in the right panel of Figure \ref{fig:M45_varytcc} shows this trend compared to simulations, intercepting $\alpha = 0.2$ at $t_{cool,mix}/t_{cc} = 0.1$. The simulation points appear consistent with the expected trend, except for the $t_{cool,mix}/t_{cc} =$ 10 simulation where cooling is weak.

Now, what is the reason for the $\propto 1+\mathcal{M}$ scaling? \cite{Scannapieco2015} attribute this to the oblique bow shock that compresses the cloud and raises its density by a factor of 1+$\mathcal{M}$. Since the cloud maintains a constant temperature of order $10^{4}$ K, this density increase induces a pressure gradient between the head and tail of the cloud that stretches the cloud in the headwind direction. The net effect is to compress the cloud primarily orthogonal to the headwind, making the cloud's total column in the headwind direction a factor of $1+\mathcal{M}$ higher. 

A related possibility is that the shock is isothermal, thereby inducing a density increase $\propto \mathcal{M}^{2}$ instead of $\propto \mathcal{M}$. This would happen if $t_{cool,cl}\ll t_{\rm shock-cross} \sim t_{cc}$ (which is the case for all surviving clouds as $t_{\rm cool,mix} > t_{\rm cool,cl}$). If accompanied by a uniform (mass conserving) compression $r_{cl} \rightarrow r_{cl}/\mathcal{M}$ of our cylinder, rather than a predominantly transverse compression, then $t_{drag} \sim \chi \mathcal{M}^{2} r /\mathcal{M} v_{wind} \propto \mathcal{M}$.

While we do observe a transient, elevated cloud density, the dominant and most persistent geometric change in our simulations is the eventual formation of a cool tail. Combined with the apparent scaling of $t_{drag}$ with cooling efficiency, it appears that momentum transfer via cooling and mixing is at least as important as momentum transfer via ram pressure.\footnote{This makes sense as $t_{\rm drag}/t_{\rm grow}\sim v_{\rm mix} / v_{\rm wind}\sim \chi^{-1/2}/\mathcal{M}$ where we used conservatively $v_{\rm mix}\sim c_{\rm s,cold}$.} Can we recover a 1 +$\mathcal{M}$ scaling from this mixing picture? This could follow if $t_{drag} \propto t_{mix} \propto \mathcal{M}$. At least in the early evolution of our clouds, it \emph{does} appear that $t_{mix} \sim 1/v_{mix} \propto \mathcal{M}$, as seen in the bottom panel of Figure \ref{fig:velTracer_vary}, which shows $v_{mix} =  \dot{m}$/$\rho_{h}$A for each of our $t_{cool,mix}/t_{cc} = 0.1$ runs with 16 cells per cloud radius. Note that this scaling also appears in adiabatic wind simulations, but $t_{mix}$ has the opposite sign, i.e. the cloud loses mass. For example, in \cite{GoldsmithPittard2017}, $t_{mix} \propto \mathcal{M}$ for their wind-tunnel simulations with $\chi = 10$ (see their Figure 5), although it does level off at very high Mach numbers and appears constant at all Mach numbers if the overdensity is $\chi = 1000$ instead \citep{Goldsmith2018}. Note that albeit detailed simulations of turbulent mixing layers in a plane-parallel setup find a somewhat different Mach scaling of $t_{mix} \propto \mathcal{M}^{-3/4\times 4/5}$, they comment that this is dependent on the geometry employed \citep{Ji2019Mixing,Fielding2020,Tan2021}.

Our analysis here, then, appears to explain the seemingly disparate dependencies of $t_{drag}$ on $\mathcal{M}$ found by \citealt{GoldsmithPittard2017} (no dependence) and \citealt{Scannapieco2015} ($t_{drag} \propto 1+\mathcal{M}$). The key difference is cooling, not necessarily because of its ability to increase cloud survival and reshape the cloud but also because of momentum transfer via mixing, which is more efficient for stronger cooling and seemingly dominant when $t_{cool,mix}/t_{cc} \sim 0.1$. While the effects of cloud compression increase the drag time, leading to e.g. slower acceleration of cold clouds in galactic winds \citep{zhang2017} or longer deceleration of infalling clouds, additional momentum transfer in the strong cooling regime somewhat offsets this, shortening the drag time by a factor we expect is $\propto (t_{cool,cl}/t_{cc})^{1/4}$.
  
\subsection{Growth Time Dependence on Cooling and Mach Number}
\label{sec:tgrowScaling}

Following similar arguments as above, we can now reason why we find a Mach number dependence for the growth time. From Figure \ref{fig:velTracer_vary}, we see that $v_{mix}$ asymptotes to similar values regardless of $\mathcal{M}$, but the paths to get there are very different. $v_{mix}$ is initially quite low for high $\mathcal{M}$, likely due to two effects: 1) high-$\mathcal{M}$ flows induce a strong shock that crushes the cloud more dramatically than low-$\mathcal{M}$ flows. Indeed, we observe that our Leading Arm is fragmented into more chunks that disperse to larger distances away from the initial cloud; this led us to increase our simulation box width for the $\mathcal{M} =$ 4.5 and 6 runs. That is, the destructive effects of the wind that cause energy \emph{loss} are strong. This is seemingly consistent with the findings of \cite{Gronke2020b}, where clouds thrown violently out of pressure equilibrium shatter into many clumps. 2) Mixing is hindered at high-$\mathcal{M}$ since the Kelvin-Helmholtz instability is suppressed for high-$\mathcal{M}$ ($v_{mix} \sim 1/\mathcal{M}$; \citealt{GoldsmithPittard2017}). That is, the effect that causes mass \emph{gain} is weaker. Only when the clouds have become partially entrained does the mixing velocity rise to considerable values. The net effect is to create longer growth times for higher Mach number flows, with a dependence that we find to scale as $t_{grow} \propto \mathcal{M}^{3}$ (Figure \ref{fig:tgrow_area}). How can we understand this dependence? We know that the growth time and drag time are related, since mass growth is causing momentum transfer and acceleration/deceleration. $t_{\rm grow}/t_{\rm drag} \propto v_{wind}/v_{mix}$. If we assume, as is found in \cite{GoldsmithPittard2017}, that $v_{mix} \propto 1/\mathcal{M}$, then $v_{wind}/v_{mix} \propto \mathcal{M}^{2}$ and $t_{grow} \propto \mathcal{M}^{2} t_{\rm drag}$. Since we find $t_{\rm drag} \propto \mathcal{M}$, $t_{\rm grow} \propto \mathcal{M}^{3}$. Therefore, it's ultimately the suppression of Kelvin-Helmholtz instability ($v_{mix} \propto 1/\mathcal{M}$), that slows down growth. While this scaling is mainly inconsequential for the Leading Arm, which we estimate is well before the partial entrainment stage and, therefore, not considerably growing in mass in any case, this scaling could have interesting consequences for entrainment and growth of cold clouds in highly supersonic galactic winds. We speculate on these consequences in \S \ref{sec:implications}.

\section{The Trailing Magellanic Stream}
\label{sec:trailingstream}

In this section, we apply the theoretical framework of radiative turbulent mixing layers to the Trailing Stream. While the most comprehensive way to assess Stream survival and growth would be to run full simulations of the LMC-SMC interactions and infall, we again emphasize the impracticality of running these simulations at the resolution required to correctly capture mixing between hot and cold phases. Instead, we begin with quick analytic estimates (\S \ref{sec:Stream_analytic}) and then present a set of idealized simulations where we corroborate our predictions for the Stream survival criterion and mass influx rate (\S \ref{sec:Stream_sims}). The acute reader may note that much of this work reproduces results of \cite{Mandelker2020b, Mandelker2021}, and we refer the reader to those papers for more detailed simulation analysis and discussion. However, while the primary focus of those papers was on cold streams in an extragalactic context, the same principles apply to our Magellanic context, and that gap has not before been bridged. Our motivations for this section are as follows:

\begin{itemize}
    \item To revisit and challenge the seemingly pervasive notion that the Stream is dissolving and losing mass during infall. We show that is likely not the case for much of the Stream.
    \item To assess the level of Stream mass growth during infall and whether that can solve the``missing mass" problem for the Stream, i.e. that in full LMC-SMC simulations, the resulting Stream is less massive than estimates derived from observations. We show that the necessary level of mass growth is likely unachievable given the initial Stream mass and infall track assumed, necessitating either a larger pre-infall mass of the Stream \citep{Lucchini2020} or a change in LMC-SMC orbit \citep{Lucchini2021}. This information is also important for studies that attempt to use Stream properties (mass, length, etc.) to constrain the Milky Way halo density (e.g. \citealt{craig2021}). 
    \item To assess the impact of radiative mixing layers on other puzzling observations of the Stream. As an initial step begging for future follow-up, we here focus on H$\alpha$ emission, which has been used as a constraint on ram pressure interactions between the Stream and Milky Way halo \citep{BlandHawthorn2007, TepperGarcia2015}. Observations \citep{Barger2017} have detected extended emission off the main HI body of the Stream and peculiarly large H$\alpha$ emission along certain portions of the Stream. Our post-processing derives the amount of H$\alpha$ emission expected from the mixing-cooling process and compares to previous simulations \citep{TepperGarcia2015} in which the Stream dissolves instead of grows.
\end{itemize}

\subsection{Analytic Estimates}
\label{sec:Stream_analytic}

The main difference between this analysis and that of \S \ref{sec:LATheory} is the morphology of the Trailing Stream vs the Leading Arm. Based on simulations of LMC-SMC tidal interactions \citep{Besla2012, Pardy2018, Lucchini2020}, we assume the Stream is formed \emph{pre-infall into the Milky Way halo}, and we model the geometry as a cylinder with major axis parallel to the infall direction. This geometry is unstable to a number of both surface and body modes of the Kelvin-Helmholtz instability; combined with thermal instability, these modes destroy the axisymmetry of the Stream and naturally form the turbulent, clumpy Stream we observe \citep{Putman1998, Bruns2005, Stanimirovic2008, Nidever2010}.

This initial geometry is fundamentally different from that of the Leading Arm, where the cylinder axis was perpendicular to the headwind. In that case, mass growth ensues after the initial cloud is stretched parallel to the headwind, thereby increasing the surface area available for mixing. In the Stream case, no such stretching occurs, and any increase in surface area, then, is tied to radial growth of the cylinder. Equation \ref{tgrowEqn2} still applies, but with $A \approx A_{eff}$, and the survival criterion is described by Equation \ref{rcyleqn}, which as discussed in \S \ref{sec:background} is identical for our assumed parameters to Equation \ref{rclEqn} derived for clouds. 

The Trailing Stream's infall velocity can't exceed the infall velocity of the LMC/SMC, which is $\mathcal{M} \sim 2-3$ depending on lookback time \citep{Besla2012}. We will assume the Stream infall is $\mathcal{M} \sim 2$ through a halo with constant temperature $T_{hot} \sim 10^{6}$ K. We also assume the Stream is in pressure equilibrium with its surroundings and therefore has an overdensity $\chi \sim 100$ and temperature $T_{cyl} \sim 10^{4}$ K. For such a setup, with metallicity $\sim (0.1-0.3) Z_{\odot}$, \emph{all Stream sections with column density $N \gtrapprox 10^{19} \rm cm^{-2}$ should maintain or grow in mass, rather than dissolve} (Equation \ref{eqn:Ncrit}). This criterion is satisfied for a large portion of the Stream \citep{Bruns2005, Nidever2010}. 

Following \cite{Mandelker2020b}, one can also show that if the Stream satisfies this criterion at some point, \emph{it will stay above this survival threshold for its entire infall through denser parts of the halo}. Take a Stream with radius $r_{\rm cyl}(r)$ and line-mass $m(r) = \pi r_{\rm cyl}^{2}(r) \rho(r)$. Assuming both the halo and Stream are isothermal and that the Stream maintains pressure equilibrium with its surroundings, the overdensity $\chi$ is constant so $\rho(r) = \chi \rho_{h}(r)$. One can quickly show that the column density $N(r) = 2r_{\rm cyl}(r)n_{\rm cyl}(r) \propto m^{1/2}(r) \rho_{h}^{1/2}(r)$. If the initial Stream is within the growth regime, and the halo density increases during infall, then both terms increase with decreasing $r$; therefore, N(r) increases while the Stream becomes denser and narrower. 

This picture of Stream survival is quite different from previously published results. Primarily to study the origin of H$\alpha$ emission along the Stream, \citet{BlandHawthorn2007} and \citet{TepperGarcia2015} simulate a supersonic headwind slamming into a turbulent Stream. This interaction sends a cascade of shocks from the front to the back of the Stream that generates H$\alpha$ bright spots. This simulated Stream is initialized as under-pressurized compared to the background by a factor of 10, however, which results in a quick destruction of the Stream within 100-200 Myrs. Instead, if the Stream is closer to pressure equilibrium, we show that cooling is strong enough -- or, in other words, the column density of the Magellanic Stream is sufficiently large -- to allow the survival of the stream.

The Stream mass, however, is not likely to grow a considerable amount during infall. From Equation \ref{tgrowEqn2}, the growth time is of order a Gyr or greater, even at the present-day position of the Clouds at $r\sim 50-60$ kpc with $n_{h} \sim 10^{-4} \rm cm^{-3}$. Unless the LMC and SMC orbits are such that they have spent more time in the inner halo (and so has the Stream), $t_{grow} \gg t_{lookback}$, and the factor of a few gap in simulated (e.g. \citealt{Pardy2018}) vs observationally inferred \citep{ElenaReview2016} Stream mass cannot be closed by mass influx through a radiative turbulent mixing layer. Note that the stream might fragment and the individual pieces separate far enough to increase $A_{\rm eff}$ (cf. \S \ref{sec:background}), and thus, increase the cold gas growth rate. However, we expect this to be an order unity effect for the Stream.

\subsection{Simulations}
\label{sec:Stream_sims}

In this section, we present a set of simulations whereby a hot (T = $10^{6}$ K) background shears past a cold (T = $10^{4}$ K), cylindrical Stream. We use the FLASH MHD code \citep{FLASHRef}, which uses a directionally unsplit staggered mesh solver \citep{2009JCoPh.228..952L, 2013JCoPh.243..269L} based on a finite-volume, high-order Godunov scheme. 

Our setup is, in many ways, identical to that of \citep{Mandelker2020a} but with a few differences. We setup a 3D, axisymmetric cylinder with radius $r_{cyl}$ and length 24$r_{cyl}$ in a domain with periodic boundary conditions along the axis of the cylinder and inflow/outflow boundaries elsewhere. The cylinder is initially static, and the background has velocity $v = 228$ km/s, i.e. a Mach number $\mathcal{M} \sim 2$. This value is motivated by the LMC infall velocity in the outer halo from \citep{Besla2012, Salem2015RAMMEDIUM, Bustard2020}. The box dimensions are fiducially $24r_{cyl}\times 24r_{cyl}\times 24r_{cyl}$, and we employ a simple density-based static mesh refinement to focus resolution on the area surrounding the Stream. In units of $r_{cyl}/\Delta x$, our base resolution is 1.25, and our highest resolution is 20, which is achieved in the dense Stream and most of the surrounding mixing layer. As in \cite{Mandelker2020a}, we find converged Stream growth rates at this resolution; in higher resolution runs, the onset of the Kelvin-Helmholtz instability simply happens earlier (see bottom panel of Figure \ref{fig:Stream_mass_growth}).

For radiative cooling, we assume the gas is in photoionization equilibrium with the metagalactic UV background \citep{2012ApJ...746..125H} with no additional ionizing source from e.g. the Clouds or Milky Way. The cooling rate is a tabulated function of density and temperature, while the gas ionization state (which feeds into the equation of state) is a tabulated function of density and internal energy \citep{Wiersma2009ThePlasmas}. Motivated by HST-COS observations of the Stream's SMC filament (with $Z \sim 0.1 Z_{\odot}$) and LMC filament (with $Z \sim 0.5 Z_{\odot}$), as well as X-ray observations probing the metallicity of the Milky Way halo \citep{MillerBregman2015, Martynenko2021}, we assume that both the Stream and Milky Way halo have metallicity $Z = 0.3 Z_{\odot}$. The cooling (or heating rate) is simply the contribution from hydrogen and helium plus the contribution from metals, scaled down by this metallicity. Note once again that the growth time and drag time (Equations \ref{eqn:tdrag} and \ref{eqn:tgrow}) are insensitive to metallicity, while small changes between 0.1 and 0.5 $Z_{\odot}$ affect the critical column density for survival (Equation \ref{eqn:Ncrit}) by factors less than a few. Subcycling is utilized to better resolve the cooling time. A temperature floor of 300 K is included, although most of the cold Stream gas is maintained at close to $10^{4}$ K by photoionization (note that we don't include self-shielding). As for the Leading Arm simulations, we also cut off cooling for gas temperatures above $0.6T_{hot} = 6 \times 10^{5}$ K.

\begin{table}
  \centering
  \caption{Stream simulation parameters. For all simulations, we choose $\mathcal{M} \sim 2$, $T_{cyl} = 10^{4}$ K, $T_{hot} = 10^{6}$ K, and $\chi = 100$ such that the Stream and halo are initially in pressure equilibrium.}
  \begin{tabularx}{0.31\textwidth}{cccc}
  \toprule
   $r_{cyl}$ & $n_{h}$ & t$_{sc}$ & N  \\
   (kpc) & (cm$^{-3}$) & (Myrs) & (cm$^{-2}$) \\
  \hline
  0.5 & $10^{-5}$  & 44.6  & $3.1 \times 10^{18}$  \\
  2.0 & $10^{-5}$  & 172  & $1.2 \times 10^{19}$  \\
  0.5 & $10^{-4}$  & 44.6  & $3.1 \times 10^{19}$  \\
  5.0 & $10^{-5}$   & 446  & $3.1 \times 10^{19}$  \\
  10.0 & $10^{-5}$  & 892  & $6.2 \times 10^{19}$  \\
  2.0 & $10^{-4}$  & 172  & $1.2 \times 10^{20}$  \\
  5.0 & $10^{-4}$ & 446 & $3.1 \times 10^{20}$  \\
  \hline
  
  \end{tabularx}
\label{table_Stream}
\end{table}

\begin{figure}
\centering
\includegraphics[width=0.48\textwidth]{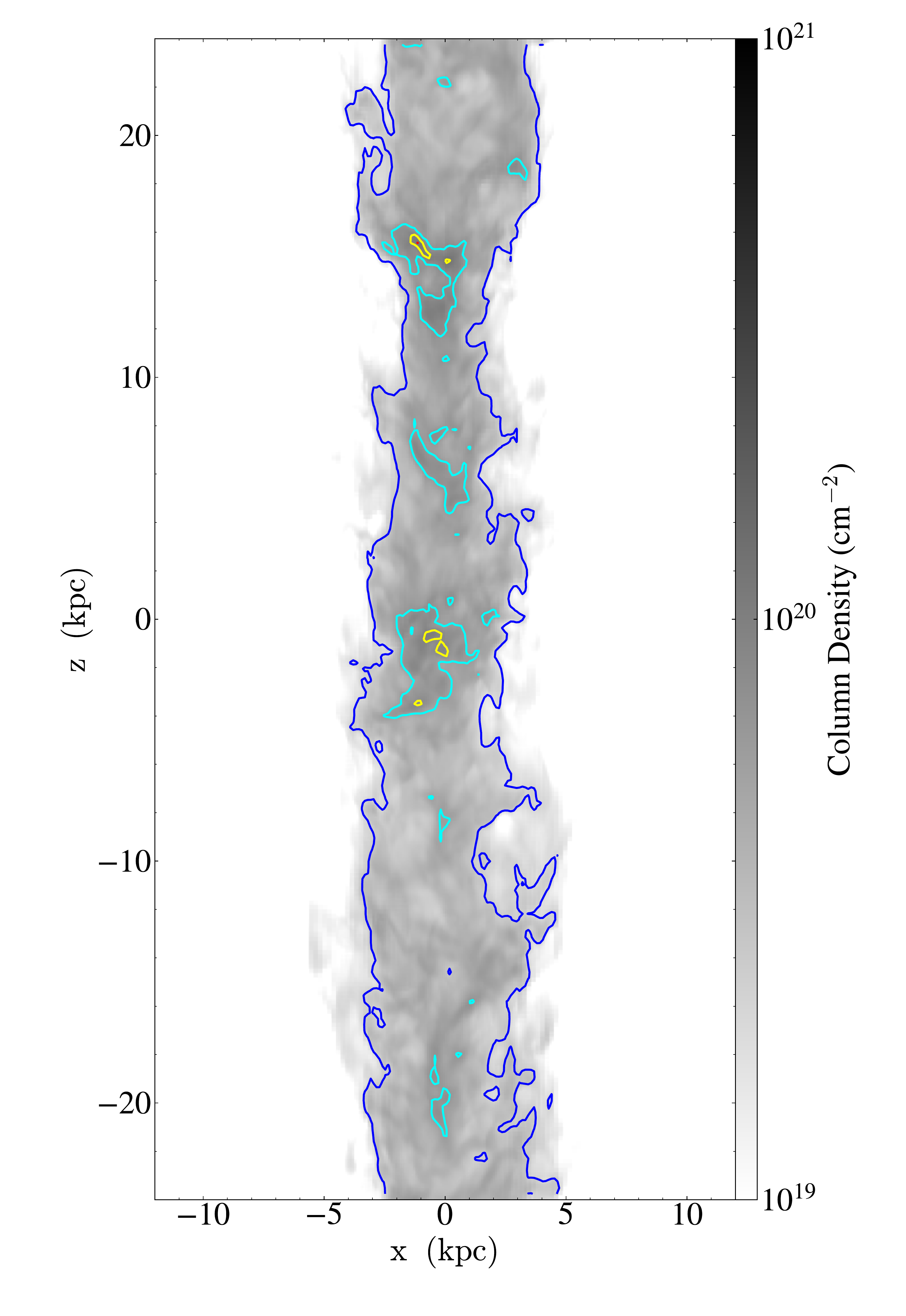}
\caption{Total Stream column density for the $r_{cyl} = 2.0$ kpc, $n_{h} = 10^{-4} \rm cm^{-3}$ simulation. Contours denote H$\alpha$ surface brightnesses (estimated as SB$_{cool}$/100) of 1 (blue), 5 (cyan), 10 (yellow), and 15 (red) mR. This snapshot is taken after 2 Gyrs $\sim 11.5 t_{sc}$ when the cold gas mass has grown by a few tens of percent. }
\label{fig:Trailing_Snapshots}
\end{figure}

Table \ref{table_Stream} shows the cylinder radius, background halo density, sound crossing time $t_{sc}$, and column density for each simulation we consider. Figure \ref{fig:Trailing_Snapshots} shows a column density snapshot after t $\sim 11.5 t_{sc}$ perpendicular to the cylinder axis for our simulation with initial radius $r_{cyl} = 2$ kpc in a background halo density of $n_{h} \sim 10^{-4}\,{\rm cm}^{-3}$. This gives an initial Stream column density of $1.2 \times 10^{20}$ cm$^{-2}$, well within the survival regime. Overplotted contours show estimated H$\alpha$ emission (see \S \ref{sec:Halpha})

Figure \ref{fig:Stream_mass_growth} shows the mass evolution for each of our simulations (top panel). The cut-off in growth vs destruction, calculated from Equation \ref{eqn:Ncrit} to be $\sim 10^{19} \rm cm^{-2}$ for low-metallicity gas appears robust: after a brief delay, in which the Kelvin-Helmholtz instability is growing from numerical noise, Streams with $N < 10^{19} \rm cm^{-2}$ gradually lose mass, while those with $N > $ few $\times 10^{19} \rm cm^{-2}$ gain mass. Even the $r_{cyl} = 5$ kpc, $n_{h} = 10^{-5} \rm cm^{-3}$ simulation shows an uptick in cold gas mass at late times, but we stop the simulation too early to follow the full evolution. Those with $N \sim 10^{19} \rm cm^{-2}$ appear marginal. 

The bottom panel of \ref{fig:Stream_mass_growth} shows the growth time for the two most quickly growing simulations with $r_{cyl} = 2, 5$ kpc and $n_{h} = 10^{-4} \rm cm^{-3}$. Simulations at our fiducial resolution ($r_{cyl}/\Delta x \sim 20$) develop the Kelvin-Helmholtz instability the fastest, but those with $r_{cyl}/\Delta x \sim 10$ converge eventually to the same $t_{grow} \sim (0.1-0.2)\chi t_{sc}$. This corresponds to $t_{grow} \sim 1.7$ and $4.2$ Gyrs for the $r_{cyl} = 2$ and $5$ kpc simulations, respectively. These rates are consistent with our analytic estimates (Equation \ref{tgrowEqn2}) but are, again, too long to give any considerable mass growth within a lookback time. 

\begin{figure}
\centering
\includegraphics[width=0.48\textwidth]{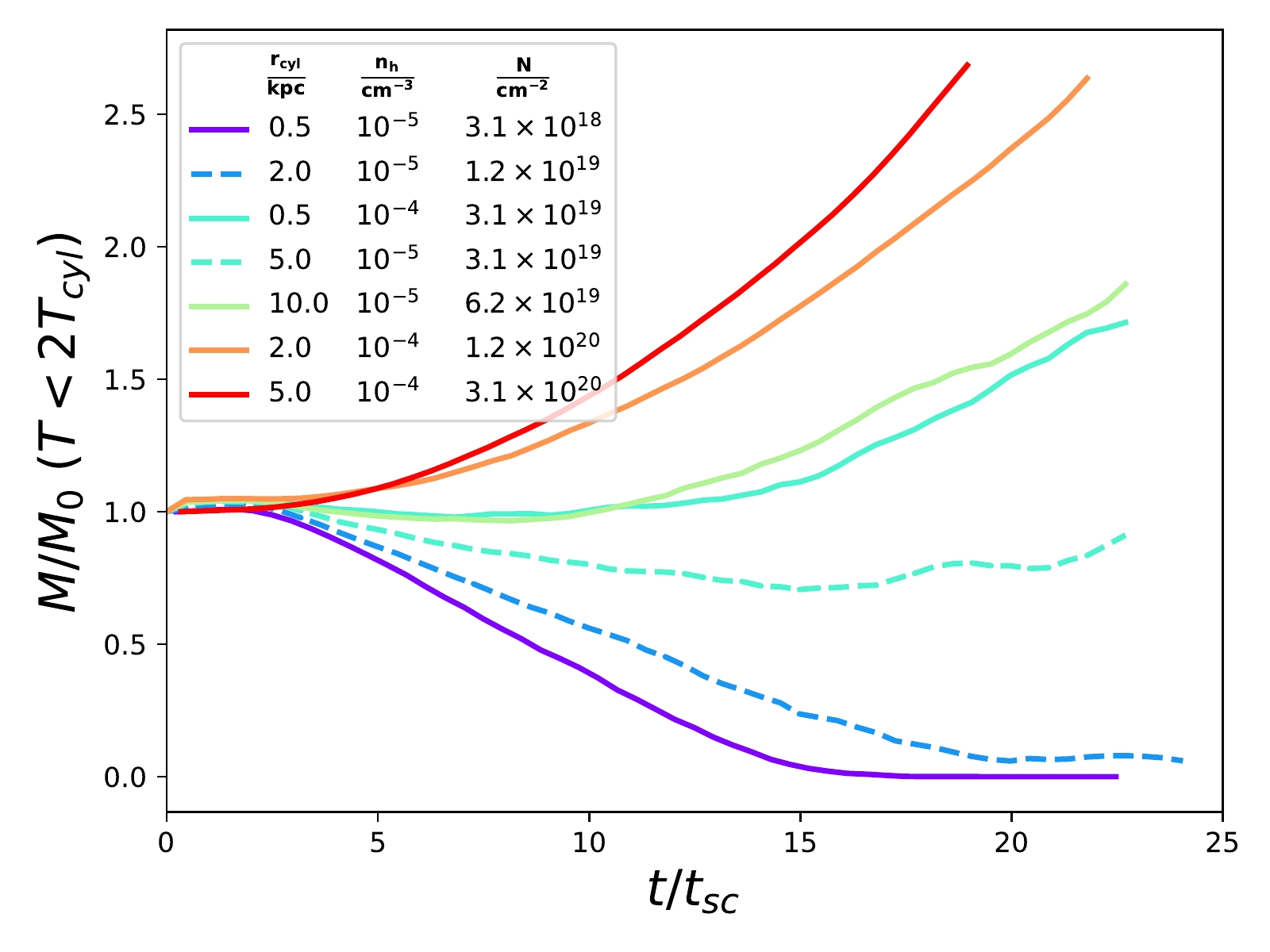}
\includegraphics[width=0.48\textwidth]{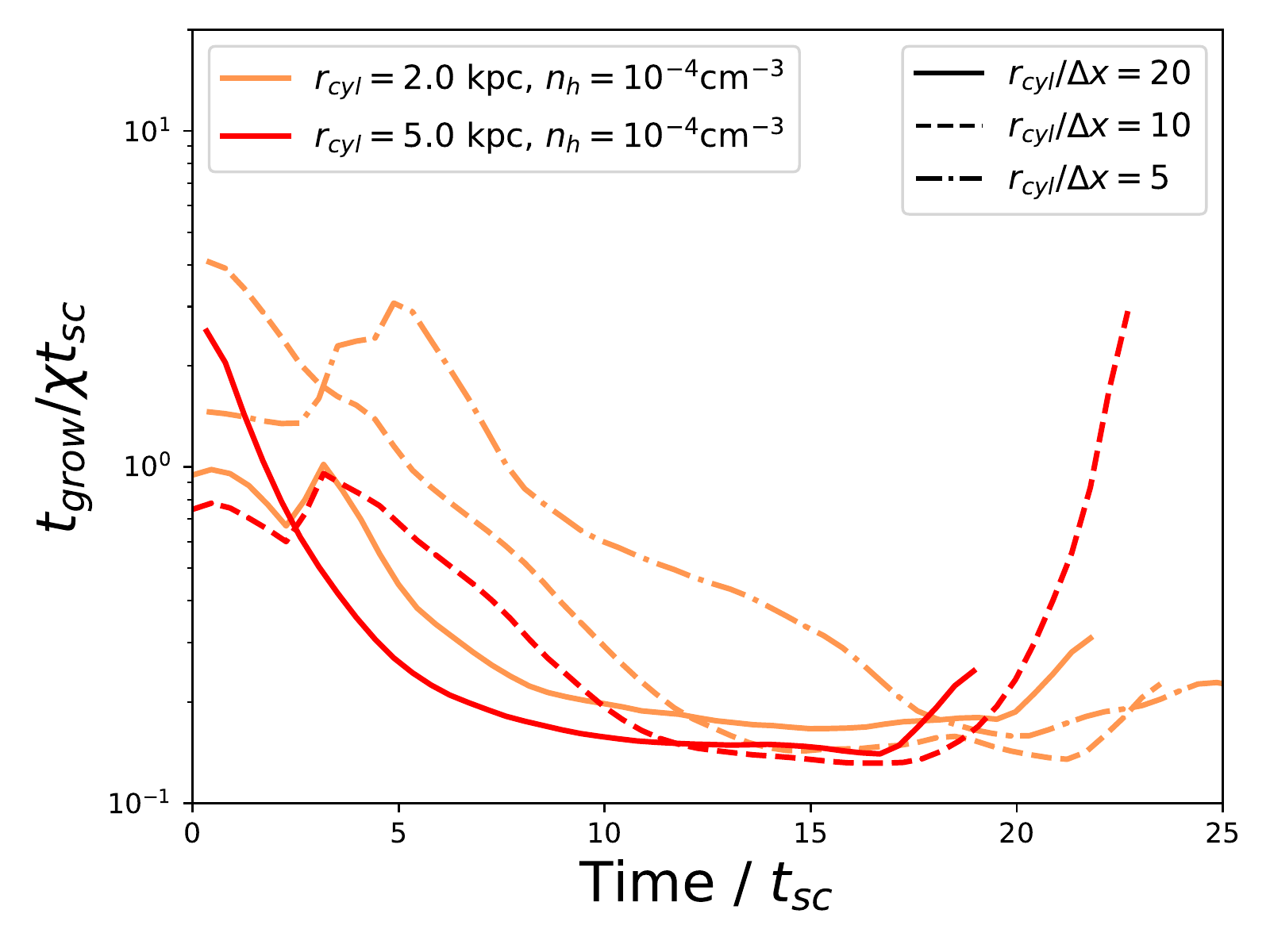}
\caption{\emph{Top:} $M/M_{0}$ for the cold gas (defined as $T < 2T_{cyl}$) for each simulation we ran. Lines are color-coded by Stream column density projected perpendicular to the Stream axis. Streams with column densities above a few $\times 10^{19} \rm cm^{-2}$ eventually start to grow in mass, roughly consistent with the expectation from Equation \ref{eqn:Ncrit} with $\mathcal{M} \sim 2$. \emph{Bottom:} Growth times for the simulations with the most rapid growth onset, and with varying resolution (different linestyles). Lower resolution leads to a delay in growth, but the growth times for $r_{cyl}/\Delta x \sim$ 10, 20 eventually converge to $\sim (0.1-0.2) \chi t_{sc}$, consistent with Equation \ref{tgrowEqn2}.}
\label{fig:Stream_mass_growth}
\end{figure}

\subsubsection{H$\alpha$ Emission}
\label{sec:Halpha}

Previous work focused on the Stream - halo interaction has been largely motivated by the peculiarly large H$\alpha$ emission observed along the Stream, 30-50 milliRayleigh (mR) along many sightlines, which is too high to originate from photoionization from the Clouds, Milky Way, and extragalactic background \citep{Weiner1996HAlpha, BlandHawthorn2013, Barger2017}. To partially explain this, a slow ``shock-cascade'' -- where stripped gas collides with Stream clouds further downstream, forming a cascade of shocks, ionizes the Stream -- was proposed in the literature \citep{BlandHawthorn2007, BlandHawthorn2009, TepperGarcia2015}. Figure 17 of \cite{Barger2017} shows the possible contribution of this shock-cascade to the measured H$\alpha$ intensity at various Magellanic Stream longitudes. Since the elevation in H$\alpha$ emission depends on the ambient density, this model is distance-dependent: \cite{Barger2017} estimate that if the Trailing Stream is positioned 75 kpc above the South Galactic Pole, the shock-cascade model could elevate the H$\alpha$ emission by 50 mR near the LMC and by 20 mR at the tail of the Stream. 

As discussed previously, though, an implication of this shock cascade model, in which the Stream is initialized far from pressure equilibrium, is a quick dissolution of Stream gas into the Milky Way halo; therefore, it's of interest to check if our simulations, which instead retain and even grow the Stream's cold gas mass, can produce similar H$\alpha$ bright spots at shocks in the radiative turbulent mixing layer.

The standard relationship between isobaric cooling luminosity and mass growth is \citep{Fabian1994}
\begin{equation}
    L = \frac{5}{2} \frac{\dot{m}}{\mu m_{p}} k_{B} T_{hot} (1+\mathcal{M}^{2})
\end{equation}
where the factor of $\mathcal{M}^{2}$ arises due to dissipation of the infalling cold stream's kinetic energy, in addition to the dissipation of thermal energy of the background cooling gas.

Using $\dot{m} \sim m_{0}/t_{grow}$, where $m_{0} = \pi r_{cyl}^{2} \ell \rho$ for a cylinder of length $\ell$, and assuming $t_{grow} \sim 0.1 \chi t_{sc} = 10t_{sc}$, the total cooling surface brightness over a cylinder of area A = $2\pi r_{cyl} \ell$ is  
\begin{equation}
    SB_{cool} = L/A \sim 100.5 \left(\frac{n_{h}}{10^{-4} cm^{-3}}\right) (1+\mathcal{M}^{2}) \quad \rm mR
    \label{eqn:SBCool}
\end{equation}

Our simulations appear to reproduce this quite well. As in \cite{Gronke2020a, Mandelker2020a}, we output the total cooling luminosity and find that it matches the expectation given $\dot{m}$ to within a factor of a few during the mass growth stage. Note that this luminosity decreases with increasing Stream distance from the Milky Way. For the same Stream radius, if we decrease the background halo density by a factor of 10, we find a uniform decrease in emission by a factor $\sim 10$ consistent with Equation \ref{eqn:SBCool}. 

The turbulent mixing layer is populated with gas at a variety of temperatures between $10^{4}$ K and $10^{6}$ K. We confirm that a large contribution to the cooling emissivity comes from $T \sim 10^{4}$ K, as seen in other mixing layer simulations \citep{Tan2021Letter}, including those that include photoionization \citep{Ji2019Mixing, Mandelker2020a}. This is because, even though photoionization significantly decreases the hydrogen bump in the cooling curve $\Lambda$ \citep{Wiersma2009ThePlasmas}, the emissivity is $\sim n^{2} \Lambda$, and the $n^{2}$ term offsets the low $\Lambda$. It's been suggested, then, that radiation in mixing layers of cold infalling streams can power Ly$\alpha$ blobs in massive halos at high redshift \citep{Goerdt2010,Mandelker2020b}; naturally, one should wonder whether detectable H$\alpha$ emission can arise too. 

We post-process our simulations to produce maps of H$\alpha$ surface brightness, calculated using Equation 13 from \citet{TepperGarcia2015}:

\begin{align*}
    \mu_{H \alpha}^{\rm (shock)} &= (1+(Y/4X)) K_{R} \alpha_{\beta}^{(H\alpha)} \int (n_{\rm H II})^{2} ds \\
    \alpha_{\beta}^{(H\alpha)}(T) &\approx \alpha_{0} \frac{(1+c)(T/10^{4} K)^{b}}{1+c(T/10^{4} K)^{d}}
\end{align*}
$K_{R} = 1.67 \times 10^{-4} \rm cm^{2} s mR$, the hydrogen and helium mass fractions are X = 0.7154, Y = 0.2703 appropriate for solar abundance, fully ionized gas \citep{Asplund2009}, and $\alpha_{\beta}^{(H\alpha)}(T)$ is the effective H$\alpha$ recombination coefficient following the fitting formula of \cite{Pequignot1991}. The constants are $\alpha_{0} = 1.169 \times 10^{-13} \rm cm^{3} s^{-1}$, c = 1.315, b = -0.6166, d = 0.523. To get the H II density, we employ Trident \citep{TridentRef} assuming photoionization equilibrium with the same metagalactic UV background we assume for our cooling function.

Contours in Figure \ref{fig:Trailing_Snapshots} show that H$\alpha$ emission is, in large regions, only of order 1 mR. During the period of $t \gtrapprox 7.5t_{sc}$, we measure the volume-averaged H$\alpha$ emission to be 0.5-1 mR with a standard deviation $\sim 1-2$ mR. Maxima fluctuate between 15 and 20 mR, somewhat lower than the findings of \cite{TepperGarcia2015}, where detectable ($> 30$ mR) surface brightnesses appear in small bright spots. As in \cite{TepperGarcia2015}, our measured average and maximum surface brightnesses are dependent on background halo density, and with a higher halo density, we may achieve detectable maxima. 

We caution, though, that we don't include self-shielding, so much of the Stream interior has unrealistically high H II densities. Our grid resolution is also higher than the $1\degree$ resolution of WHAM \citep{Barger2017}, and we don't account for beam smearing that would decrease the surface brightness. Nevertheless, this may be worth future study given that our evolved Stream is clearly turbulent, with bright H$\alpha$ knots similar to those in \cite{BlandHawthorn2007} and \cite{TepperGarcia2015} \emph{but without a fast disruption of the Stream}. Future work could also include the contribution from collisional excitation that peaks near $T \sim 2 \times 10^{4}$ K, whereas here we've only considered H$\alpha$ recombination. The difference is likely a small factor, with still only of order 1\% of the hydrogen cooling luminosity being due to H$\alpha$; the greatest contribution is Ly$\alpha$ \citep{Goerdt2010}. 

While shock ionization may be sufficient to explain the underlying H$\alpha$ emission along the Stream, these values of tens of mR fall far short of the observed Stream emission above the South Galactic Pole, where there is an especially large bump in H$\alpha$ emission ($>$ 300 mR). The most plausible explanation for this, so far, is recent activity from the Galactic Center. A bipolar, radiative ionization cone from a Seyfert nucleus associated with SGR A* can not only explain the elevated H$\alpha$ emission but also higher hydrogen ionization fraction, higher C IV/ CII and Si IV/ Si II ratios, and high C IV and Si IV column densities along that segment of the Stream \citep{BlandHawthorn2013,BlandHawthorn2019, Fox2020}. 

Another possible explanation for elevated H$\alpha$ emission is a change in Clouds orbit, leading to a Stream that is closer to the Milky Way. Recent simulations suggest this might be the case, with a column density-weighted average Stream distance of only $d \approx 24.7$ kpc from the Sun \citep{Lucchini2021}. In this scenario, the H$\alpha$ surface brightness needed to explain the observed luminosity is lower ($\propto d^{2}$). Also, not only would the larger UV radiation field from the Milky Way increase the H$\alpha$ emission, as noted in \cite{Lucchini2021}, but also the mass growth rate onto the Stream would increase. The total cooling surface brightness (Equation \ref{eqn:SBCool}) could then be as large as 1,000-10,000 mR, with an average H$\alpha$ contribution of order 10-100 mR, if $n_{h} \sim 10^{-3}$ at 20 kpc \citep{stanimirovic2002, Bregman2018}.

\section{Discussion}
\label{sec:discussion}

While the main purpose of our work was to apply results from radiative turbulent mixing to the Magellanic Stream in particular, our results also inform our broader understanding of extragalactic multiphase gas more broadly. We begin our discussion with implications for cold filaments, galactic winds, and ram pressure stripped tails behind infalling dwarf galaxies, specifically the LMC and SMC (\S \ref{sec:implications}. We then discuss refinement guidelines to accurately capture mass entrainment in large-scale LMC-SMC simulations (\S \ref{sec:resolutionGuidelines}) and discuss the caveats of our hydrodynamic treatment (\S \ref{sec:missingPhysics}).

\subsection{Broader Extragalactic Implications}
\label{sec:implications}
 
While we focused in this work on the Milky Way halo, our findings also have implications for extragalactic systems. Most notably, they can be applied to multiphase galactic winds, ram pressure stripping of galaxies, and cosmological filamentary inflows. As the latter has been discussed extensively in the literature \citep{Mandelker2020a,Mandelker2020b}, we focus on the former two.

\emph{Galactic winds:} Observations of fast-moving, cold gas in galactic winds \citep{Veilleux2020} have driven a surge of interest in the cloud-crushing problem. A main development in recent years has been the aforementioned survival and growth of cold gas in certain regimes ($t_{cool,mix} < t_{cc}$), but most studies to-date have focused on transonic or mildly supersonic flows. For a spherically symmetric cloud, previous simulations in the mass growth regime showed poor convergence in high-$\mathcal{M}$ flows \citep{Gronke2020a}, possibly due to shattering of cold gas into smaller and smaller clumps down to the cooling length $c_{s}t_{cool}$ \citep{McCourt2018} expected for $\mathcal{M} \gtrsim 1.6$ flows at $\chi \gtrsim 100$ \citep{Gronke2020b}. This scale is not resolved in our work or in any published cloud-crushing simulations for overdensities of $\chi \sim 100$. Nevertheless, with a slight shift from a spherically symmetric cloud to an axisymmetric 2.5D cloud, we find reasonable convergence, likely due to a ``shielding" effect, i.e., a more quiescent region in the tail leading to the re-coagulation of the clumps instead of further fragmentation (as seen in \citealp{McCourt2018, Melso2019}, see also the setup of \citealp{Banda2020}). 

High-$\mathcal{M}$ flows do, however, have longer drag times $\propto 1+\mathcal{M}$ and longer growth times (\S \ref{sec:dragExplanation} and Figure \ref{fig:tgrow_area}). These aspects deserve future study since they call into question the feasibility of cold cloud acceleration and growth in highly supersonic outflows. We speculate that this is why the high resolution, global galaxy simulations of \cite{Schneider2020}, which model a highly supersonic M82-like wind, don't find clear evidence of mass influx from the hot to cold phase. Instead, especially at large radii, mass flows from the cold to hot phase. This may be explained by the higher drag times and higher growth times we find. Interestingly, the cold mass outflow rate in \cite{Schneider2020} rises until a distance of r $\sim 3$ kpc, which could reflect either an increase in total mass outflow rate or efficient cloud growth in this region where the hot phase is instead only mildly supersonic (see their Figure 6).

These additional dependencies on the Mach number are particularly important for subgrid models of cold gas in galactic winds such as \citet{PHEW} as they affect the cold gas mass and velocity, and hence most potential observables. Furthermore, if geometry dependent effects such as the `shielding' can indeed change the behavior dramatically (as suggested in this work), they would need to be taken into account by such models.

\emph{Ram pressure stripped tails:} In addition to LMC-SMC tidal stripping, a popular explanation for the twisting, bifurcated filaments of the Stream is ram pressure stripping of the Clouds by the Milky Way corona (e.g. \citealt{Moore1994, Hammer2015THECLOUDS, Wang2019}). While this is achieved in some simulations with dense Milky Way coronae, it's problematic that detailed simulations of LMC ram pressure stripping, which are tuned to fit the truncation radius of the LMC's leading edge, find that the mass stripped from the galaxy is small compared to the Stream mass \citep{Salem2015RAMMEDIUM}. It increases somewhat when LMC outflows are flung into the headwind \citep{Bustard2018, Bustard2020}, but seemingly not enough to create the observed LMC filament  \citep{Nidever2008TheArm}. However, these simulations, which focus resolution on the dense LMC disk, fall short of the refinement criterion (\S \ref{sec:resolutionGuidelines}) to capture cold gas survival \emph{behind} the LMC. The result is that recent ejecta is quickly mixed into the Milky Way halo, but this situation should actually be quite ripe for mass growth: the LMC/SMC infall is only $\mathcal{M} \sim 2-3$, and stripped gas forming a long, cooling tail undergoes areal expansion that is key to rapid mass growth. Such an effect is, indeed, realized in recent simulations of ``jellyfish" galaxies ram pressure stripped by the intracluster medium \citep{Tonnesen2021}. 

Similar coagulation and mass growth in long, cooling tails behind the Clouds may be an enticing way to produce the bifurcated structure of the Stream without requiring more direct ejection of gas from the Clouds via an increase in halo density (i.e. an increase in $P_{ram} \sim \rho v^{2}$).

\subsection{Refinement Guidelines}
\label{sec:resolutionGuidelines}

As discussed in \S~\ref{sec:LATheory}, one posibility for the inability of larger scale simulations to reproduce the Leading Arm is numerical overmixing due to insufficient resolution. In this section, we estimate what the resolution or refinement requirements are in order to see the survival and growth of the Leading Arm discussed in this work.

A cloud of radius $r_{cl}$ and density $\rho_{cl}$ has a mass $M_{cl} = 4/3 \pi \rho_{cl} r_{cl}^{3}$. From here on, let's assume the approximately $10^{4}$ K cloud is in pressure equilibrium with its surroundings, meaning its overdensity $\chi \sim 100$ relative to the background. For a mean mass of $\bar{m} = 1.67 \times 10^{-24}$ g, $\rho_{cl} = 100 \bar{m} \times 10^{-4} n_{-4}$ where $n_{-4}$ is the background halo density normalized to $10^{-4}$ cm$^{-3}$. We then re-write the cloud radius in terms of the total gas column density and the background halo density: $r_{cl} \approx N / 2 n_{cl} = 50 N_{19} / n_{-4}$ where $N_{19}$ is the column density normalized to $10^{19}$ cm$^{-2}$. The cloud mass is then 

\begin{equation}
    M_{cl} \approx 4 \times 10^{3} M_{\odot} \frac{N_{19}^{3}}{n_{-4}^{2}}
\end{equation}

To resolve such a cloud by $l$ cells per dimension at a mass resolution of $m_{res}$, one needs $\frac{M_{cl}}{m_{res}} = \frac{4}{3} \pi l^{3}$. We re-write this in terms of the column density, i.e. what column density can be resolved by $l$ cells at a mass resolution of $m_{res}$:
\begin{equation}
    N_{19} \approx 0.1 n_{-4}^{2/3} l m_{res}^{1/3} = n_{-4}^{2/3} l m_{res,3}^{1/3}
\end{equation}
where $m_{res,3} = m_{res}/10^{3} M_{\odot}$. 

Simulations suggest convergence of mass growth when the cloud is resolved by a minimum of $l =$ 8 cells \citep{Gronke2020a, kanjilal2021}; therefore, at the mass resolution of $\sim 4 \times 10^{3} M_{\odot}$ typical of state-of-the-art Magellanic system simulations \citep{Pardy2018, Wang2019, Lucchini2020}, only gas structures with column densities $\gtrapprox 10^{20} \rm cm^{-2}$ at $n_{-4} \sim 1$ or column densities $\gtrapprox 2.5 \times 10^{19} \rm cm^{-2}$ at $n_{-4} \sim 0.1$ are expected to have converged mass flux through the cloud-halo interface. In the outer halo, where the latter value of $n_{-4} \sim 0.1$ is probably appropriate, this means the evolution of clouds with masses $\gtrapprox 6 \times 10^{6} M_{\odot}$, less than e.g. the current Leading Arm mass of $3 \times 10^{7} M_{\odot}$, are well-reproduced; however, for higher background densities deeper into the Milky Way halo, convergence is unlikely even for structures as massive as the Leading Arm. 

Similarly, \cite{TepperGarcia2019} use a grid-based code and quote a maximum spatial resolution of $\sim 60\,$pc, which is sufficient to resolve (by $l = 8$ cells) a column of $N \sim 10^{20} \rm cm^{-2}$ at $n_{-4} \sim 1 $ or a column of $N \sim 10^{19} \rm cm^{-2}$ at $n_{-4} \sim 0.1$. While they do not mention their refinement criterion, we can presume for the sake of argument that they have a density threshold in order to keep the mass per cell $\sim $constant. 

For the cloud-crushing problem, in particular, these refinement schemes are likely problematic. Much of the mass growth occurs in the tail behind the disrupting cloud, and with coarse-resolution in low-density gas, the cloud fragments that would coagulate and grow at higher resolution would instead irreversibly mix into the background. This issue of numerical convergence plagues all hydrodynamic, galaxy-scale simulations to some extent, prompting the adoption of new refinement schemes \citep{vandeVoort2019, Hummels2019, Peeples2019, Nelson2020, Mandelker2021} that force uniform refinement in the CGM or are based on pressure or cooling length instead of density. Such schemes are key to more accurate modeling of cold gas in galaxy halos, and they may have great utility for simulations of the Magellanic System as well.

\subsection{Missing Physics}
\label{sec:missingPhysics}

In this study, we used a simplified setup and omitted several physical effects -- most notably the impact of magnetic fields and thermal conduction. While it has been shown in similar `wind tunnel' studies that magnetic draping can facilitate entrainment through magnetic drag \citep{McCourt15}, for overdensitites of $\chi \sim 100$ fairly strong magnetic fields ($\beta< 1$ in the hot medium, cf. equation (23) in \citealp{Gronke2020a}) are required to allow entrainment. Such values of $\beta$, while likely not realized on average in the CGM \citep{Prochaska2019}, may be created by magnetic field amplification at cloud interfaces. For instance, \cite{McClureGriffiths2010} measure a coherent field strength $> 6 \mu$G in a Leading Arm HVC, which is strong enough to have an influence, but the corrections to drag time and cold mass growth appear to be order unity even in simulations with $\beta$ approaching $10^{-2}$ due to compressional amplification \citep{Gronke2020a,Butsky2020,Ji2016}.

Similarly, thermal conduction does impact the dynamics and can prolong the lifetime of cold gas by compressing the clouds \citep{Bruggen2016THECONDUCTION}. Here, we focused on the $r_{\rm cl} > r_{\rm cl,crit}$ regime in which cold gas can survive altogether. The Leading Arm and Trailing Stream comfortably fit into this regime. Furthermore, we note that it has recently been shown that the inclusion of thermal conduction does not affect the cold gas mass growth rate in turbulent mixing layers \citep{Tan2021} which also explains the convergence we find in $t_{\rm grow}$. This is because the mass transfer is not dominated by numerical diffusion but rather by mixing, and we do resolve the outer mixing scale $\sim r_{\rm cl}$ reasonably well.

However, we note that both the inclusion of magnetic fields as well as thermal conduction can affect the morphology of the cold gas \citep[e.g.,][]{Gronke2020a,Bruggen2016THECONDUCTION} as well as observables \citep{Tan2021Letter} and would need to be considered in a study focusing on the long-term evolution of the Leading Arm. Naturally, in such future work one would also need to take into account the hot gas halo profile as well as the effect of gravity.

\section{Conclusions}
\label{sec:conclusions}

In this paper, we studied the effects of radiative turbulent mixing on the $\sim 10^4\,$K gas originating from the Magellanic clouds. 
We were motivated by two questions concerning the formation, survival, and possible growth of Magellanic components: 1) Can the Leading Arm survive its (possibly high Mach number) ``swing-out'' from the Clouds and sufficiently separate itself from the Clouds in opposition to ram pressure? 2) What are the rate and sign of the mass flux through the Trailing Stream's radiative turbulent mixing layer, and is this sufficient to grow previously simulated Streams (e.g. \citealt{Pardy2018}) to the observationally estimated mass $ > 2 \times 10^{9} M_{\odot}$ \citep{ElenaReview2016}?

\subsection{Survival and Growth of Cold Gas in High Mach Number Flows}
Our findings from the simulations (\S \ref{sec:LASimulations}) are as follows:

\begin{itemize}
    \item For Mach numbers tested (up to $\mathcal{M} = 6$), we find cloud survival and subsequent growth via radiative, turbulent mixing that has been previously shown for lower Mach number flows. The survival criterion ($r_{cl} > r_{crit}$, or $t_{cool,mix}/t_{cc} \sim 1$) (Equation \ref{rclEqn}) appears robust, and we find good convergence with our assumed 2.5D cloud morphology (see Appendix). This is contrary to 3D, spherical cloud simulations in the mass growth regime, which show poor convergence at high Mach number \citep{Gronke2020a}.
    
    \item We find $t_{drag} \sim \alpha (1+\mathcal{M})\chi^{1/2}t_{cc}$, with $\alpha \sim 0.2-0.3$ for our strongly cooling ($t_{cool,mix}/t_{cc} \sim 0.1$) simulations. The Mach number dependence follows from a combination of geometric effects and momentum transfer via mixing -- both critically depend on cooling (see Appendix and Figure \ref{fig:M45_varytcc}), unlike adiabatic simulations where $t_{drag}$ is independent of $\mathcal{M}$ (\S \ref{sec:dragExplanation}). 
    
    \item Longer drag times for high-$\mathcal{M}$ flows are accompanied by longer cloud tails in the strong cooling regime: $l_{tail}(t)/r_{cl,0} \sim \alpha (1+\mathcal{M}) \chi \rm ln(1+t/t_{drag})$. This increases the surface area and eventually offsets the destructive effects of high Mach number cloud crushing, leading to considerable mass growth. In the weak cooling regime ($t_{cool,mix}/t_{cc} \sim 1$), mass growth is slower, and the tail length is shorter (Figure \ref{fig:Mach45_Projections}).
    
    \item In addition to longer drag times for high-$\mathcal{M}$ flows, we also find longer growth times $t_{grow} \propto \mathcal{M}^{3}$ (Figure \ref{fig:tgrow_area}). This suggests that cloud entrainment and growth in highly supersonic outflows, such as starburst winds like M82, is more difficult than in transonic flows typically studied in many cloud-crushing simulations. Cloud growth along a wind flux tube is, in fact, likely distance-dependent. This would need to be taken into account when comparing to larger scale galactic wind simulations.
    
\end{itemize}

\subsection{Implications for the Leading Arm}
These results have implications, most clearly, for the Leading Arm, whose Magellanic origin has recently been questioned based on arguments of short drag times and fast Leading Arm breakup in the Milky Way halo \citep{TepperGarcia2019}; the proposed alternative is that the Leading Arm is debris from a fore-runner dwarf galaxy. Our complementary findings, based on idealized but high-resolution simulations, are as follows:

\begin{itemize}
    \item Reasonably massive ($ \sim 3 \times 10^{7} M_{\odot}$, \citealt{Bruns2005}) Leading Arms have sufficient column densities ($N > N_{crit} \approx 4.8 \times 10^{18} \mathcal{M}$ cm$^{-2}$) to survive infall through the Milky Way halo. In this scenario, the filaments and clumps of the Leading Arm could be part of a long debris field in the wake of the progenitor cloud (see Figure \ref{fig:ProjectionPlots}), which survives infall and even grows in mass due to radiative cooling at the cloud-halo interface. The distance variation along the Leading Arm inferred from observations  supports this picture.
    \item Whether the Leading Arm can overcome drag and stay ahead of the Clouds is most sensitive to the Leading Arm's velocity at swing-out. A more violent tidal stripping event ($\mathcal{M} > 4$), especially in the lower-density outer halo, is preferable (Figure \ref{fig:LA_analytic}). Some more recent formation scenarios are permissible but require higher $\mathcal{M}$ swing-outs.
    \item The drag time and mass doubling time for these allowed scenarios are at least as long as the infall time, however, meaning we could be seeing the Leading Arm in its more destructive, stretching phase rather than its rapid growth phase (Figure \ref{fig:LA_profile}). The Leading Arm, then, likely had to be at least as massive at swing-out as it is today.
\end{itemize}

While our study focuses on the origin of the Leading Arm due to an earlier LMC-SMC interaction, the results presented here would also affect the `forerunner' theory (in which the Leading Arm is due to gas stripped from other dwarf galaxies). 
In each case, the Leading Arm comprises the disrupted clouds and filaments of a larger progenitor object. However, our results suggest that the Leading Arm may not need to originate in a dwarf galaxy protected by a dark matter halo, as suggested by \cite{TepperGarcia2019}. Instead, cooling and coagulation allows the clumps to survive and even grow in mass despite fast breakup of the progenitor cloud. Sufficiently resolved simulations of Leading Arm infall subject to gravity, a changing density profile, etc. are a logical next step to confirm this.

\subsection{Implications for the Trailing Magellanic Stream}
We also briefly estimate the effects of radiative turbulent mixing layers on the Trailing Stream (\S \ref{sec:trailingstream}).

\begin{itemize}
    \item We estimate that the majority of the Stream (with $N > N_{crit} \approx 10^{19}$ cm$^{-2}$) is surviving infall and gaining mass, but the mass growth rate is too long (a few Gyrs) to significantly increase the Stream mass. 
    \item The associated cooling luminosity and estimated H$\alpha$ emission in the turbulent mixing layer are likewise fairly small on average, but bright spots (tens of mR) may partially explain the observed H$\alpha$ emission along pointings separated from the main HI stream \citep{Barger2017}. This shock ionization is qualitatively similar to the shock cascade model put forth by \cite{BlandHawthorn2007} but here operates in a Stream that grows in mass instead of quickly dissolves.
    \end{itemize}

\section*{Acknowledgments}
The authors gratefully acknowledge the anonymous referee and Nir Mandelker for detailed feedback, as well as Elena D'Onghia, Andy Fox, Scott Lucchini, Peng Oh, Ellen Zweibel and the organizers and participants of the KITP ``Fundamentals of Gaseous Halos" workshop. CB was supported in part by the National Science Foundation under Grant No. NSF PHY-1748958 and by the Gordon and Betty Moore Foundation through Grant No. GBMF7392. 
MG was supported by NASA through the NASA Hubble Fellowship grant HST-HF2-51409 and acknowledges support from HST grants HST-GO-15643.017-A, and HST-AR15039.003-A.
Computations were performed on the Stampede2 supercomputer under allocations TG-PHY210004, TG-AST190019 and TG-AST180036 provided by the Extreme Science and Engineering Discovery Environment (XSEDE), which is supported by National Science Foundation grant number ACI-1548562 \citep{xsede}. 

\software{Athena (v4.0; \citealt{Athena2008}), yt \citep{ytPaper}, Trident \citep{TridentRef}, FLASH (v4.2; \citealt{FLASHRef}), Matplotlib \citep{matplotlibRef}}

\appendix

\subsection{Varying Halo Density Profile}
\begin{figure*}
\centering
\includegraphics[width=0.45\textwidth]{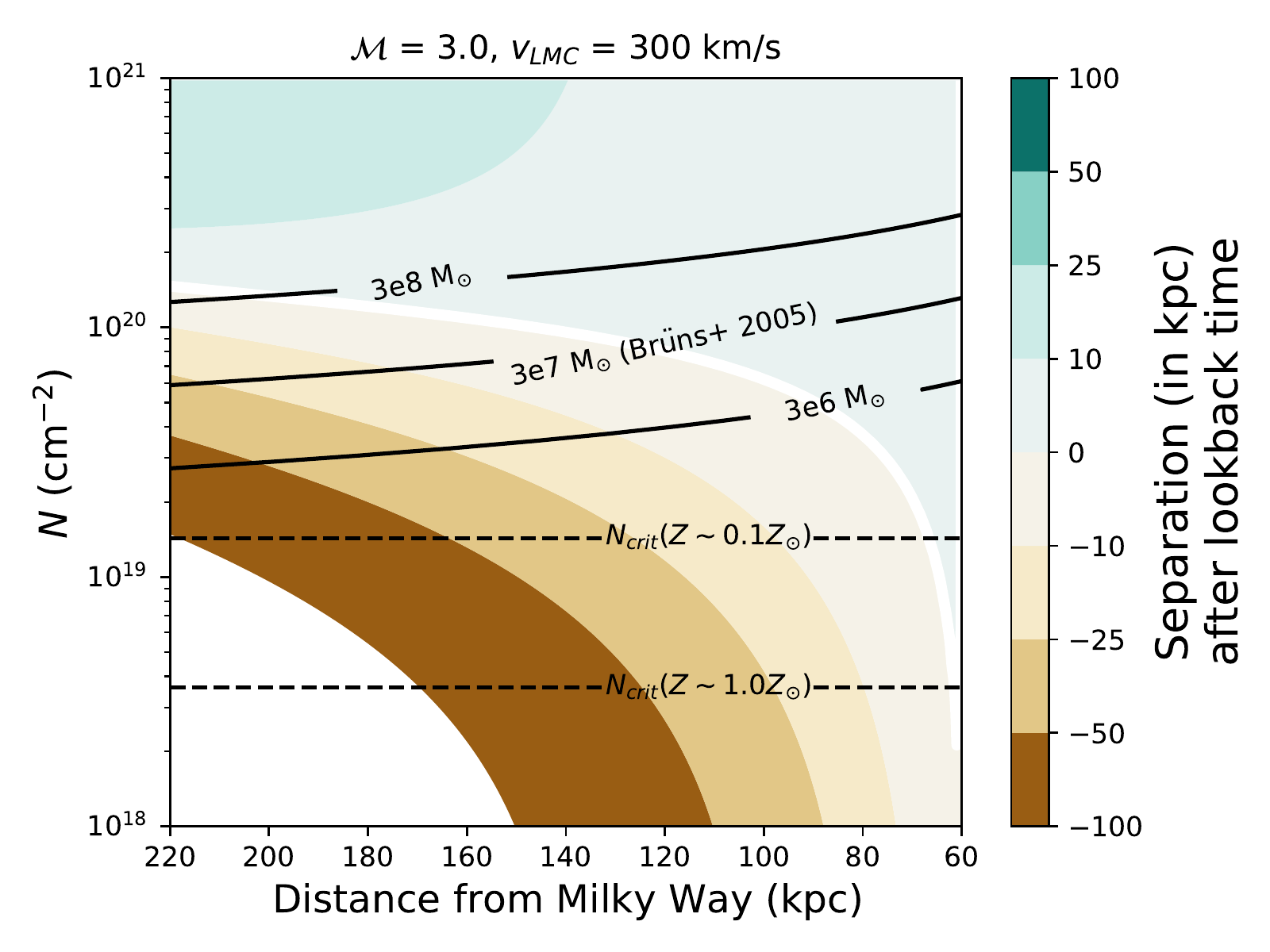}
\includegraphics[width=0.45\textwidth]{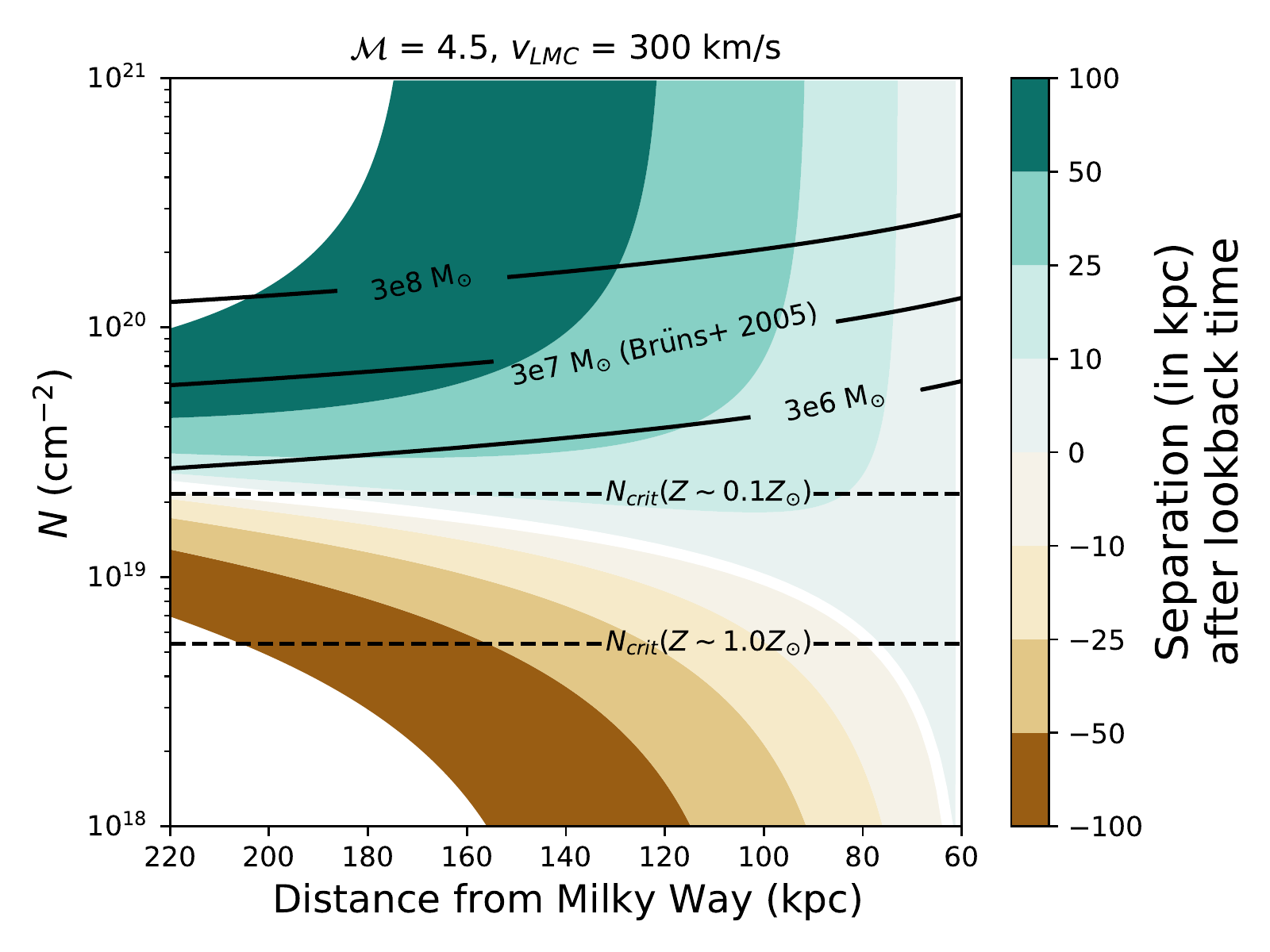}
\caption{Final separation between Leading Arm and LMC/SMC, now assuming the Milky Way halo density profile from \cite{Faerman2020}. Mappings between LMC/SMC radius and lookback time are unchanged. The results are nearly identical to that shown in Figure \ref{fig:LA_analytic}.}
\label{fig:LA_analytic_Appendix}
\end{figure*}

In \S \ref{sec:LATheory}, we probed the survival of the Leading Arm and its separation from the Clouds using a toy analytic problem. Our results are presented in Figure \ref{fig:LA_analytic} assuming the \citealt{Salem2015RAMMEDIUM} Milky Way halo density profile (Equation \ref{eqn:Salem}). Here, Figure \ref{fig:LA_analytic_Appendix} shows the same analysis using instead the \citealt{Faerman2020} density profile (Equation \ref{eqn:Faerman}), which gives a denser Milky Way corona beyond $\sim 100$ kpc. The resulting figure is almost identical to Figure \ref{fig:LA_analytic}, showing our results are insensitive to small changes in halo density; instead, the separation is most sensitive to the velocity of the Leading Arm swing-out.

\subsection{Resolution Study}
\begin{figure*}
\centering
\includegraphics[width=0.4\textwidth]{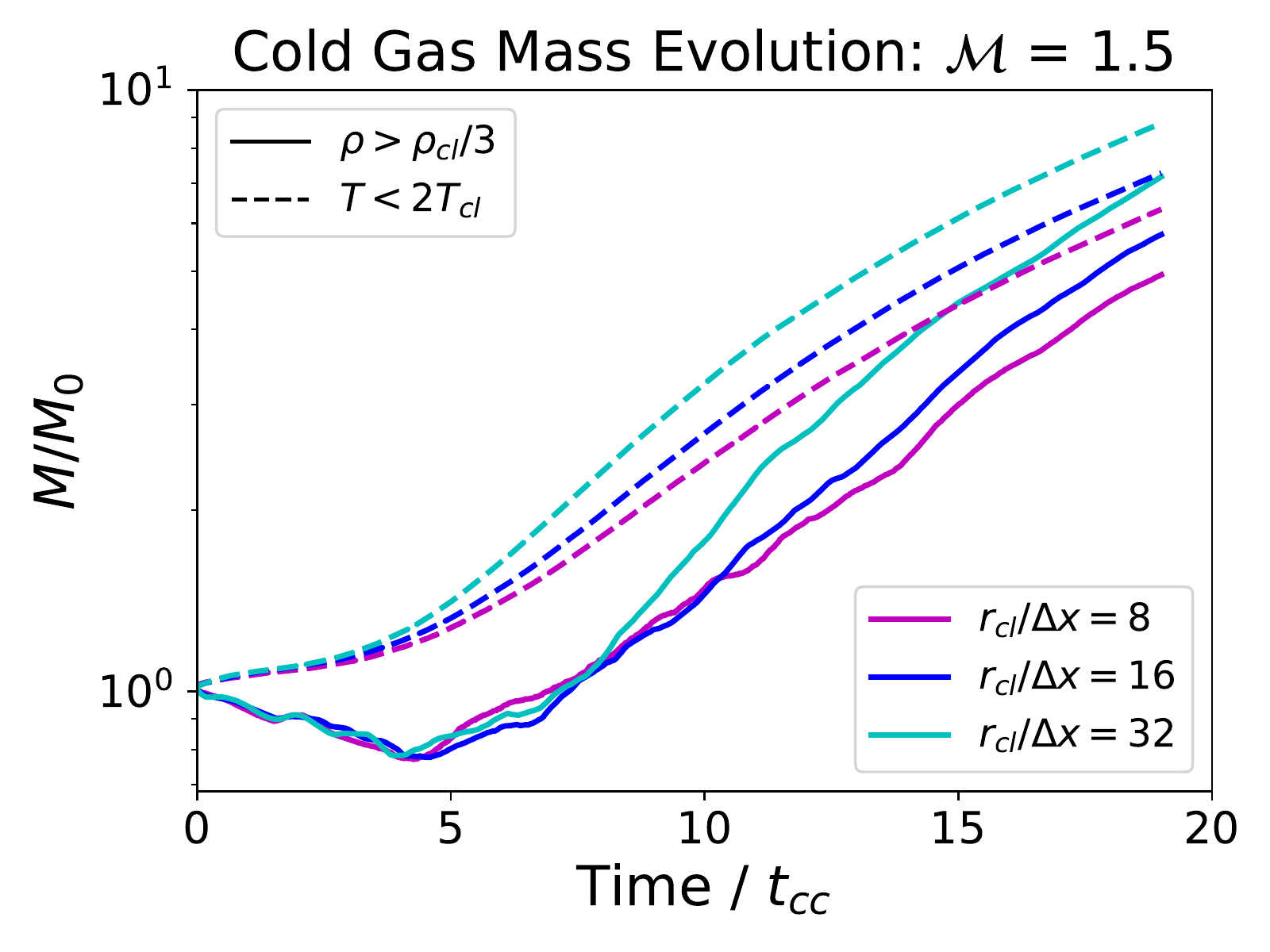}
\includegraphics[width=0.4\textwidth]{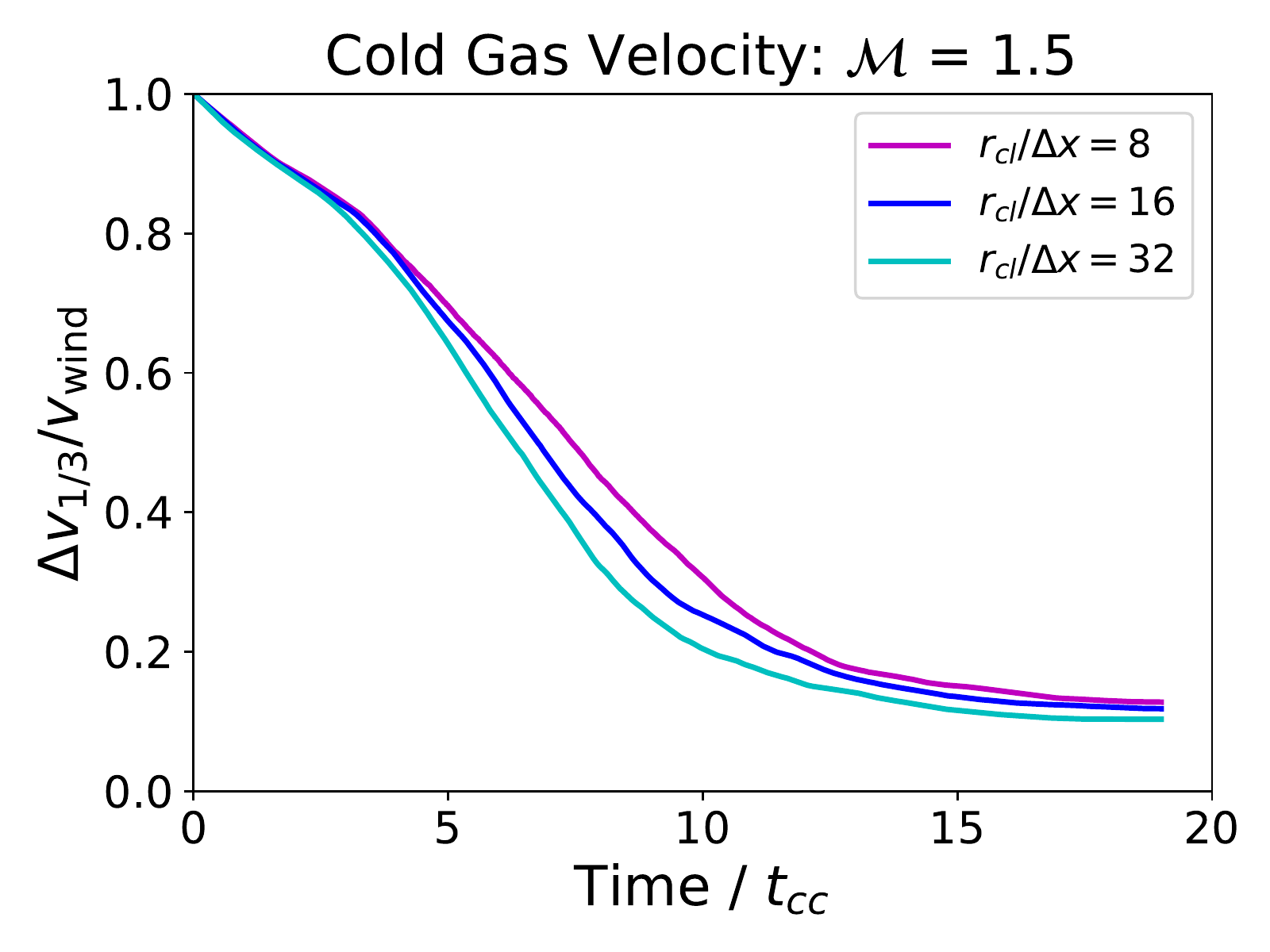}
\includegraphics[width=0.4\textwidth]{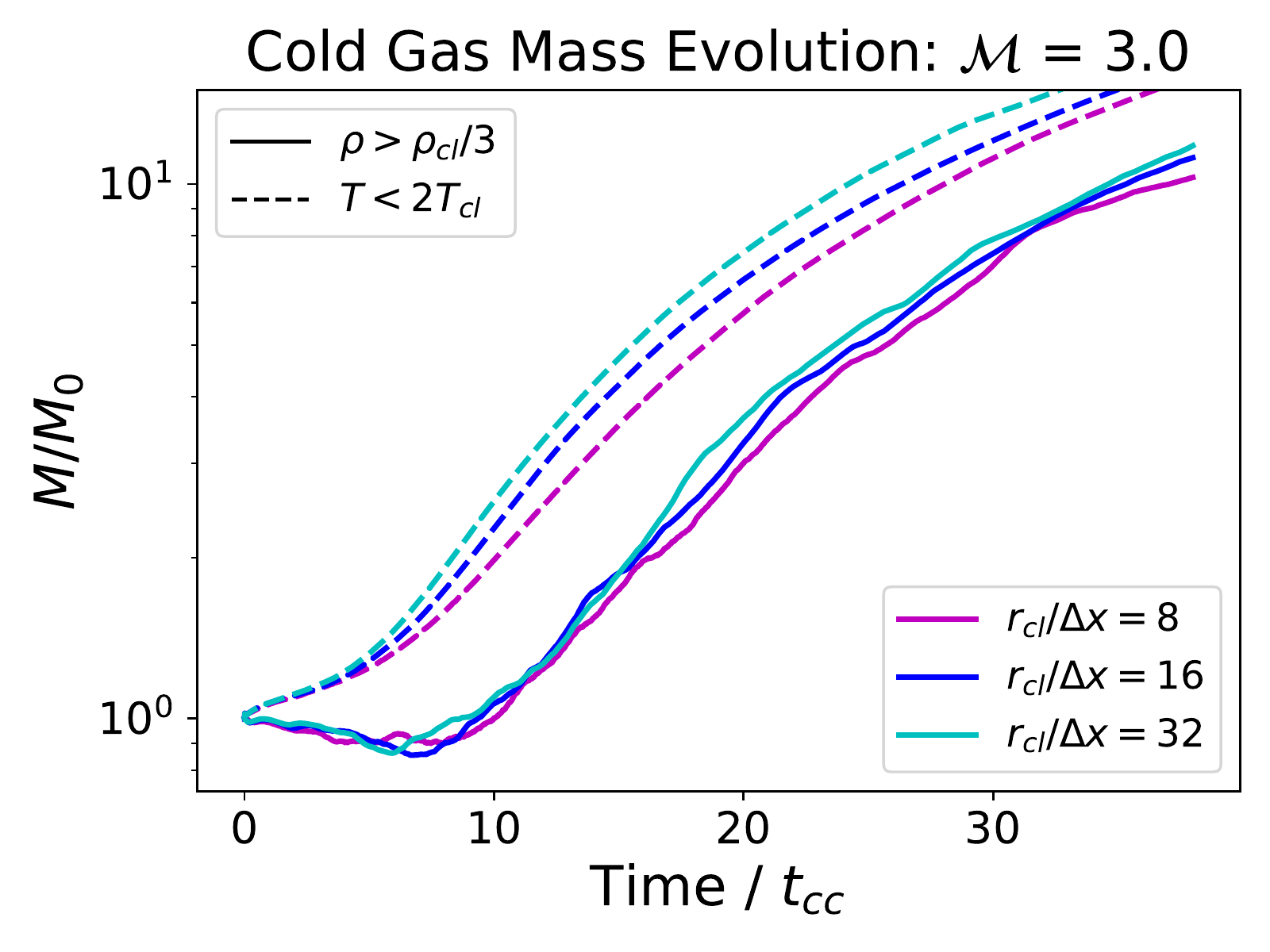}
\includegraphics[width=0.4\textwidth]{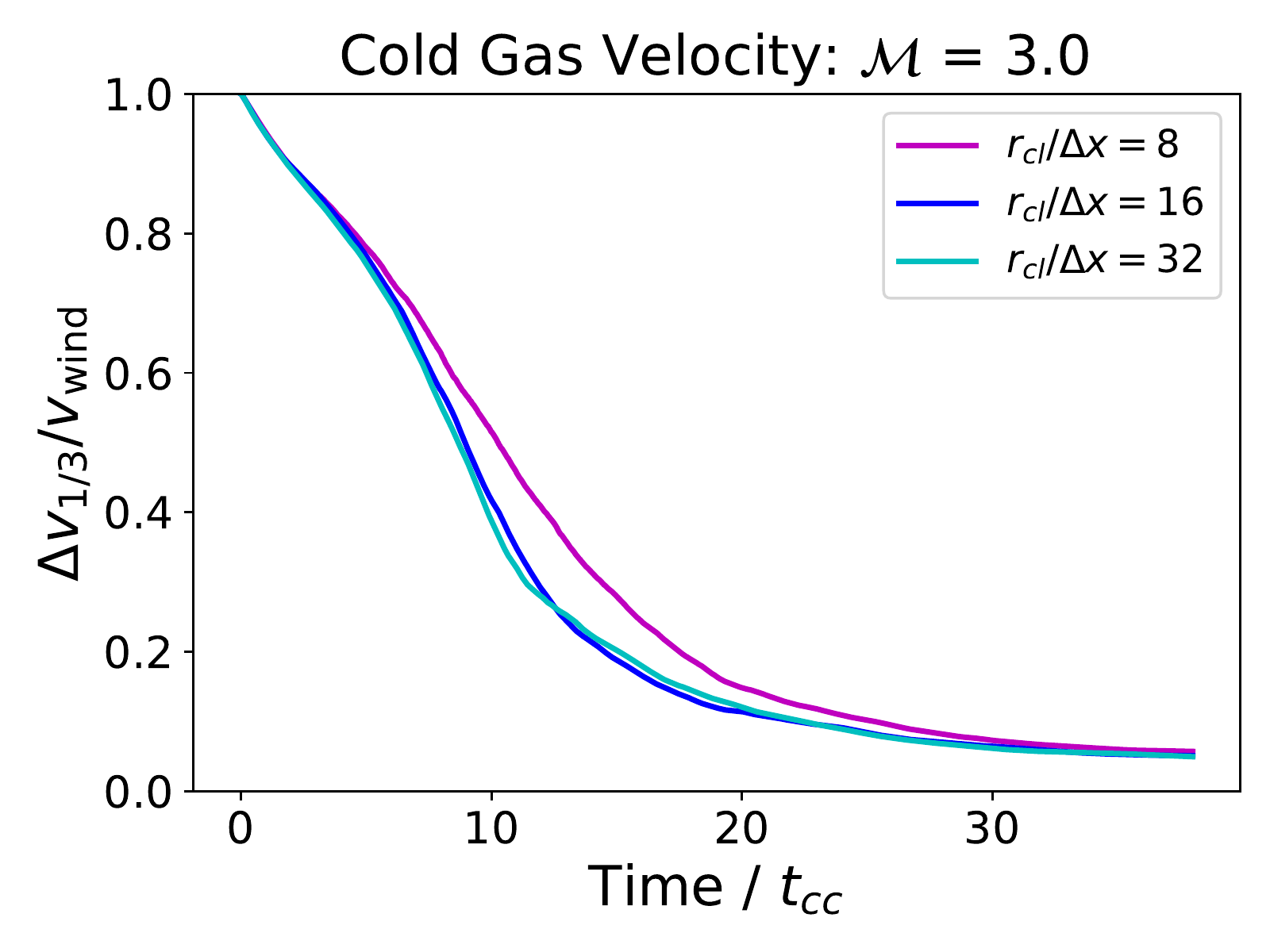}
\includegraphics[width=0.4\textwidth]{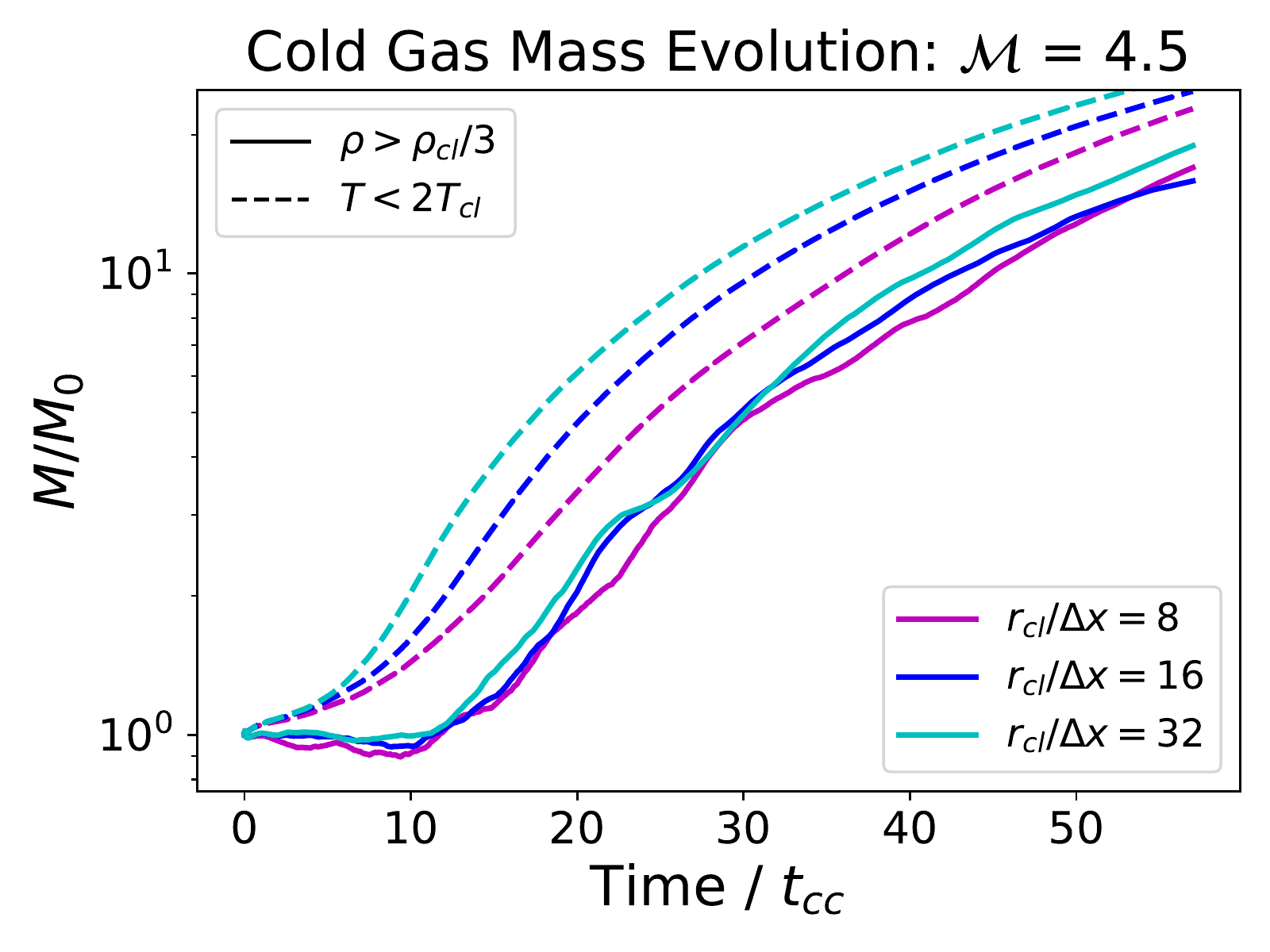}
\includegraphics[width=0.4\textwidth]{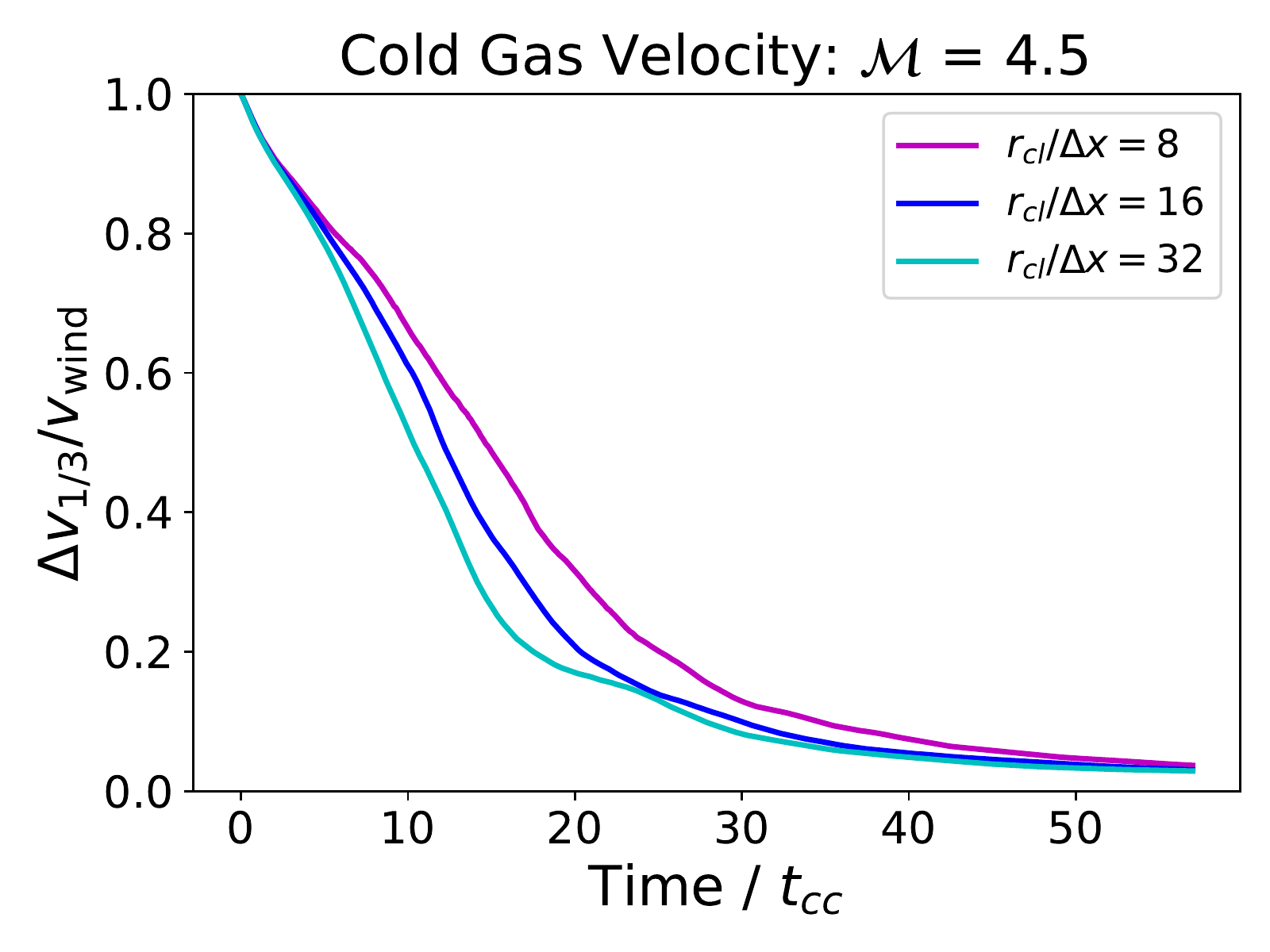}
\caption{Resolution study for the cold gas mass growth (left) and cold gas velocity (right). All simulations have $t_{cool,mix}/t_{cc} = 0.1$ but have different Mach numbers $\mathcal{M} =$ 1.5, 3, 4.5. Contrary to previous high Mach number simulations for a 3D spherical cloud, which were clearly not converged \citep{Gronke2020a}, we find that M ($\rho > \rho_{cl}/3$) is well-converged for these simulations with an effectively 2.5-D morphology. On the other hand, M ($T < 2 T_{cl}$) continues to rise with increasing resolution, likely reflecting the greater number of cells that lie at the temperature floor. The cold gas velocities appear reasonably converged at 32 cells per cloud radius, though there is a clear trend for slightly shorter drag times at higher resolution. A separate study is needed to assess convergence at even higher resolution, for higher $\mathcal{M}$ flows, and for spherical cloud geometries.}
\label{fig:resStudy}
\end{figure*}

We ran each of our $\mathcal{M} = 1.5$, $3$, and $4.5$ simulations with $t_{cool,mix}/t_{cc} = 0.1$ at resolutions of 8, 16, and 32 cells per cloud radius. Figure \ref{fig:resStudy} shows the resulting cold gas masses and average cold gas velocities. The $\mathcal{M} = 6$ simulations were only run at 8 and 16 cells per cloud radius and aren't shown here. Somewhat surprisingly, our results show good convergence, especially in the cold gas mass measured as $\rho > \rho_{cl}/3$. The $T > 2T_{cl}$ curves trend higher with increased resolution, likely reflecting increased (better-resolved) turbulent pressure support. The drag time becomes shorter at high resolution, which is possibly another indication that momentum transfer in the tail crucially affects the drag time. When compared to previous simulations of high-$\mathcal{M}$ cloud-crushing \citep{Gronke2020a}, though, these differences seem miniscule. We believe this is related to the choice of ``cloud" geometry. \cite{Gronke2020a} modeled spherical 3D clouds and found poor convergence for simulations with $\mathcal{M} > 1.5$ (see their Figure 13). Our ``2.5D" cloud geometry, motivated by our envisioned geometry of the Leading Arm at swing-out from the Clouds, appears to facilitate a ``shielding" effect that prevents shear instabilities from further stripping cloud fragments. This effect is seen as well in \cite{Melso2019} in simulations of infalling gas clouds and in \cite{Banda2020}, which models shock-multicloud interactions and reports that most bulk properties are converged even for $\mathcal{M} = 10$ simulations. That ``shielding" causes convergence, however, is pure speculation, and we plan to study this further in future work. Indeed, in these supersonic simulations, a bow shock develops, and the post-shock interactions between hot and cold phases are primarily subsonic. A clump finding routine, combined with an analysis of local shear, hence local $t_{\rm cool,mix}/t_{cc}$, will help reveal the true cause of convergence and differences compared to the 3D clouds modeled in \cite{Gronke2020a}. 

While the mass growth rate and entrainment time are reasonably well converged in these simulations, we caution that the mixing layer itself and any associated observables are \emph{not} converged. These depend also on the inclusion of thermal conduction, which does not affect the mass influx rate in turbulent mixing layer simulations \citep{Tan2021} but does affect ion column densities \citep{Tan2021Letter}.

\bibliographystyle{apj}
\bibliography{bibliography}

\end{document}